\documentclass[usegraphicx,useAMS,usenatbib]{mn2e}
\newcommand{\hb}{H$\beta$}
\newcommand{\hg}{H$\gamma$}
\newcommand{\hi}{H\,{\sc i}}

\newcommand{\kms}{km~s$^{-1}$}

\newcommand{\aip}{AI$^+$}
\newcommand{\ait}{AI$^{\rm Tot}$}
\newcommand{\cpca}{C\,{\sc iii}] PCA}
\newcommand{\mpca}{Mg\,{\sc ii} PCA}
\newcommand{\rcon}{r_{\rm con}}

\newcommand{\rwind}{r_{\rm wind}}
\newcommand{\ushape}{$\mathsf U$}
\newcommand{\wshape}{$\mathsf W$}
\newcommand{\vshape}{{\sffamily \slshape V}}
\newcommand{\vshaped}{{\sffamily \slshape V}-shaped}
\newcommand{\vshapeW}{{\sffamily \slshape V}/$\mathsf W?$}

\newcommand{\alii}{Al\,{\sc ii}}
\newcommand{\AlII}{Al\,{\sc ii}\,$\lambda$1670}
\newcommand{\aliii}{Al\,{\sc iii}}
\newcommand{\AlIII}{Al\,{\sc iii}\,$\lambda\lambda$1854,1862}

\newcommand{\cii}{C\,{\sc ii}}
\newcommand{\CII}{C\,{\sc ii}\,$\lambda$1334}
\newcommand{\ciii}{C\,{\sc iii}]}
\newcommand{\CIII}{C\,{\sc iii}]\,$\lambda$1908}
\newcommand{\Cthree}{C\,{\sc iii}*}
\newcommand{\CTHREE}{C\,{\sc iii}*\,$\lambda$1175}
\newcommand{\civ}{C\,{\sc iv}}
\newcommand{\CIV}{C\,{\sc iv}\,$\lambda\lambda$1548,1550}

\newcommand{\caii}{Ca\,{\sc ii}}

\newcommand{\feii}{Fe\,{\sc ii}}
\newcommand{\feiii}{Fe\,{\sc iii}}
\newcommand{\feiv}{Fe\,{\sc iv}}

\newcommand{\heii}{He\,{\sc ii}}
\newcommand{\HeIIsf}{He\,{\sc ii}\,$\lambda$1640}

\newcommand{\lalala}{$\lambda$$\lambda$$\lambda$}
\newcommand{\lya}{Ly$\alpha$}

\newcommand{\mgii}{Mg\,{\sc ii}}

\newcommand{\MgII}{Mg\,{\sc ii}\,$\lambda\lambda$2796,2803}

\newcommand{\Niii}{N\,{\sc iv}]}

\newcommand{\Nv}{N\,{\sc v}}
\newcommand{\NV}{N\,{\sc v}\,$\lambda\lambda$1238,1242}

\newcommand{\oi}{O\,{\sc i}}
\newcommand{\OI}{O\,{\sc i}\,$\lambda$1302}
\newcommand{\oii}{[O\,{\sc ii}]}
\newcommand{\OII}{[O\,{\sc ii}]\,$\lambda\lambda$3727,3729}
\newcommand{\oiii}{[O\,{\sc iii}]}

\newcommand{\pv}{P\,{\sc v}}
\newcommand{\PV}{P\,{\sc v}\,$\lambda\lambda$1118,1128}

\newcommand{\ovi}{O\,{\sc vi}}
\newcommand{\OVI}{O\,{\sc vi}\,$\lambda\lambda$1031,1037}
\newcommand{\SIV}{S\,{\sc iv}\,$\lambda$1062+S\,{\sc iv}*\,$\lambda$1072}
\newcommand{\Siv}{S\,{\sc iv}}

\newcommand{\SIii}{Si\,{\sc ii}}

\newcommand{\Siiii}{Si\,{\sc iii}]}
\newcommand{\SIiv}{Si\,{\sc iv}}

\newcommand{\SiII}{Si\,{\sc ii}\,$\lambda$1263}
\newcommand{\SiIII}{Si\,{\sc iii}]\,$\lambda$1892}
\newcommand{\SiIV}{Si\,{\sc iv}\,$\lambda\lambda$1393,1402}

\newcommand{\SiO}{Si\,{\sc iv}/O\,{\sc iv}]\,$\lambda$1400}

\newcommand{\aj}{Astron.~J.}% Astronomical Journal
\newcommand{\araa}{Annu. Rev. Astron. Astrophys.}% Annual Review of Astron and Astrophys
\newcommand{\apj}{Astrophys.~J.}% Astrophysical Journal
\newcommand{\apjl}{Astrophys. J. Lett.}% Astrophysical Journal, Letters
\newcommand{\apjs}{Astrophys. J. Suppl. Ser.}% Astrophysical Journal, Supplement
\newcommand{\aap}{Astron. Astrophys.}% Astronomy and Astrophysics
\newcommand{\mnras}{Mon. Not. R. Astron. Soc.}% Monthly Notices of the RAS
\newcommand{\pasj}{Publ. Astron. Soc. Jpn.}% Publications of the ASJ
\newcommand{\ssr}{Space Sci. Rev.}% Space Science Reviews
\begin{document}

\title[Quasars with Redshifted BAL Troughs]{
Broad Absorption Line Quasars with Redshifted Troughs:
High-Velocity Infall or Rotationally Dominated Outflows?
}
\author[Hall et al.]{P. B. Hall$^{1}$\thanks{E-mail: phall@yorku.ca (PBH)}, 
W. N. Brandt$^{2,3}$,
P. Petitjean$^4$,
I. P\^aris$^4$,
N. Filiz Ak$^{2,3,5}$,
Yue Shen,$^{6,16}$
\newauthor
R. R. Gibson$^7$,
\'E. Aubourg$^8$,
S. F. Anderson$^7$,
D. P. Schneider$^{2,3}$,
D. Bizyaev$^9$,
\newauthor
J. Brinkmann,$^9$,
E. Malanushenko$^9$,
V. Malanushenko$^9$,
A. D. Myers$^{10}$,
\newauthor
D. J. Oravetz$^9$,
N. P. Ross$^{11}$,
A. Shelden$^9$,
A. E. Simmons$^9$,
\newauthor
A. Streblyanska$^{12}$,
B. A. Weaver$^{13}$,
D. G. York$^{14,15}$\\ %,
$^1$Department of Physics and Astronomy, York University, Toronto, ON M3J 1P3, Canada\\
$^2$Department of Astronomy \& Astrophysics, Pennsylvania State University, University Park, PA 16802, USA\\
$^3$Institute for Gravitation and the Cosmos, Pennsylvania State University, University Park, PA 16802, USA\\
$^4$Universit\'e Paris 6, Institut d'Astrophysique de Paris, 75014, Paris, France\\
$^5$Faculty of Sciences, Department of Astronomy and Space Sciences, Erciyes University, 38039 Kayseri, Turkey\\
$^6$Carnegie Observatories, 813 Santa Barbara Street, Pasadena, CA 91101, USA\\
$^7$Department of Astronomy, University of Washington, Seattle, WA 98195, USA\\
$^8$Universit\'e Paris 7, APC, 75205, Paris, France\\
$^9$Apache Point Observatory, Sunspot, NM 88349, USA\\
$^{10}$Department of Physics and Astronomy, University of Wyoming, Laramie, WY 82071, USA\\
$^{11}$Lawrence Berkeley National Laboratory, 1 Cyclotron Road, Berkeley, CA 92420, USA\\
$^{12}$Instituto de Astrof\'{\i}sica de Canarias (IAC), E-38200 La Laguna, Tenerife, Spain\\
$^{13}$Center for Cosmology and Particle Physics, New York University, New York, NY 10003, USA\\
$^{14}$The University of Chicago, Department of Astronomy and Astrophysics, Chicago, IL 60637, USA\\
$^{15}$The University of Chicago, Enrico Fermi Institute, Chicago, IL 60637, USA\\
$^{16}$Hubble Fellow}

%\date{Accepted 1988 December 15. Received 1988 December 14; in original form 1988 October 11}

\pagerange{\pageref{firstpage}--\pageref{lastpage}} \pubyear{2012}

\maketitle

\label{firstpage}

\begin{abstract}
We report the discovery in the Sloan Digital Sky Survey
and the SDSS-III Baryon Oscillation Spectroscopic Survey
of seventeen broad absorption line (BAL) quasars with high-ionization troughs 
that include absorption redshifted relative to the quasar rest frame.
The redshifted troughs extend to velocities up to $v\simeq 12,000$~\kms\ and
the trough widths exceed 3000~\kms\ in all but one case.
Approximately 1 in 1000 BAL quasars with blueshifted \civ\ absorption also
has redshifted \civ\ absorption;
objects with \civ\ absorption present only at redshifted velocities
are roughly four times rarer.
	In more than half of our objects, redshifted absorption is seen
	in \cii\ or \aliii\ as well as \civ, making low-ionization absorption
	at least ten times more common among BAL quasars with
	redshifted troughs than among standard BAL quasars.
However, the \civ\ absorption equivalent widths in our objects are on average
smaller than those of standard BAL quasars with low-ionization absorption.

We consider several possible ways of generating redshifted absorption.
The two most likely possibilities 
may be at work simultaneously, in the same objects or in different ones.
Rotationally dominated outflows seen against a quasar's extended continuum 
source can produce redshifted and blueshifted absorption, but variability
consistent with this scenario is seen in only one of the four objects with
multiple spectra.  %two epochs of high signal-to-noise ratio spectra.
The infall of relatively dense and low-ionization gas 
to radii as small as 400 Schwarzschild radii
can in principle explain the observed range of trough profiles,
but current models do not easily explain the origin and survival of such gas.
Whatever the origin(s) of the absorbing gas in these objects, it must be located
at small radii to explain its large redshifted velocities, and thus offers a
novel probe of the inner regions of quasars.
\end{abstract}


\section{Introduction} \label{intro}

The most luminous active galactic nuclei (AGN) are known as quasars.
Broad absorption line (BAL) quasars are those quasars which show ultraviolet
absorption troughs thousands of km\,s$^{-1}$ wide 
\nocite{lyn67,allenbal}(e.g., {Lynds} 1967; {Allen} {et~al.} 2011), widths for which 
the accretion process in quasars is thought to be ultimately responsible.
The traditional minimum velocity width for a BAL trough is 2000~\kms\ at 10\%
depth below the continuum \nocite{wea91}({Weymann} {et~al.} 1991).  Objects with narrower intrinsic
absorption troughs, down to 500~\kms\ wide, are often referred to as
mini-BAL quasars \nocite{hs04}(e.g., {Hamann} \& {Sabra} 2004).
Objects with intrinsic troughs $<$500~\kms\ wide are
referred to as narrow absorption line (NAL) quasars \nocite{gb08}(see, e.g., {Ganguly} \& {Brotherton} 2008).
Intrinsic troughs are those which arise from gas 
connected with the accretion process onto the quasar,
as opposed to arising elsewhere in the host galaxy,
but it is not always possible to determine the origin of a given
absorption system. %(e.g., \nocite{hs04}{Hamann} \& {Sabra} 2004).
If studying intrinsic quasar absorption regardless of velocity width is the goal, 
less restrictive minimum width definitions for BAL troughs can be established
\nocite{sdss123,trump06}({Hall} {et~al.} 2002; {Trump} {et~al.} 2006).  However, such definitions also run the risk of
increased contamination from intervening absorption systems \nocite{ksgc08}({Knigge} {et~al.} 2008).
Intervening absorption systems unrelated to the quasar will generally have
widths $<$500~\kms, but blending and clustering can produce apparently broader
absorption profiles, especially in spectra where the resolution or 
signal-to-noise ratio (or both) is relatively low.

BAL quasars are often subdivided into three subtypes depending on the 
ionization stages seen in absorption \nocite{sdss123}(e.g., {Hall} {et~al.} 2002).
High-ionization BAL quasars (HiBALs) have absorption from
\CIV\ (all wavelengths in \AA), \NV, and \OVI,
and are the most common subtype.
Low-ionization BAL quasars (LoBALs) have high-ionization absorption
plus absorption from \CII, \AlIII, and/or \MgII.
Iron low-ionization BAL quasars (FeLoBALs) have high- and low-ionization 
absorption plus absorption from excited states of \feii\ and/or \feiii,
and are the least common subtype.

BAL troughs are found over a wide range of velocities relative to the quasar,
both collectively and in individual objects.  The traditional
velocity range over which the strength of BAL troughs detected in \CIV\ is 
evaluated is $-$25000~\kms\ to $-$3000~\kms\ \nocite{wea91}({Weymann} {et~al.} 1991), where we adopt 
the convention that negative velocities denote absorption
blueshifted from the quasar redshift.
The high-velocity cutoff was established
to avoid confusion with \SiO\ emission and \SiIV\ BAL troughs;
the low-velocity 
to avoid strong associated narrow-line \civ\ complexes. 

In this paper we adopt the convention that quasars with troughs
$>$2000~\kms\ wide, regardless of the trough velocity offset,
are BAL quasars.
In that case, the highest velocity known for a BAL outflow is either
$-$56000~\kms\ in PG~2302+029 \nocite{jea96}({Jannuzi} {et~al.} 1996)
or $-$66000~\kms\ in H~1414+089 \nocite{fol83}({Foltz} {et~al.} 1983).

Extending the velocity limits within which BAL troughs are measured to
the widest possible wavelength range accessible in a given quasar sample is
necessary if the goal is to study all broad intrinsic absorption in quasars.
One drawback of that approach is contamination of low-velocity troughs with
complexes of associated narrow-line absorption at the systemic redshift;
studies at spectral resolution sufficient to resolve such complexes
can help account for this effect.
Another drawback is contamination of high-velocity troughs in one transition
with low-velocity troughs of a shorter-wavelength transition
(e.g., high-velocity \CIV\ with low-velocity \SiIV).
Physical considerations based on elemental abundances and ionization fractions
can be used to correcly identify troughs in such cases; for example, 
\civ\ absorption can be seen without accompanying \SIiv, but the reverse
is not true (see, e.g., \nocite{2010ApJ...713..220G}{Gibson} {et~al.} 2010).
As long as care is taken to account for the above effects,
the ideal starting velocity for measuring the strength of BAL troughs
is easy to define: the starting velocity for a given trough
should be chosen to encompass the entire trough in that transition.

In principle, studying BAL troughs may require starting
velocities which are redshifted relative to the quasar rest frame.
\nocite{sdss123}{Hall} {et~al.} (2002) presented two %$z\sim 0.9$ 
quasars with \mgii\ troughs showing absorption both blueshifted
and redshifted\footnote{In \nocite{sdss123}{Hall} {et~al.} (2002) we referred to these objects as
having longward-of-systemic absorption.  Here we drop the use of `longward'
in favor of `redshifted'.  The latter term can be ambiguous, but is valid here
in the sense that along our line of sight at least some of 
the ultraviolet (UV) absorption present in each of these quasars
is redshifted relative to the quasar's systemic redshift.}
relative to the quasar rest frame: 
SDSS J112526.12$+$002901.3 (J1125) %(0281-51614-427) nmf
and SDSS J112828.31$+$011337.9 %(0281-51614-523 and 0512-51992-123) nmf,nmf
(J1128).
We suggested that these are systems where the quasar's extended ultraviolet
continuum source is seen through an outflow 
which is dominated along our line of sight by its rotational velocity.
This possibility had earlier been raised in the context of NAL outflows by
\nocite{gan01}{Ganguly} {et~al.} (2001, their \S 5.3).

Here we use data from the Sloan Digital Sky Survey (SDSS; \nocite{yor00}{York} {et~al.} 2000)
and the SDSS-III \nocite{sdss3}({Eisenstein} {et~al.} 2011) Baryon Oscillation Spectroscopic Survey 
(BOSS; \nocite{bossover}{Dawson} {et~al.} 2013),\footnote{http://www.sdss3.org/surveys/boss.php}
discussed in \S\,\ref{data}, to present seventeen quasars with 
broad absorption troughs along our line of sight which are redshifted 
relative to the quasar's systemic redshift and are 
seen in multiple transitions including \civ, \SIiv\ and \Nv\ (\S\,\ref{cands}).
We consider various properties of this population of objects in \S\,\ref{props},
discuss several possible explanations for these objects in \S\,\ref{poss},
and discuss some implications and tests of those explanations in \S\,\ref{end}.
Notes on confirmed, candidate and rejected objects are presented in the
Appendices.

\section{Data}\label{data}

The SDSS-I and SDSS-II surveys
used two fiber-fed, double spectrographs to obtain resolution $R\sim$2100
spectra over 3800--9200\,\AA\ for 
$\sim$10$^6$ galaxies and $\sim$10$^5$ quasar candidates \nocite{sdss85,dr7}({Stoughton} {et~al.} 2002; {Abazajian} {et~al.} 2009).
The SDSS-III initiative consists of four surveys including the BOSS. 
The BOSS will obtain spectra of $\sim$1.5$\times$10$^6$ 
luminous red galaxies and $\sim$1.5$\times$10$^5$ quasars at $z>2.2$
\nocite{bossover}({Dawson} {et~al.} 2013).
All known $z>2.2$ quasars in the BOSS footprint which are point sources in
SDSS imaging \nocite{fuk96,gun98,dr8,dr8err}({Fukugita} {et~al.} 1996; {Gunn} {et~al.} 1998; {Aihara} {et~al.} 2011a,b), 
including BAL quasars, are being targeted \nocite{bossQTS}({Ross} {et~al.} 2012).
In addition, $\sim$2000 known BAL quasars %at $z<2.2$ 
with SDSS spectra\footnote{We refer to spectra taken during SDSS-I 
or SDSS-II as SDSS spectra, and spectra taken during SDSS-III as BOSS spectra.}
are being targeted for reobservation via a BOSS ancillary project
\nocite{bossover}({Dawson} {et~al.} 2013).  Together,
the above observations will enable unprecedented studies of BAL quasar
variability on multi-year timescales (e.g., \nocite{2012ApJ...757..114F}{Filiz Ak} {et~al.} 2012).

The BOSS uses the same 2.5m Sloan Foundation telescope \nocite{gun06}({Gunn} {et~al.} 2006) 
as SDSS-I/II did, but the fiber-fed spectrographs have been upgraded with 
1000 2$\arcsec$ optical diameter fibers instead of 640 3$\arcsec$ ones,
improved optics, higher throughput gratings over a wider spectral range of 
3600--10400\,\AA\ at a resolution $1300<R<3000$, 
and new CCDs with improved blue and red response \nocite{bosssmee}({Smee} {et~al.} 2012).
The first BOSS spectra have been publicly released as part of 
the SDSS Data Release Nine \nocite{dr9}(DR9; {Ahn} {et~al.} 2012).

The BOSS spectra shown herein are from the BOSS reduction pipeline version
v5\_4\_45 \nocite{bosspipe}({Bolton} {et~al.} 2012) for all but one object.\footnote{The exception is
J1628.  Its spectrum herein is from version v5\_6\_0, the version in use when 
J1628 was observed in Sept. 2012.} %(long after DR9 was made public).}
That version of the pipeline has a systematic flux excess shortward of 
4100\,\AA\ at a level of $\simeq$2.5\% of the flux at 5600\,\AA, increasing to
$\simeq$12.5\% at 3600\,AA\ \nocite{bossdr9q}(\S\,2.4.1 of {P{\^a}ris} {et~al.} 2012); 
this fact rarely affects the conclusions we draw from our spectra, 
but should be kept in mind.
All SDSS spectra shown herein are the improved sky-subtraction versions of \nocite{wh10}{Wild} \& {Hewett} (2010).
The BOSS and SDSS spectra shown herein have not been corrected for Galactic extinction.
In all spectra, we interpolated over narrow regions near strong night sky lines 
if large flux residuals from the sky subtraction were present.

\section{BAL Quasars with Redshifted Absorption}\label{cands}

\setcounter{table}{0}
\begin{table*}
 \centering
 \begin{minipage}{177mm}
 \caption{Confirmed and Candidate BAL Quasars with Redshifted Absorption}
  \begin{tabular}{@{}lrrccclccc@{}}
\hline
Name~(SDSS~J)        &       RA   &     DEC   &$i_{PSF}$ & Redshift          & Redshift & BAL  & Trough      & FIRST           & $M_i$ \\
                     &  (J2000)   & (J2000)   &  mag.    & $z$               & source   & type & shape       & mJy beam$^{-1}$ &       \\
\hline
\multicolumn{10}{c}{Quasars with redshifted \civ\ absorption}\\
\hline
002825.02$+$010604.2 &   7.104254 &    1.101175 & 20.509 & 4.1152$\pm$0.0102 & HW10     & Lo   & $\mathsf U$ & $<$0.73 &   $-$27.00 \\
014829.81$+$013015.0 &  27.124227 &    1.504184 & 20.415 &  3.061$\pm$0.008  & \cpca\   & Lo   & \vshape     & $<$1.06 &   $-$26.40 \\
080544.99$+$264102.9 & 121.437461 &   26.684147 & 21.745 &  2.703$\pm$0.008  & \cpca\   & Lo   & \vshape     & $<$0.96 &   $-$24.81 \\
082818.81$+$362758.6 & 127.078399 &   36.466304 & 19.767 &  2.366$\pm$0.005  & full PCA & Lo   & $\mathsf W$ & $<$0.93 &   $-$26.52 \\
083030.26$+$165444.7 & 127.626083 &   16.912417 & 19.160 & 2.4345$\pm$0.0005 & inspection & Lo   & $\mathsf U$ & $<$0.99 &   $-$27.17 \\ %-390,0 
094108.92$-$022944.7 & 145.287167 & $-$2.495776 & 20.197 &  3.446$\pm$0.002  & \cpca\   & Hi   & $\mathsf W$ & $<$1.01 &   $-$26.90 \\
101946.08$+$051523.7 & 154.942000 &    5.256583 & 20.978 &  2.452$\pm$0.001  & \cpca\   & FeLo & $\mathsf U$ & $<$1.00 &   $-$25.34 \\
103412.33$+$072003.6 & 158.551375 &    7.334333 & 18.161 & 1.6893$\pm$0.0018 & HW10     & Lo$^*$ & $\mathsf W$ & $<$1.01 &   $-$26.85 \\ %1.6893 0.0018
114655.05$+$330750.1 & 176.729393 &   33.130585 & 19.294 &  2.780$\pm$0.001  & \cpca\   & Lo?  & \vshape     & $<$0.95 &   $-$27.26 \\
114756.00$-$025023.4 & 176.983335 & $-$2.839839 & 19.278 & 2.5559$\pm$0.0056 & HW10     & Lo   & \vshapeW    & $<$1.03 &   $-$27.17 \\
132333.01$+$004633.8 & 200.887578 &    0.776082 & 20.286 &  2.455$\pm$0.038  & inspection & Lo?  & $\mathsf U$ & $<$0.98 &   $-$26.02 \\
143945.28$+$044409.2 & 219.938667 &    4.735889 & 20.850 &  2.492$\pm$0.001  & \cpca\   & Lo?  & \vshape     & $<$1.00 &   $-$25.56 \\
144055.59$+$315051.7 & 220.231641 &   31.847709 & 20.295 &  2.954$\pm$0.015  & \cpca\   & Lo   & $\mathsf W$ & $<$0.96 &   $-$26.42 \\
162805.80$+$474415.6 & 247.024167 &   47.737667 & 18.481 & 1.5949$\pm$0.0019 & HW10     & Hi   & $\mathsf W$ & $<$0.93 &   $-$26.38 \\ %1.5949 0.0019 \\
170953.28$+$270516.6 & 257.472000 &   27.087944 & 20.560 &  3.126$\pm$0.003  & \cpca\   & Lo   & $\mathsf U$ & $<$0.98 &   $-$26.36 \\
172404.44$+$313539.6 & 261.018500 &   31.594333 & 19.600 &  2.516$\pm$0.001  & \cpca\   & Lo   & $\mathsf U$ & $<$0.95 &   $-$26.86 \\
215704.26$-$002217.7 & 329.267750 & $-$0.371583 & 20.071 &  2.240$\pm$0.002  & \cpca\   & Lo   & \vshape     & $<$0.73 &   $-$26.18 \\
\hline
\multicolumn{10}{c}{Quasars with redshifted \mgii\ absorption}\\
\hline
112526.12$+$002901.3 & 171.358833 &    0.483694 & 17.896 & 0.8633$\pm$0.0007 & HW10     & FeLo & \vshape     & $<$0.99 &   $-$25.54 \\ %0.8633 0.0007 %me: 0.8654/1 
112828.31$+$011337.9 & 172.117958 &    1.227194 & 18.366 & 0.8932$\pm$0.0007 & HW10     & FeLo & \vshape     & $<$0.98 &   $-$25.16 \\ %0.8932 0.0007 %me: 0.8931/1
\hline
\multicolumn{10}{c}{Quasars with candidate redshifted \civ\ or \mgii\ absorption}\\
\hline
005030.13$+$023915.0 &  12.625542 &    2.654167 & 19.987 &  2.118$\pm$0.001  & full PCA & Lo   & $\mathsf U$ & $<$0.93 &   $-$25.94 \\
123901.00$+$014813.4 & 189.754167 &    1.803722 & 19.973 &  2.413$\pm$0.001  & inspection & Hi   & \vshape     & $<$0.76 &   $-$25.85 \\
131637.26$-$003636.0 & 199.155275 & $-$0.610007 & 18.049 & 0.9304$\pm$0.0007 & HW10     & Fe?Lo & $\mathsf W$ & 2.08$\pm$0.14 &   $-$25.55 \\ 
134243.87$+$362301.9 & 205.682805 &   36.383885 & 21.340 & 2.6917$\pm$0.0004 & inspection & Hi   & $\mathsf W$ & $<$0.95 &   $-$25.14 \\ 
163319.76$+$190856.7 & 248.332333 &   19.149083 & 17.412 &  1.935$\pm$0.001  & \mpca\   & Lo?  & \vshape     & 2.34$\pm$0.15 & $-$28.31 \\
170456.42$+$232825.7 & 256.235083 &   23.473806 & 18.918 & 3.4500$\pm$0.0005 & inspection & Lo   & $\mathsf U$ & $<$0.94 &   $-$28.22	\\ %3.4100 0.0088 \\
213342.06$+$071408.7 & 323.425278 &    7.235772 & 21.416 &  2.165$\pm$0.005  & inspection & Hi   & \vshape     & $<$0.83 &   $-$24.55 \\
\hline
\end{tabular}\\
The $i_{PSF}$ column gives the PSF magnitude in the $i$ band.
Redshifts (see \S\,\ref{zsys}) are given with uncertainties derived from the emission lines 
used to measure the redshifts; {\em not} included are the systematic uncertainties related
to the shifts of those lines from the systemic redshift of each quasar.
Redshift sources are
inspection (for our own inspection redshifts),
full, \ciii\ or \mgii\ PCA (Principal Component Analysis; \nocite{bossdr9q}{P{\^a}ris} {et~al.} 2012),
and HW10 \nocite{hw10}({Hewett} \& {Wild} 2010).
The BAL type is Hi for high-ionization, Lo for low-ionization,
and FeLo for iron low-ionization (\S~\ref{intro}).
Trough shapes are discussed in \S~\ref{uvw}.
The FIRST column gives the peak flux or limit in the FIRST catalog \nocite{bwh95,firstplus}({Becker}, {White} \& {Helfand} 1995; {Hodge} {et~al.} 2011).
The $M_i$ column gives the absolute $i$-band magnitude from 
\nocite{bossdr9q}{P{\^a}ris} {et~al.} (2012) (\nocite{dr7q}{Schneider} {et~al.} (2010) for J1628).\\
$^*$ J1034 has blueshifted low-ionization absorption but no clear redshifted low-ionization absorption.
\label{tab1}
\end{minipage}
\end{table*}

\begin{figure*} %%\vspace*{174pt} \makebox[\textwidth]{
\includegraphics[angle=0, width=0.490\textwidth]{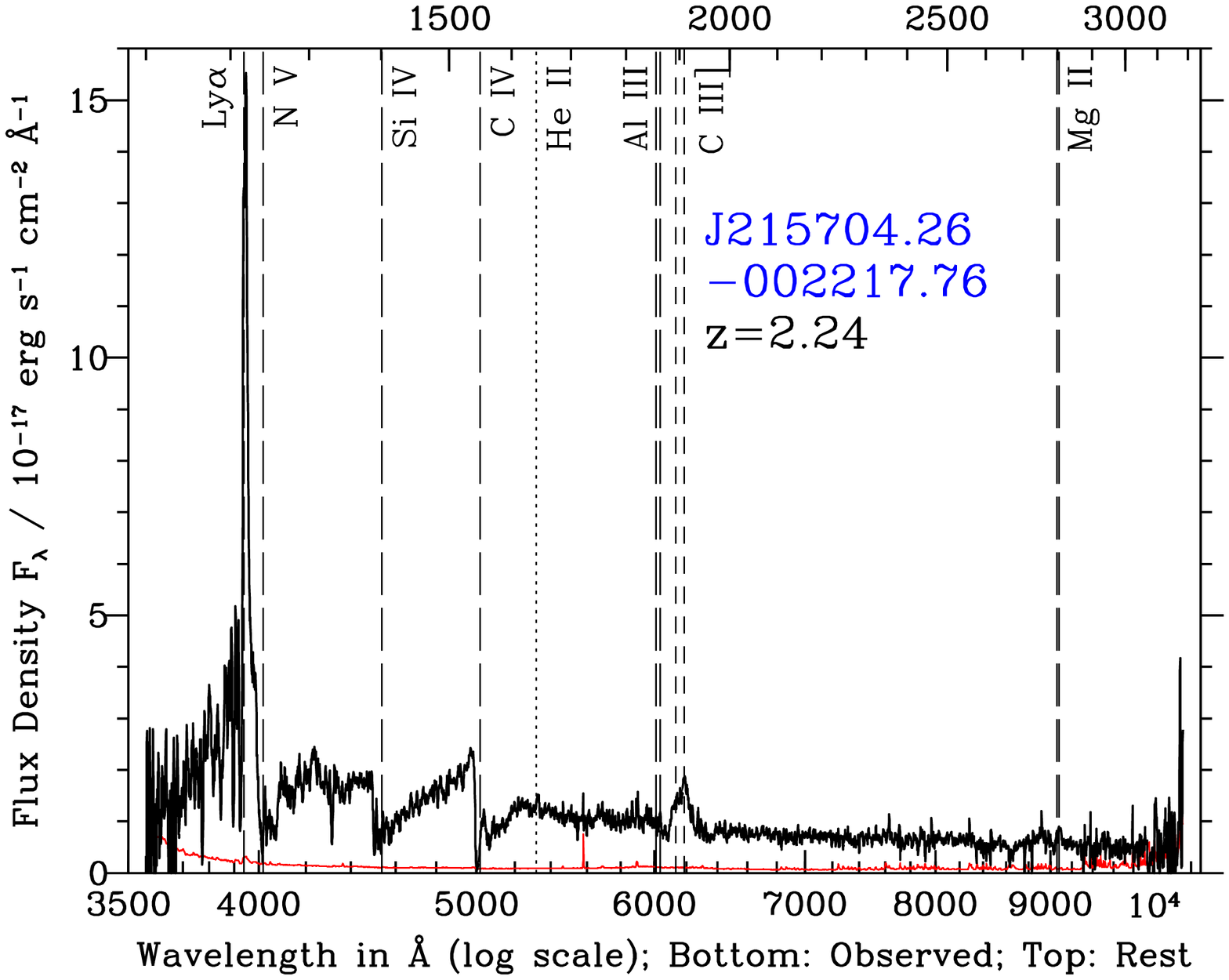} %}
\includegraphics[angle=0, width=0.490\textwidth]{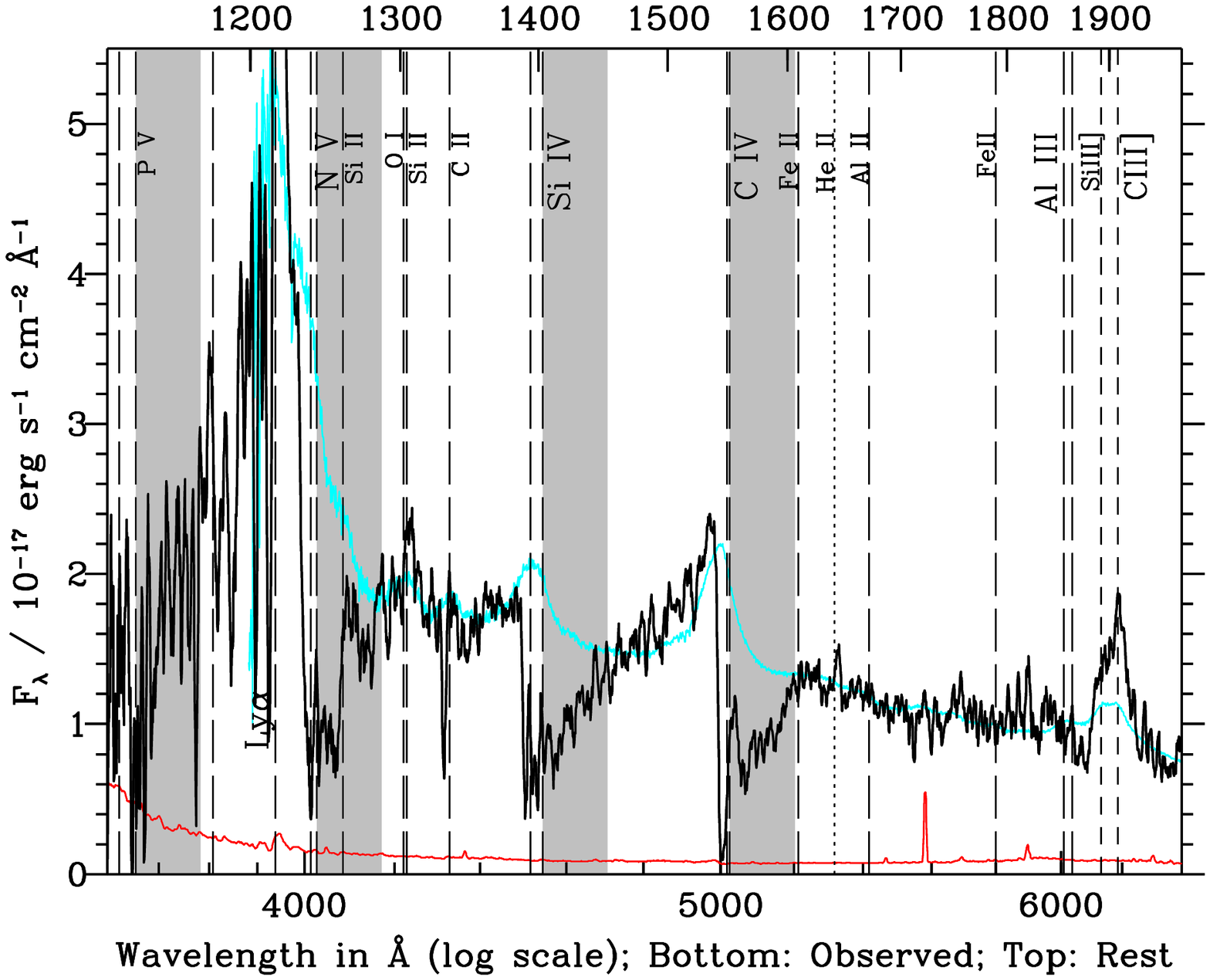} %}
\caption{(Left) The full BOSS spectrum of J2157 
(black, with error array in red); both the spectrum
and error array have been smoothed with a five-pixel boxcar filter.  
(Right) Detailed view of the spectrum.
For visual comparison, the cyan line plots the largest-\civ-blueshift 
composite spectrum from Richards et~al. (2002), 
with its spectral slope adjusted to match the bluer slope of J2157.
Compared to the composite, J2157 has stronger \ciii+\Siiii\ emission 
and stronger blueshifted \civ\ and \lya+\Nv\ emission.
The long-dashed vertical lines show the locations in the quasar rest frame
of potentially absorbing transitions, each labeled at the top of the panel.
The short-dashed lines show the locations of transitions seen only in emission
(\SiIII\ and \CIII).
The dotted line shows the wavelength of \HeIIsf, which in practice is only
seen in emission but could in principle be seen in absorption and be confused
with redshifted \civ\ absorption.
The shaded regions show the wavelengths where redshifted absorption is expected
in the long-wavelength member of the doublets \PV, \NV, \SiIV\ and \CIV,
based on the relative velocities over which such absorption is observed 
in \civ.  No alternative identification of the troughs is plausible.
The wavelengths of the two strongest troughs between the \lya\ and 
\civ\ emission lines are not a good match to blueshifted \SIiv\ and \civ; 
moreover, the trough longward of the 
\civ\ emission line has no plausible identification in that interpretation.}
\label{f_zoom} \end{figure*}

\begin{figure*} %%\vspace*{174pt} \makebox[\textwidth]{ % 0.497
\includegraphics[angle=0, width=0.497\textwidth]{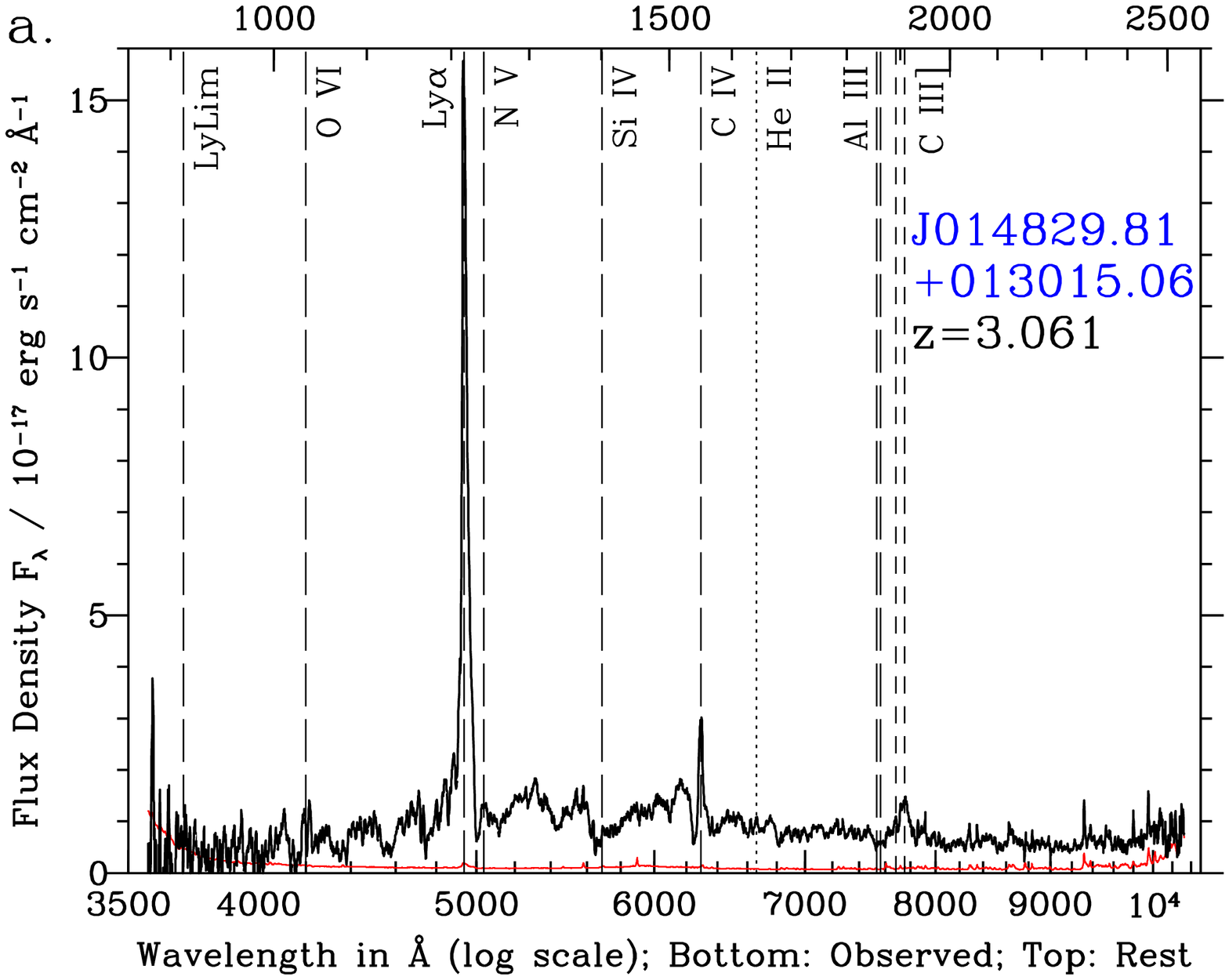} %35--105
\includegraphics[angle=0, width=0.497\textwidth]{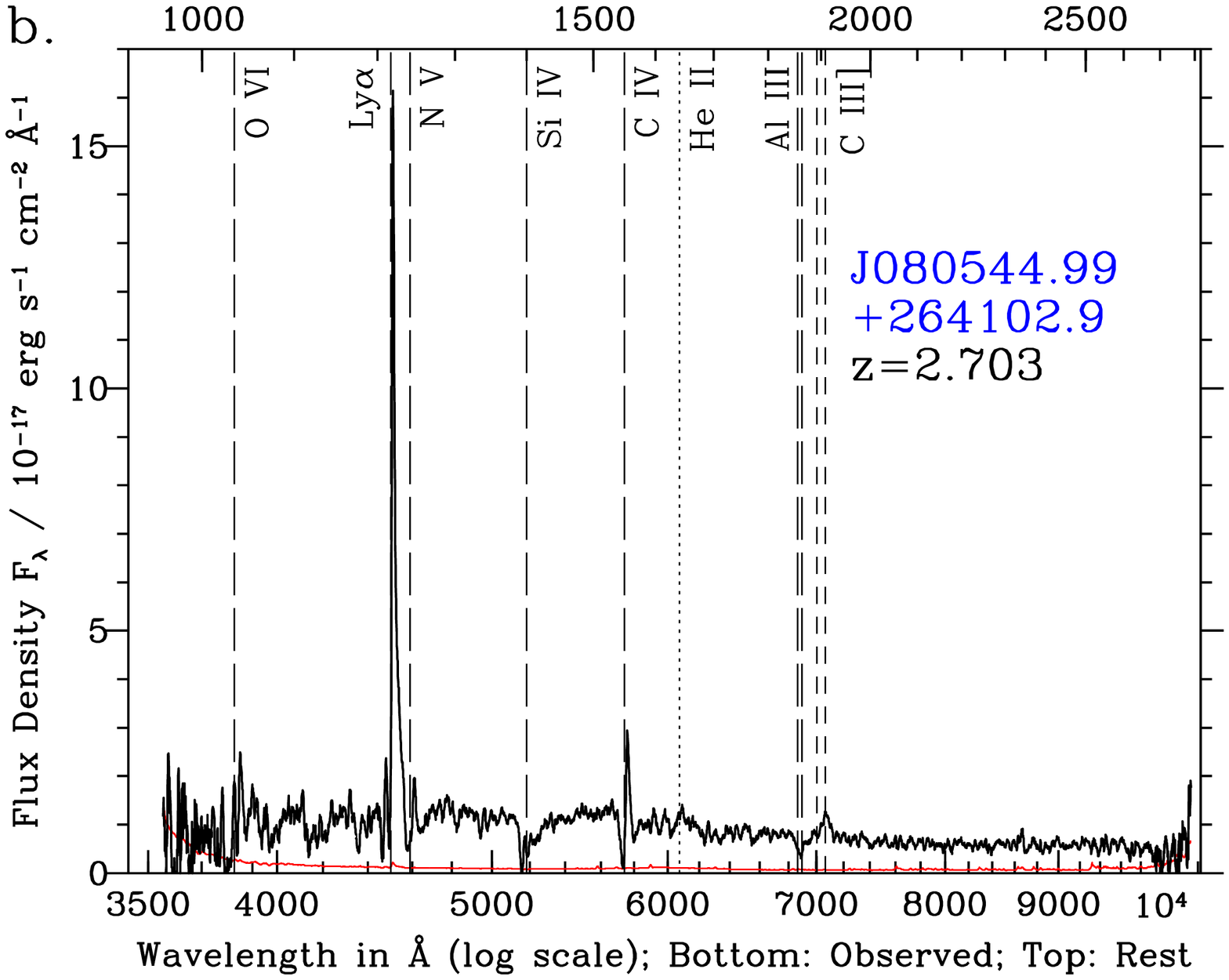} %35--105
\includegraphics[angle=0, width=0.497\textwidth]{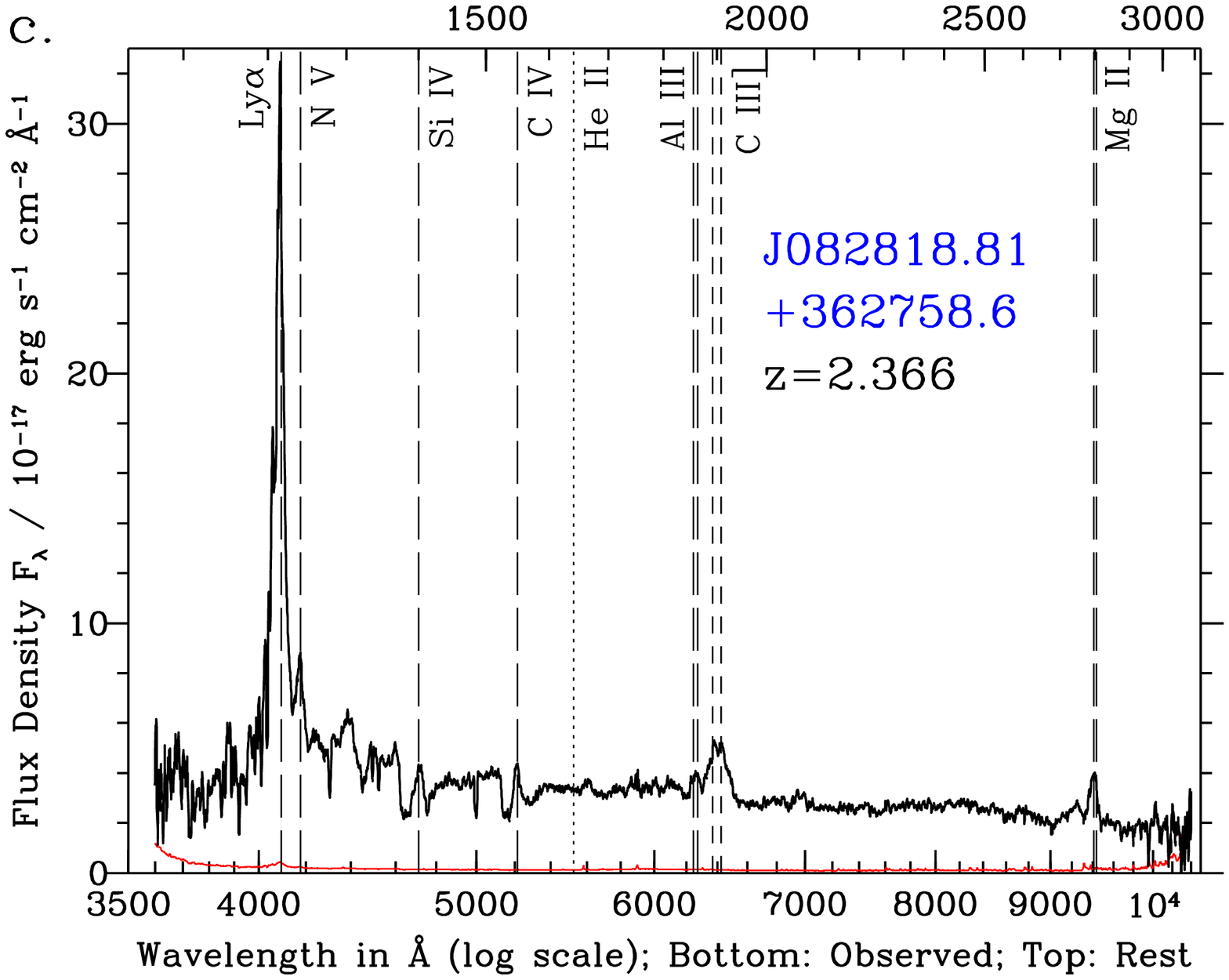} %35--105
\includegraphics[angle=0, width=0.497\textwidth]{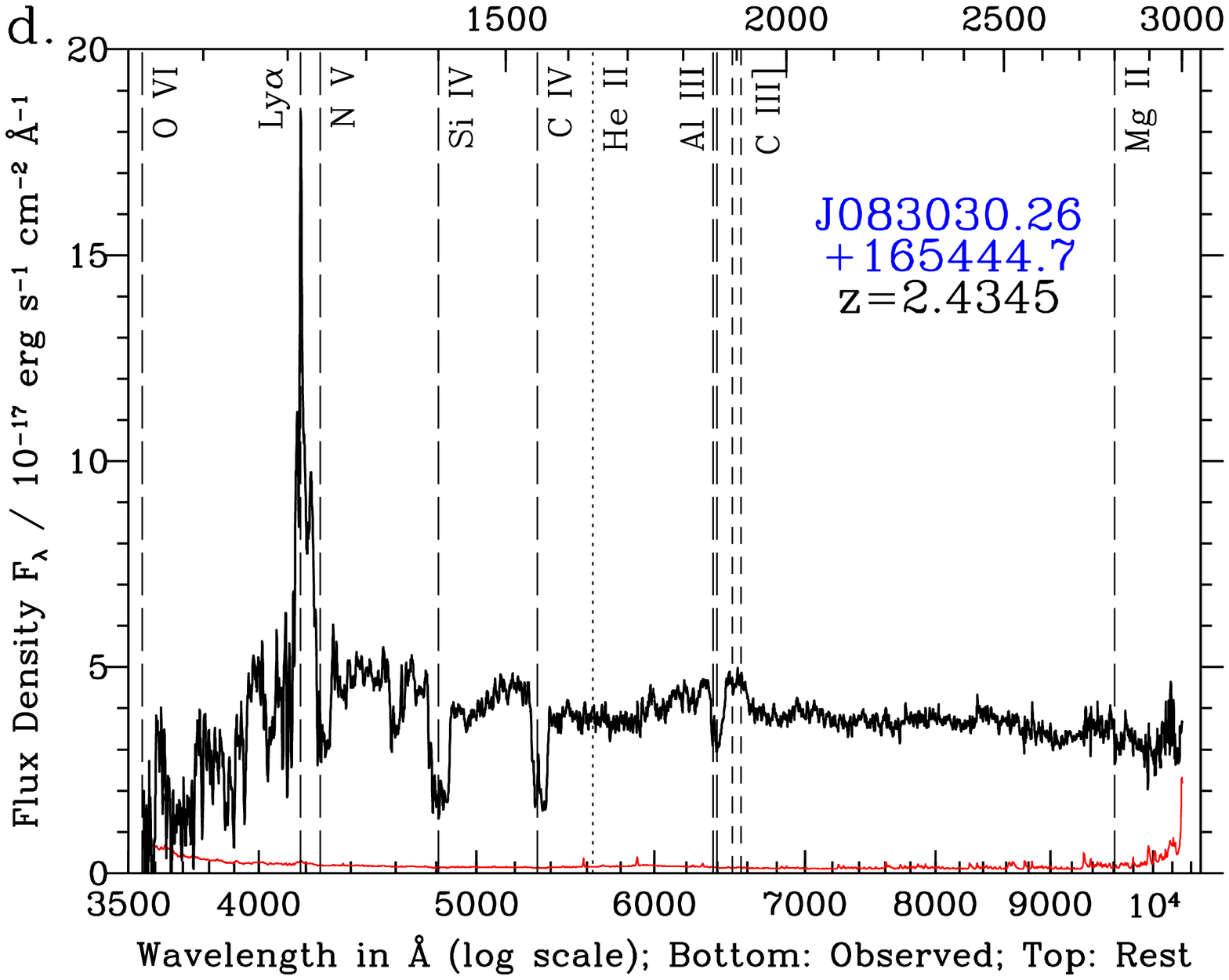} %35--105
\includegraphics[angle=0, width=0.497\textwidth]{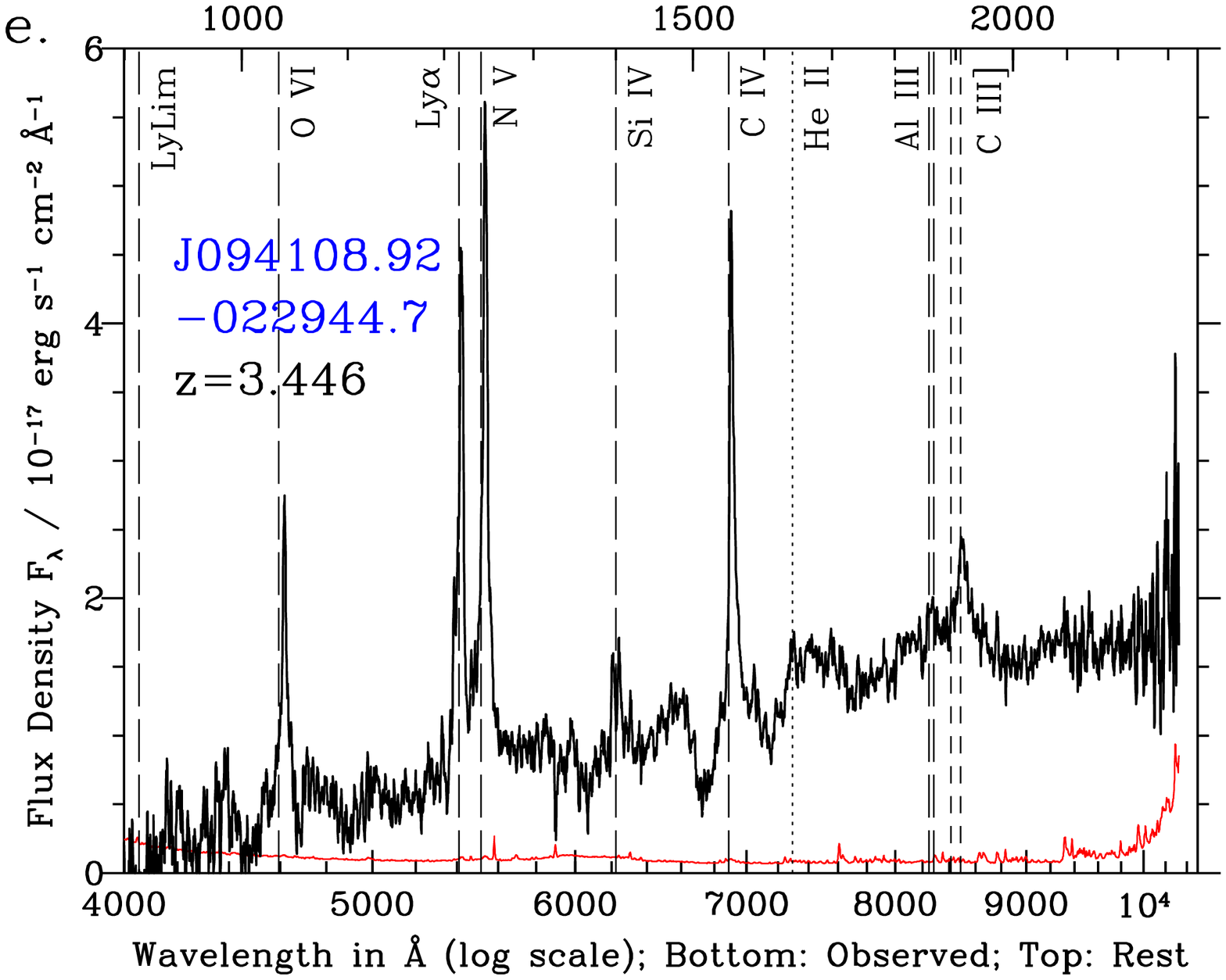} %35--105
\includegraphics[angle=0, width=0.497\textwidth]{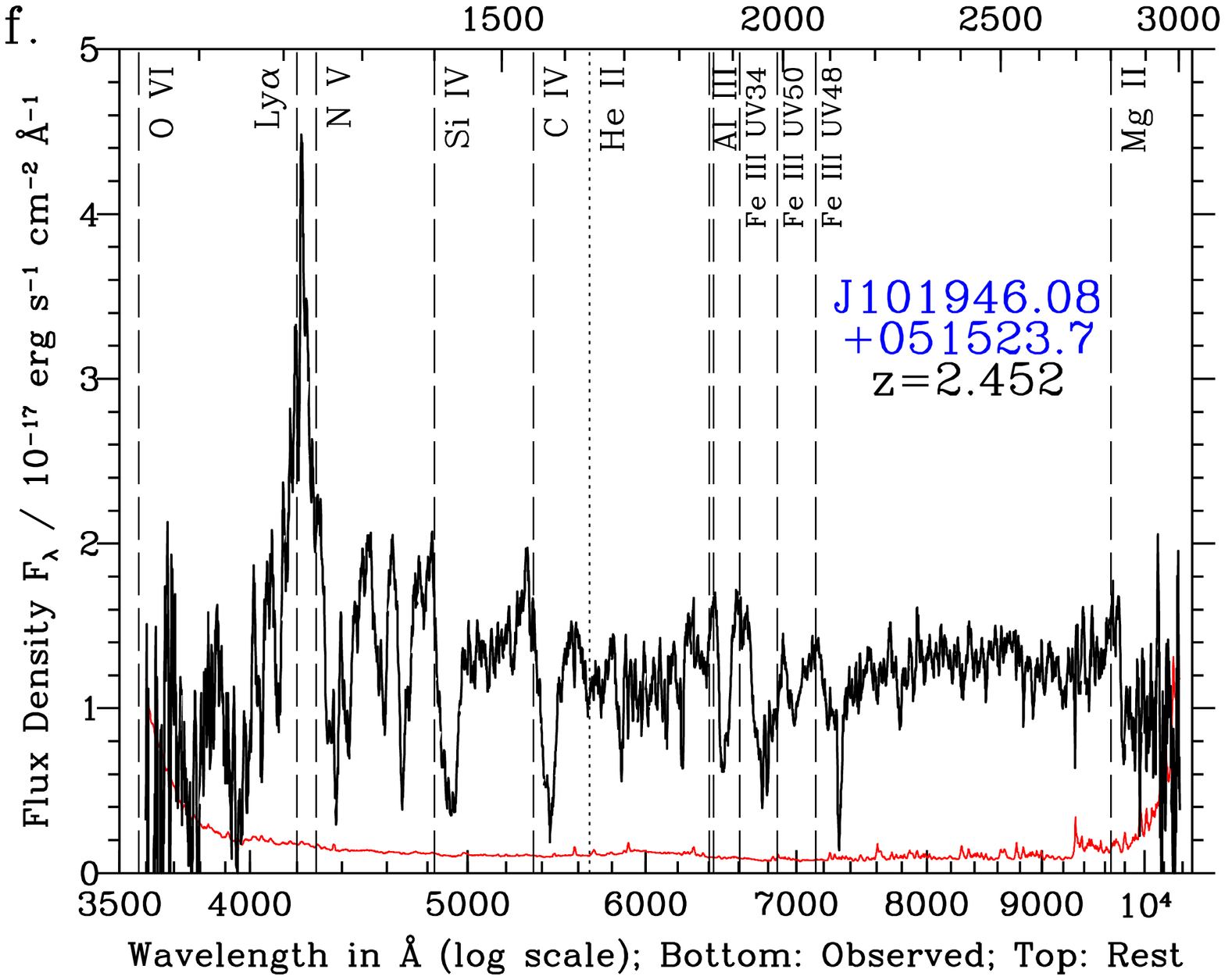} %35--105
\caption{BOSS discovery spectra of other quasars with redshifted high-ionization absorption.
The spectra and error arrays (shown in red) are smoothed with weighted-average
boxcar filters of a different width for each quasar,
depending on the signal-to-noise ratio of the spectrum.
The long-dashed vertical lines show the locations in the quasar rest frame
of potentially absorbing transitions, each labeled at the top of the panel.
The short-dashed lines show the locations of transitions seen only in emission
(\Siiii\ and \ciii).
The dotted line shows the wavelength of \HeIIsf, which in practice is only
seen in emission but could in principle be seen in absorption.
Note that the rise in the spectrum of J1439 (panel i)
at the longest observed wavelengths is spurious.}\label{f_cands} \end{figure*}

\begin{figure*} %%\vspace*{174pt} \makebox[\textwidth]{ % 0.497
\includegraphics[angle=0, width=0.497\textwidth]{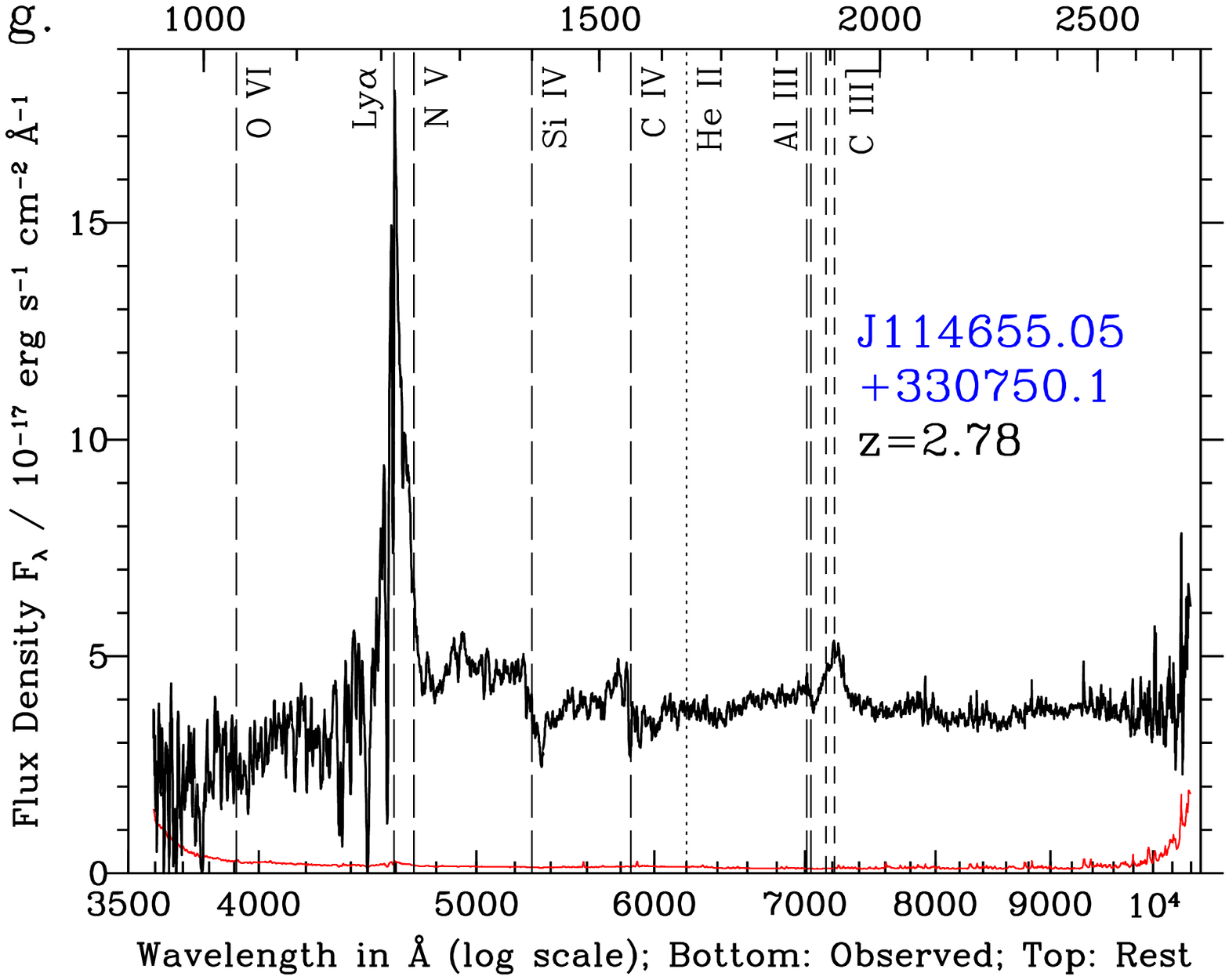} %35--105
\includegraphics[angle=0, width=0.497\textwidth]{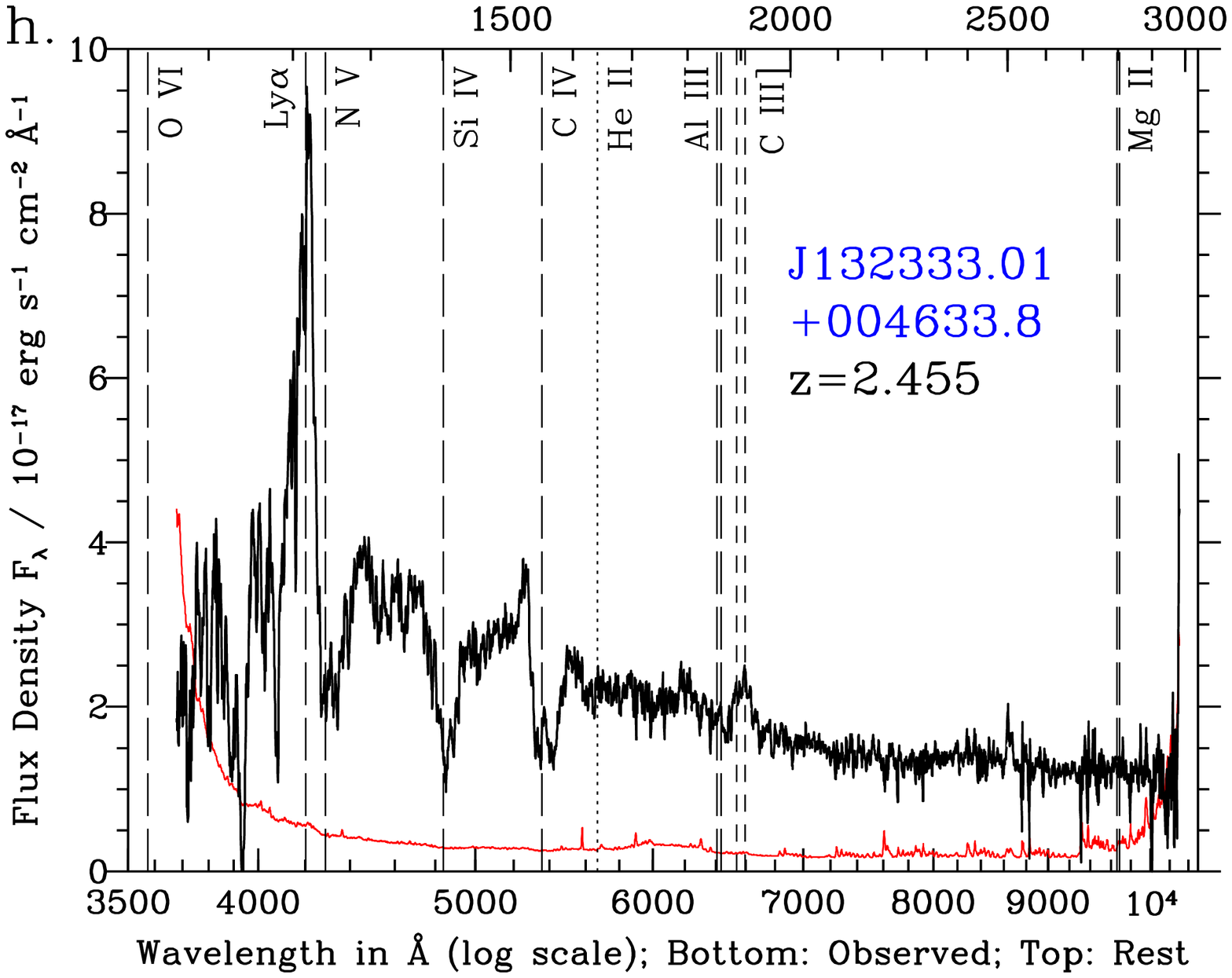} %35--105
\includegraphics[angle=0, width=0.497\textwidth]{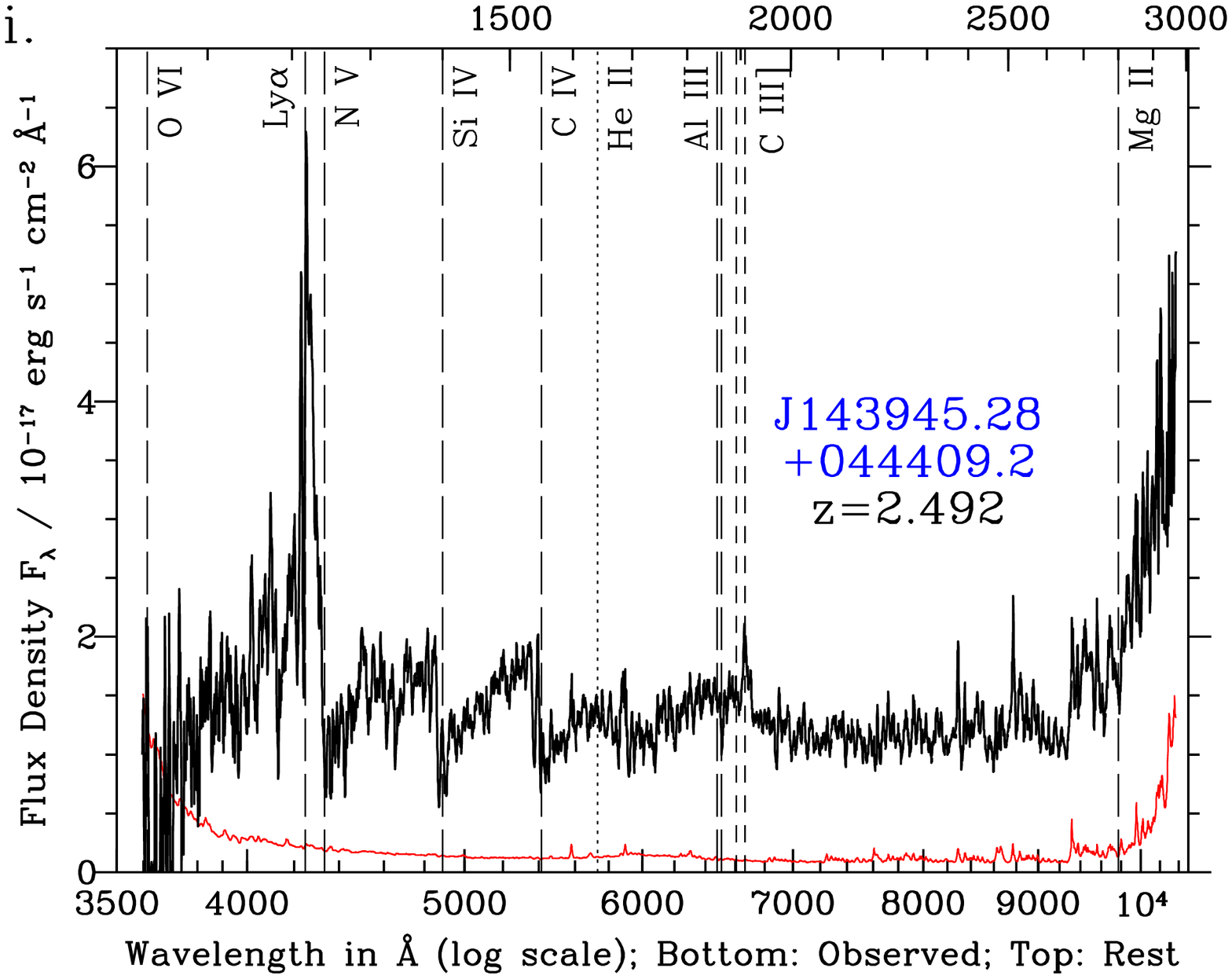} %35--105
\includegraphics[angle=0, width=0.497\textwidth]{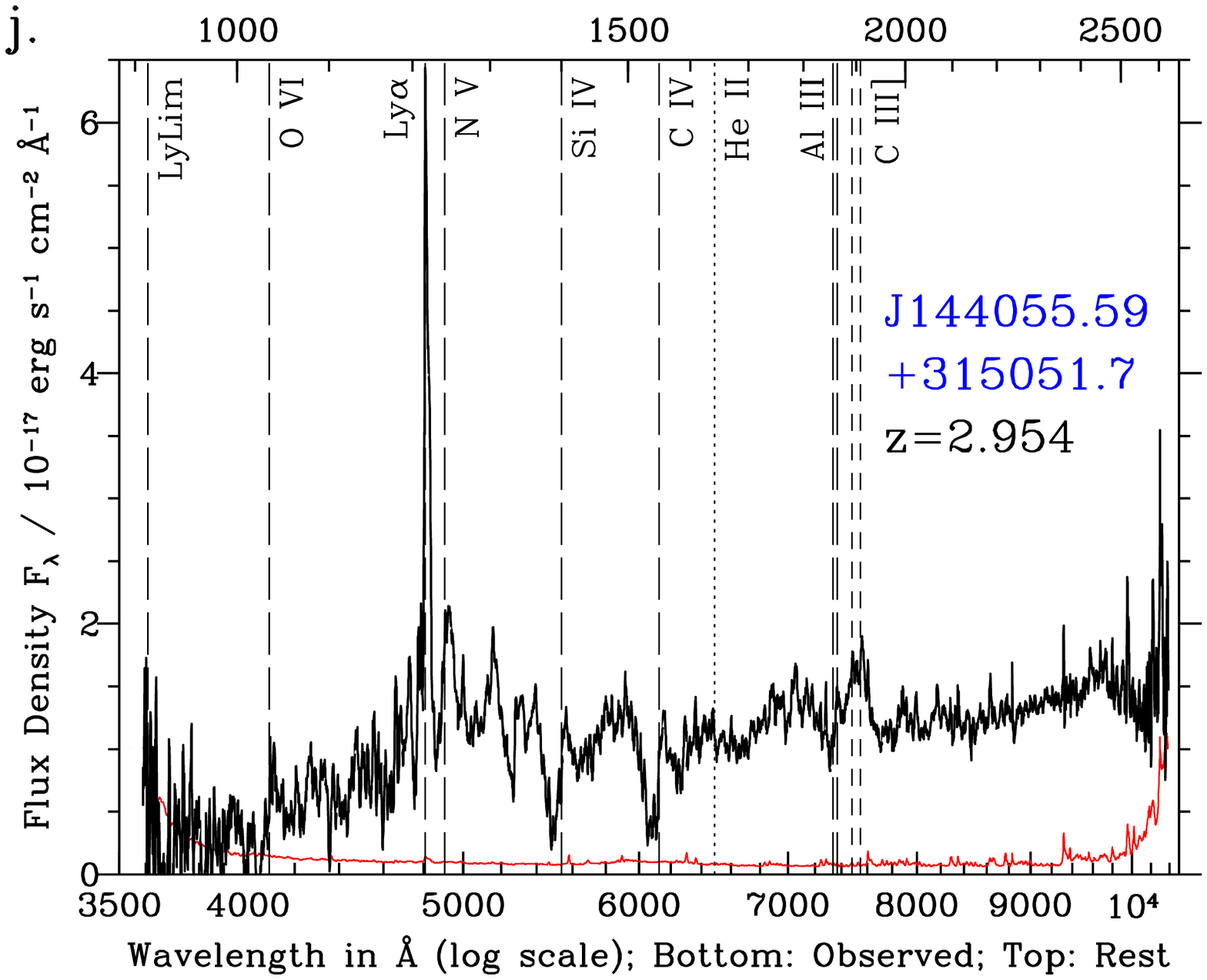} %35--105
\includegraphics[angle=0, width=0.497\textwidth]{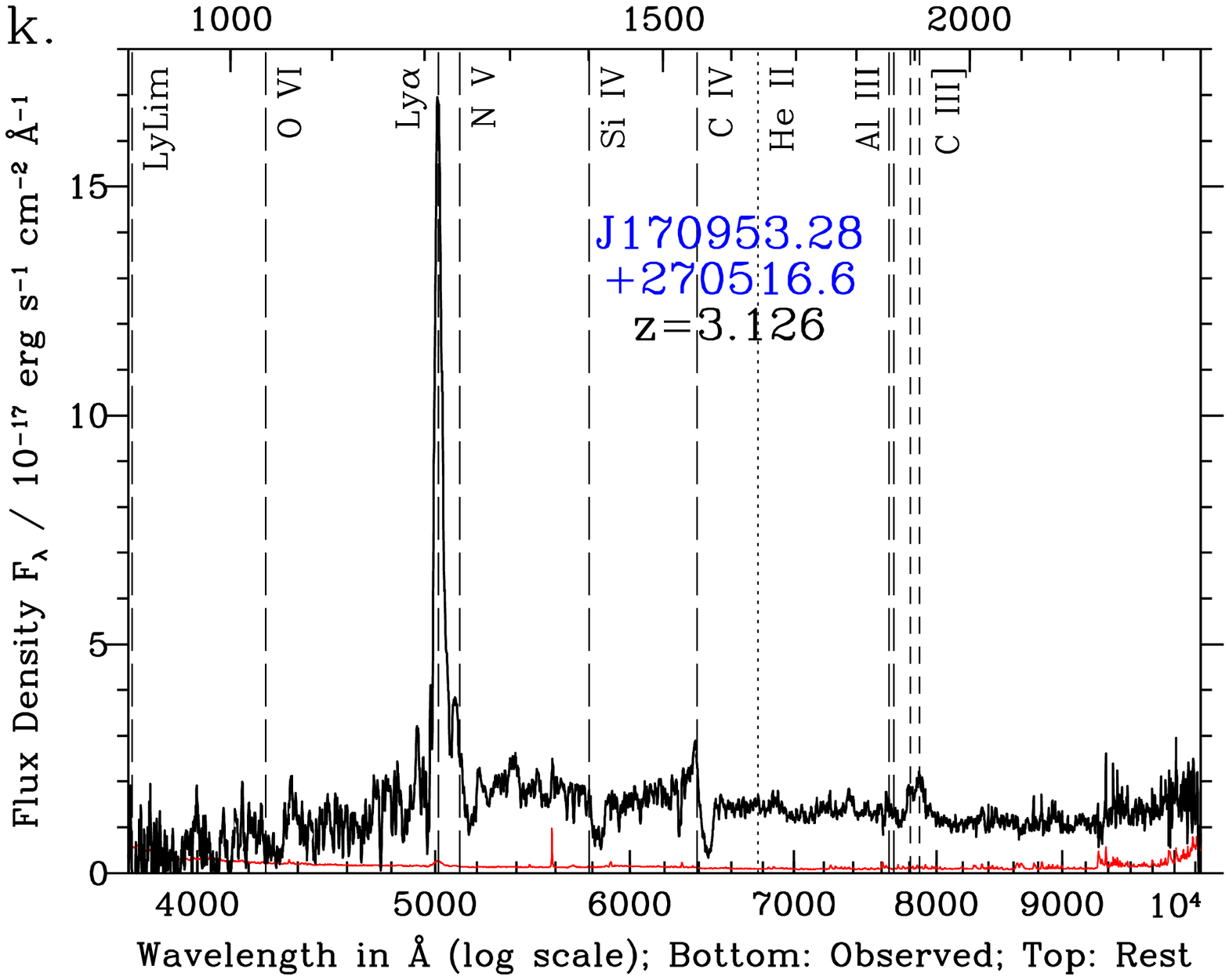}     %38--92
\includegraphics[angle=0, width=0.497\textwidth]{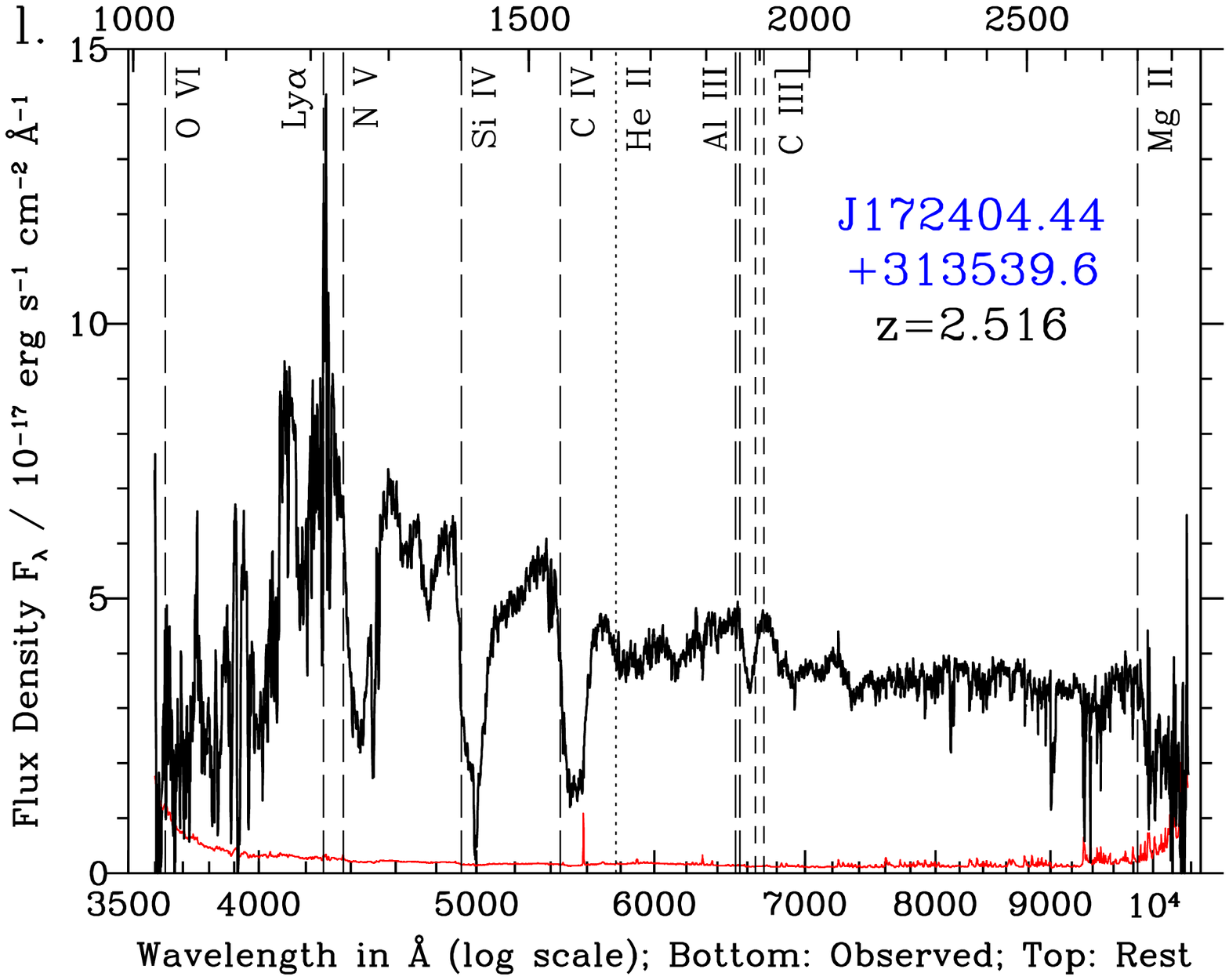} %35--105
\contcaption{} \end{figure*}

\begin{figure*} %%\vspace*{174pt} \makebox[\textwidth]{ % 0.497
\includegraphics[angle=0, width=0.497\textwidth]{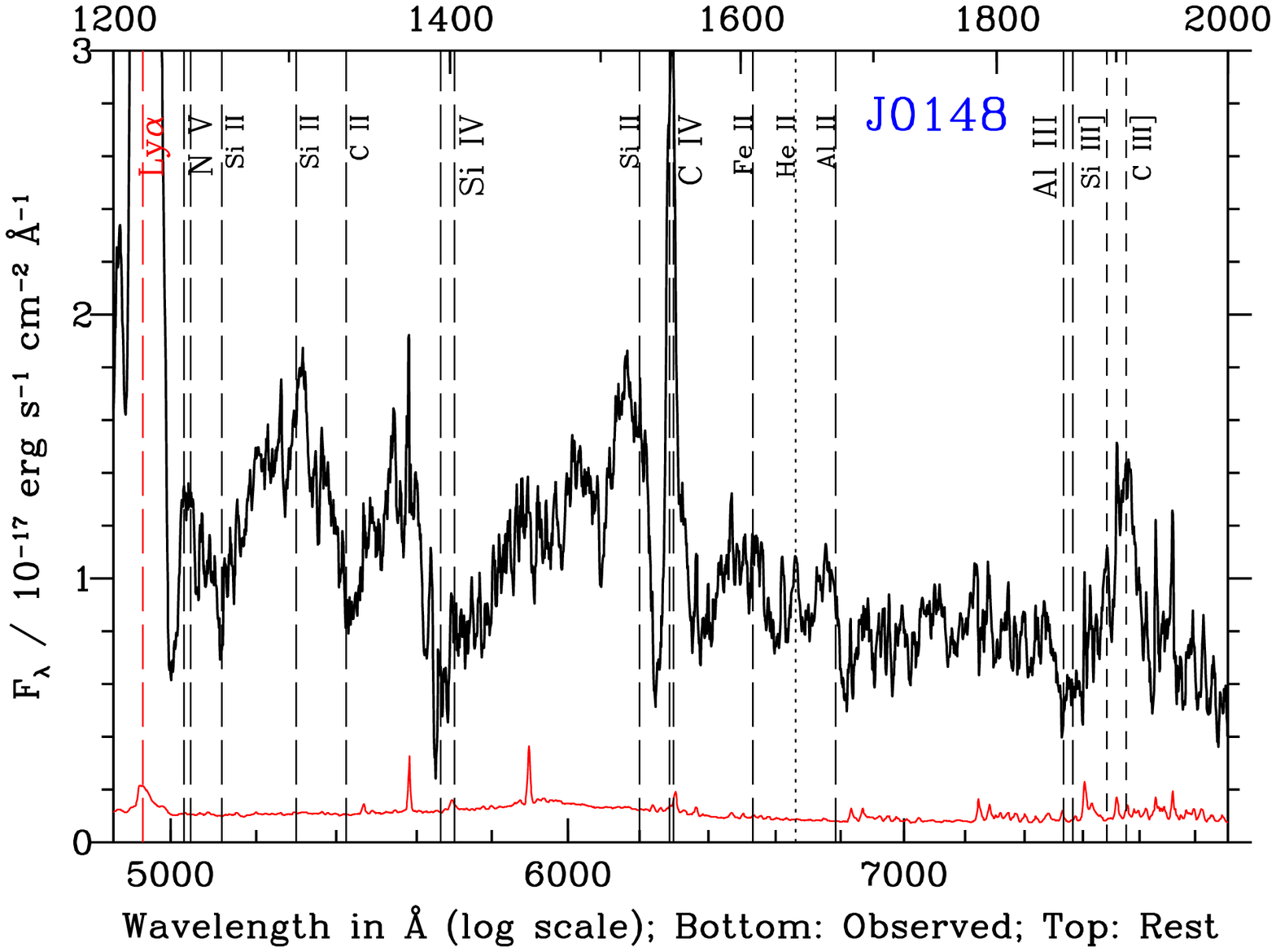} %35--105
\includegraphics[angle=0, width=0.497\textwidth]{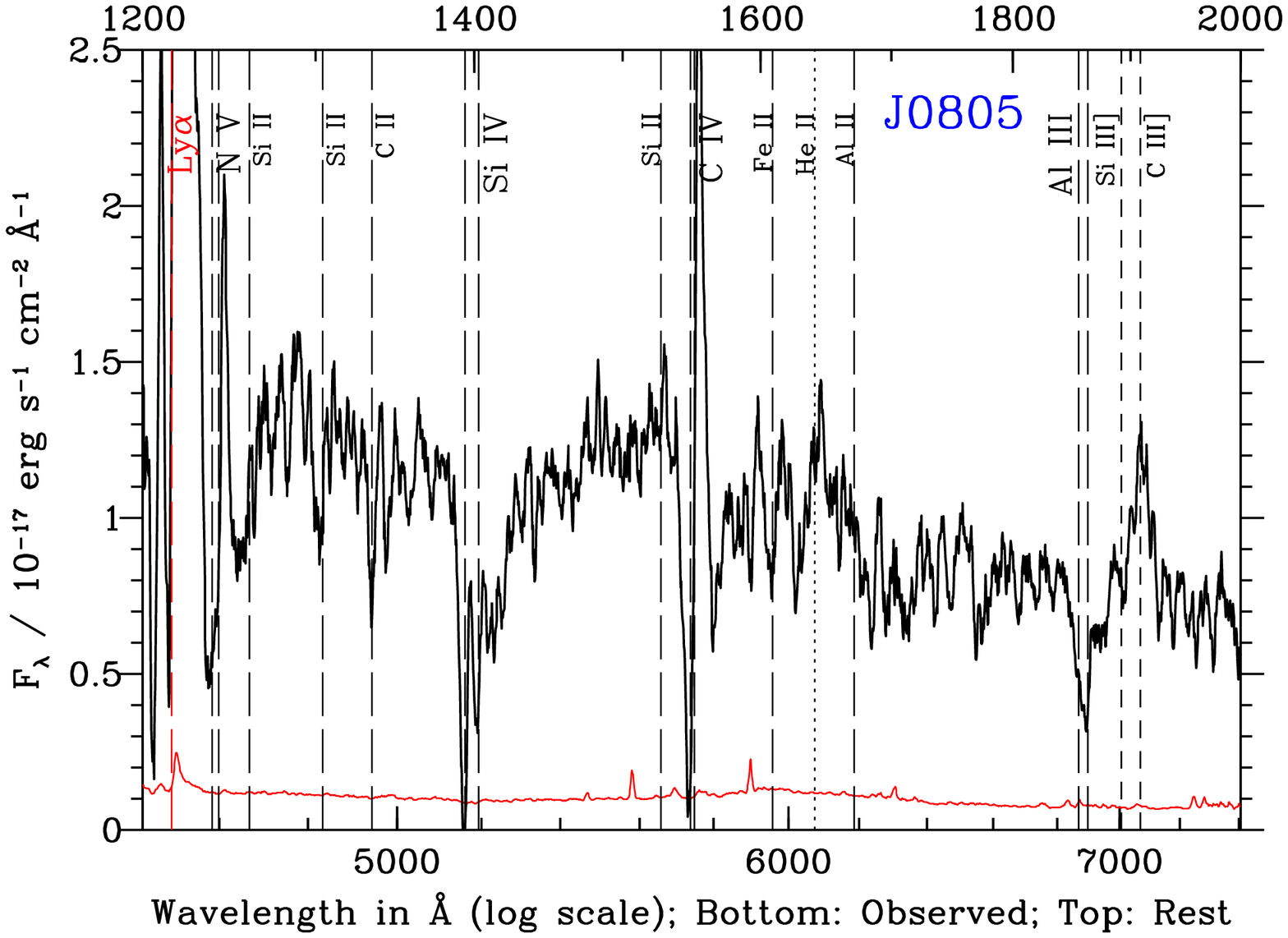} %35--105
\includegraphics[angle=0, width=0.497\textwidth]{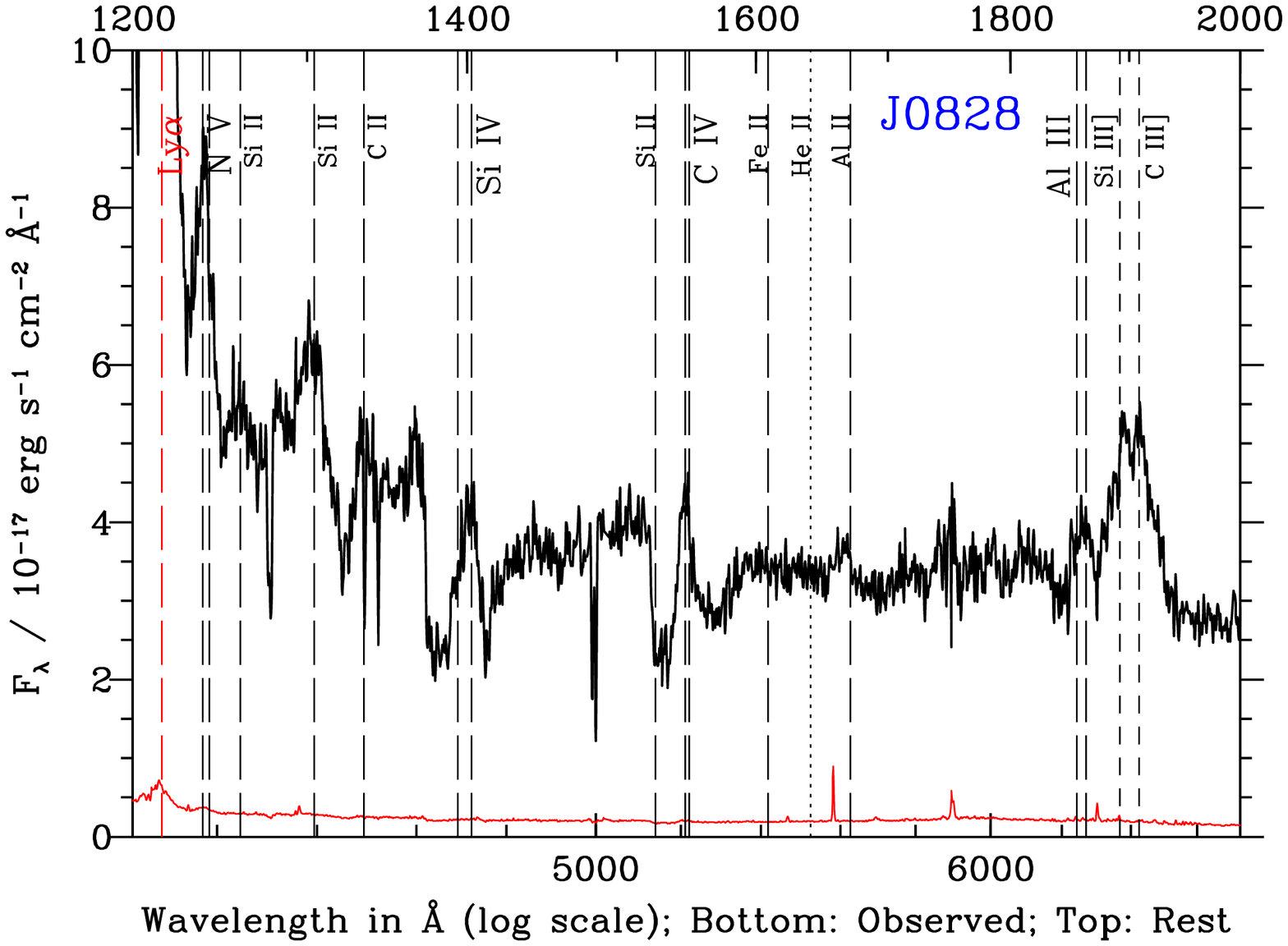} %35--105
\includegraphics[angle=0, width=0.497\textwidth]{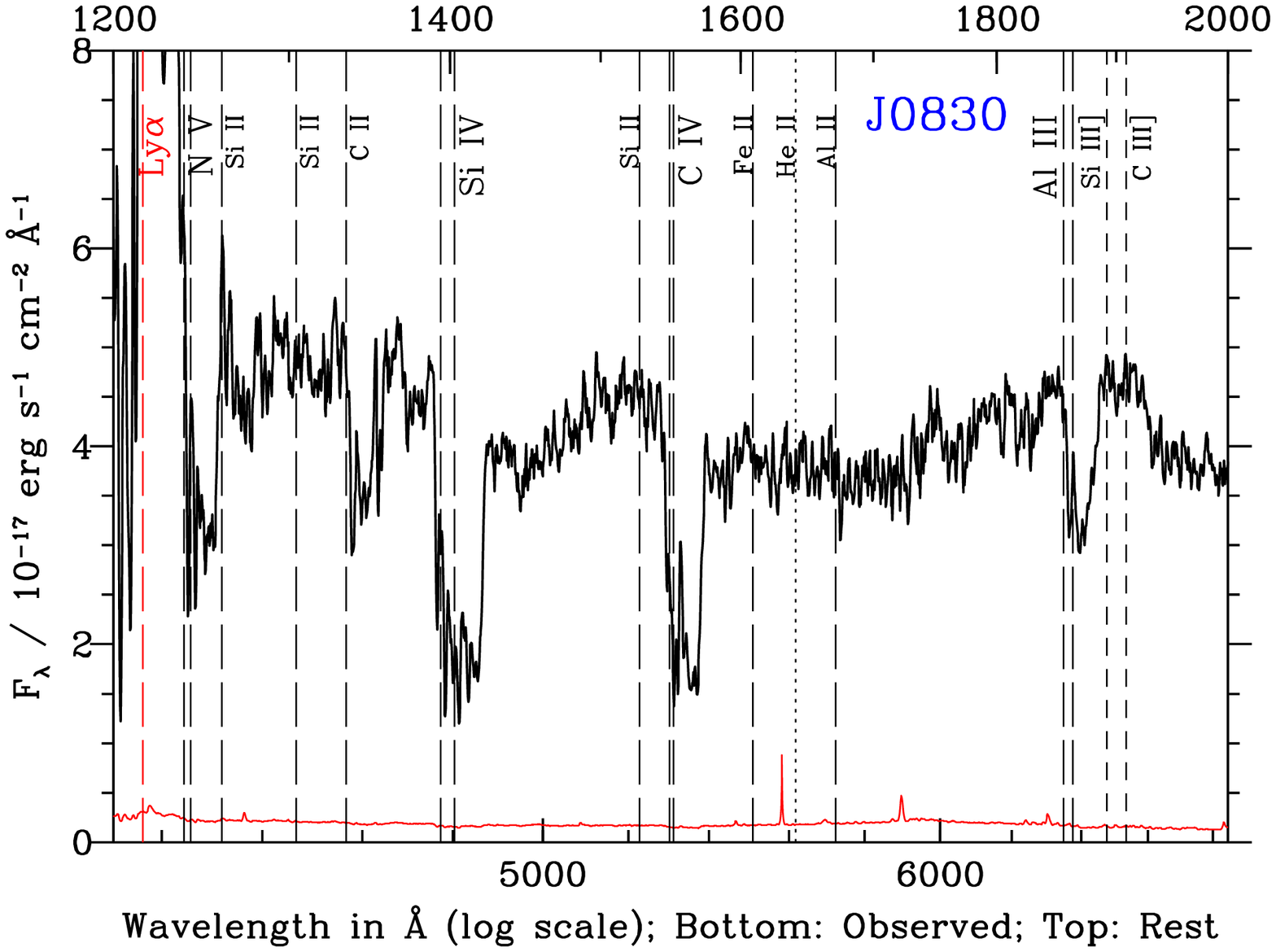} %35--105
\includegraphics[angle=0, width=0.497\textwidth]{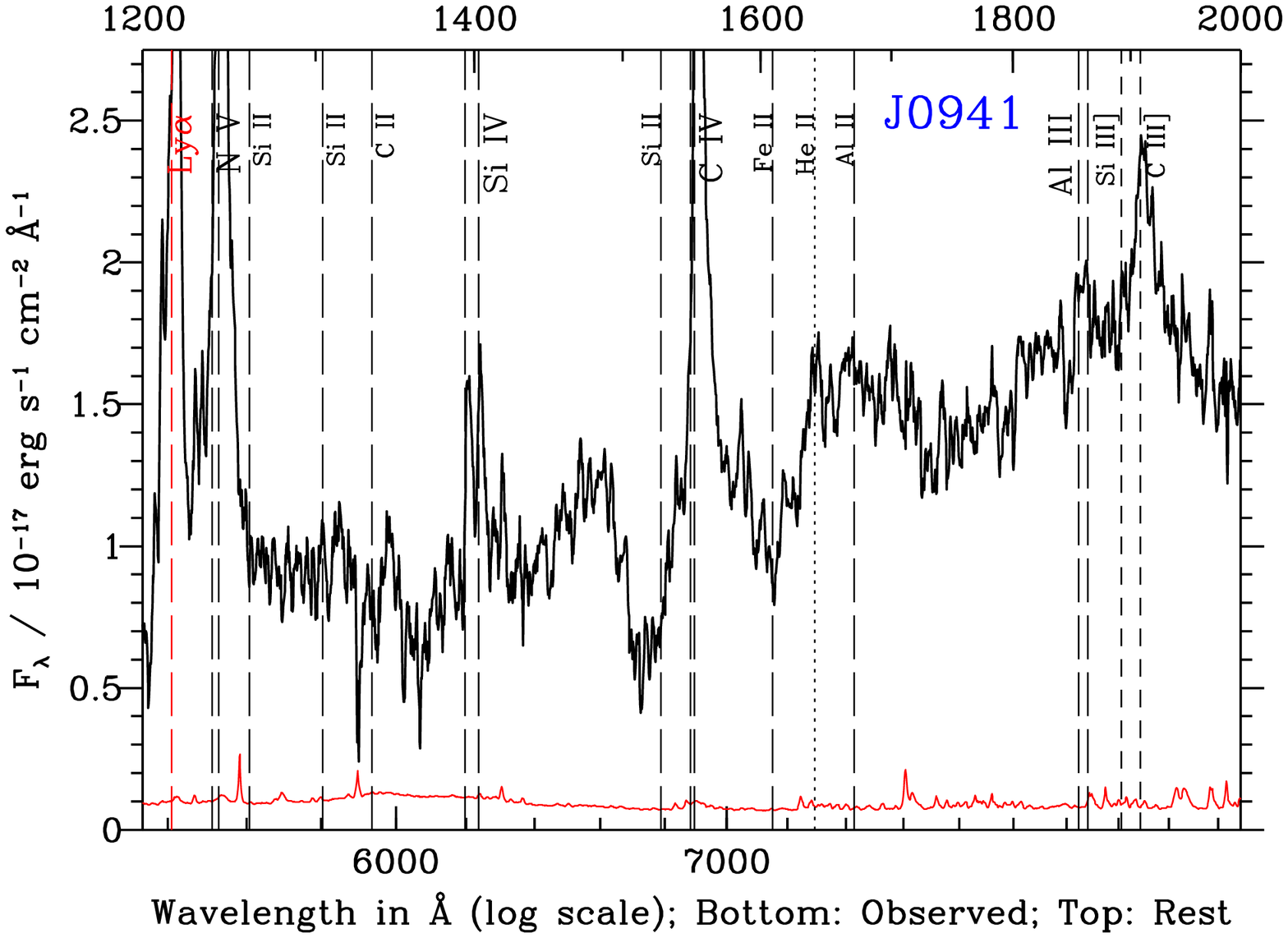} %35--105
\includegraphics[angle=0, width=0.497\textwidth]{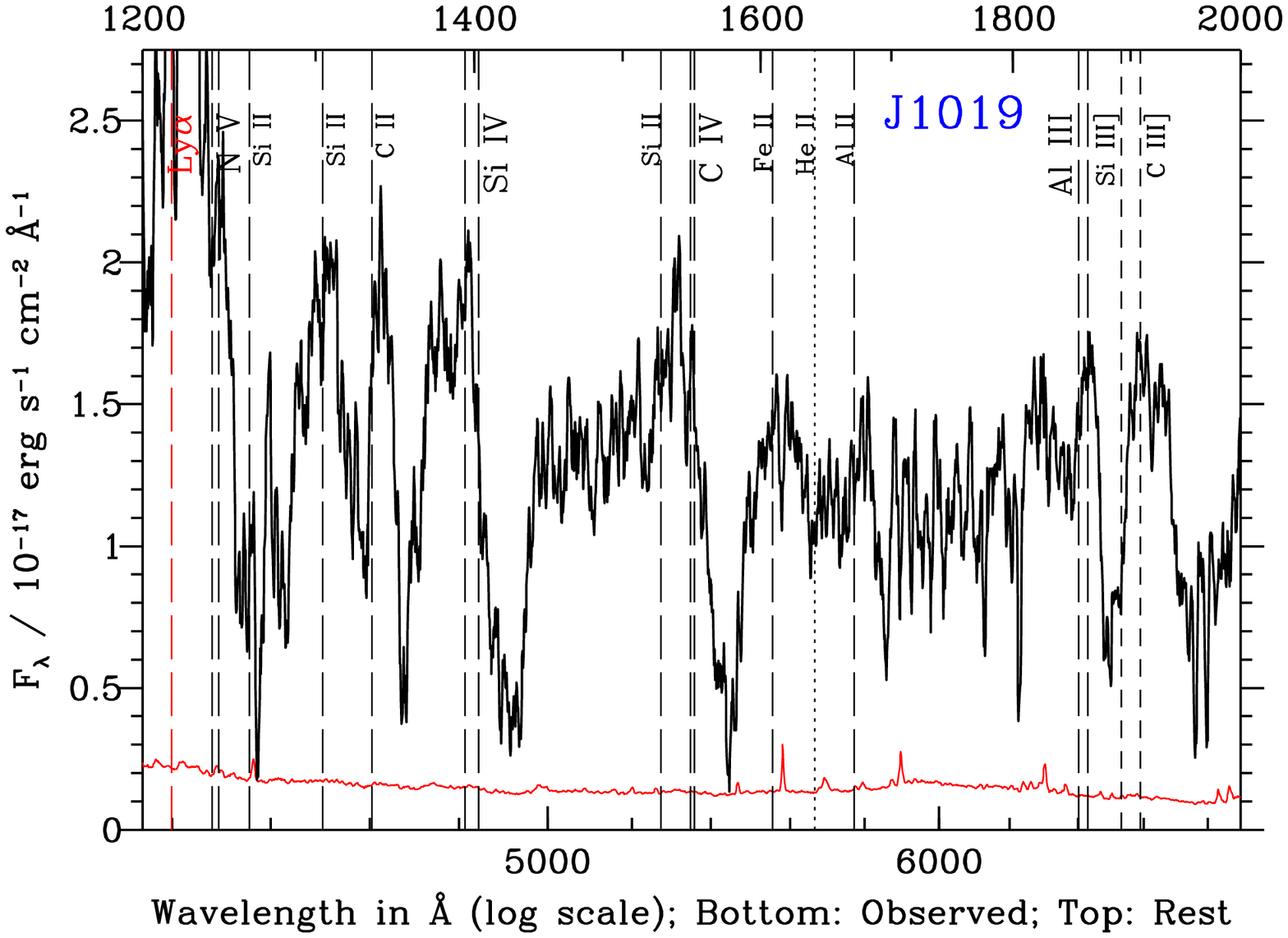} %35--105
\caption{Close-ups of the BOSS discovery spectra shown in 
Figure \ref{f_cands}.}\label{f_closeups} \end{figure*}

\begin{figure*} %%\vspace*{174pt} \makebox[\textwidth]{ % 0.497
\includegraphics[angle=0, width=0.497\textwidth]{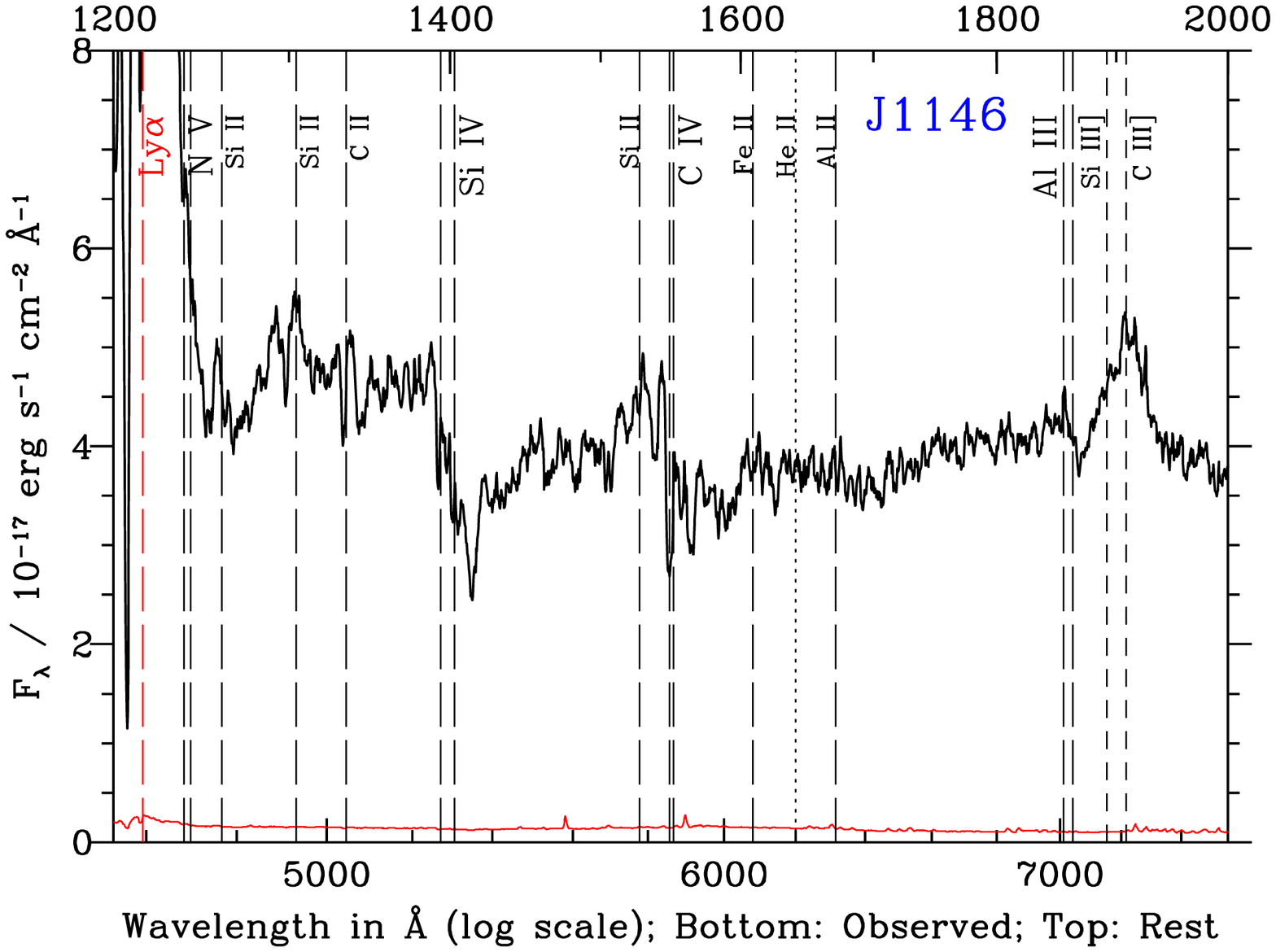} %35--105
\includegraphics[angle=0, width=0.497\textwidth]{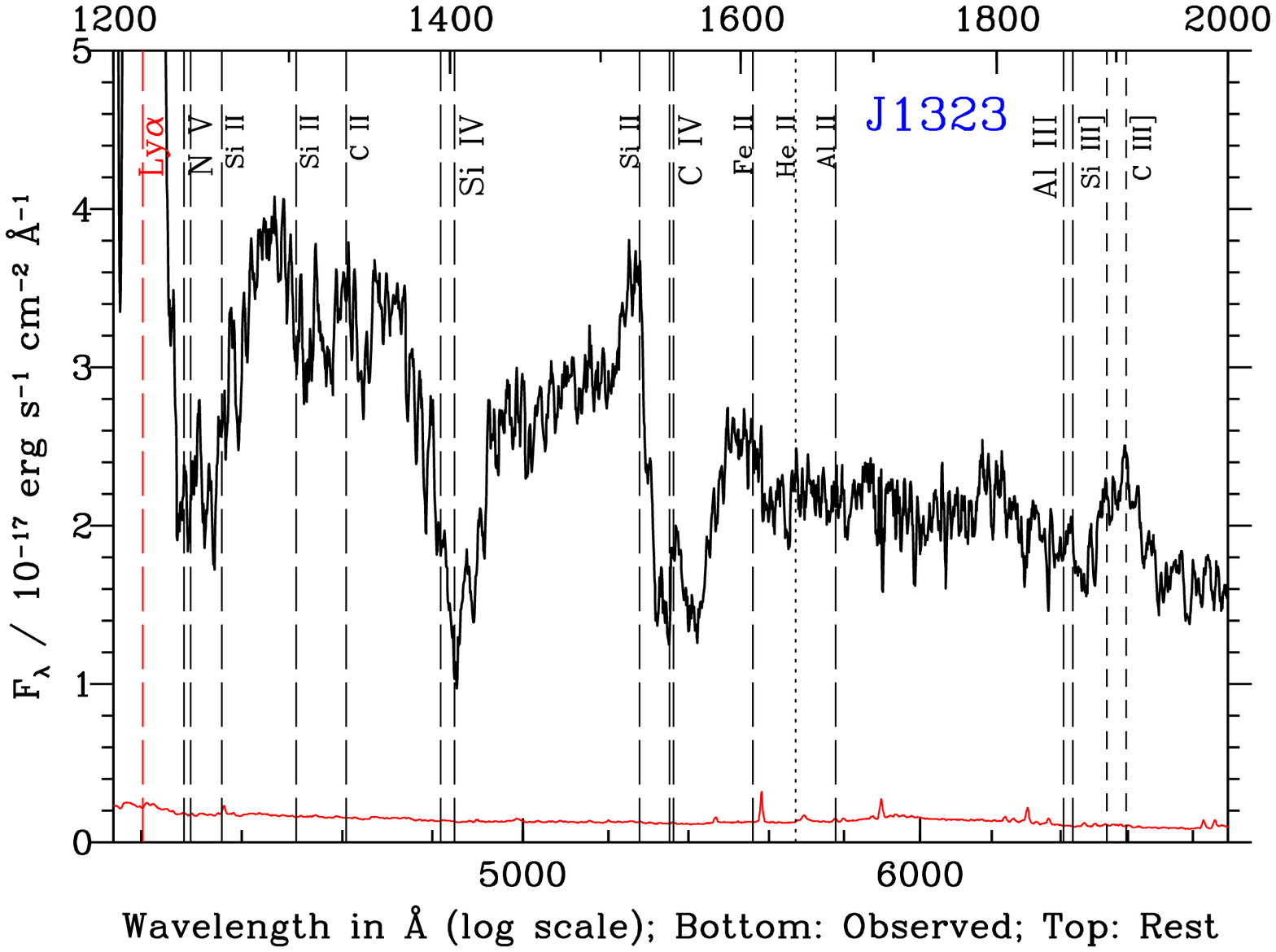} %35--105
\includegraphics[angle=0, width=0.497\textwidth]{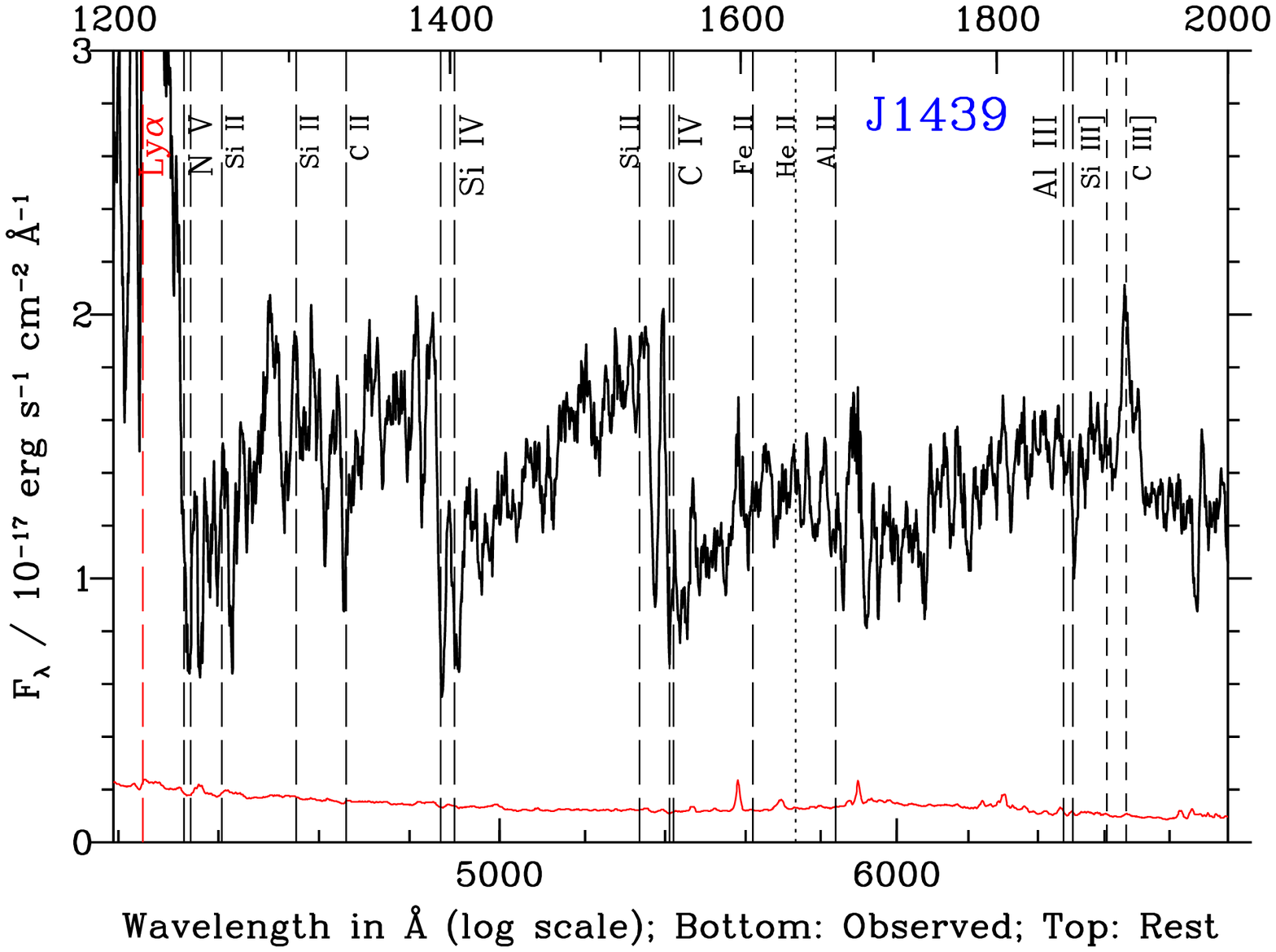} %35--105
\includegraphics[angle=0, width=0.497\textwidth]{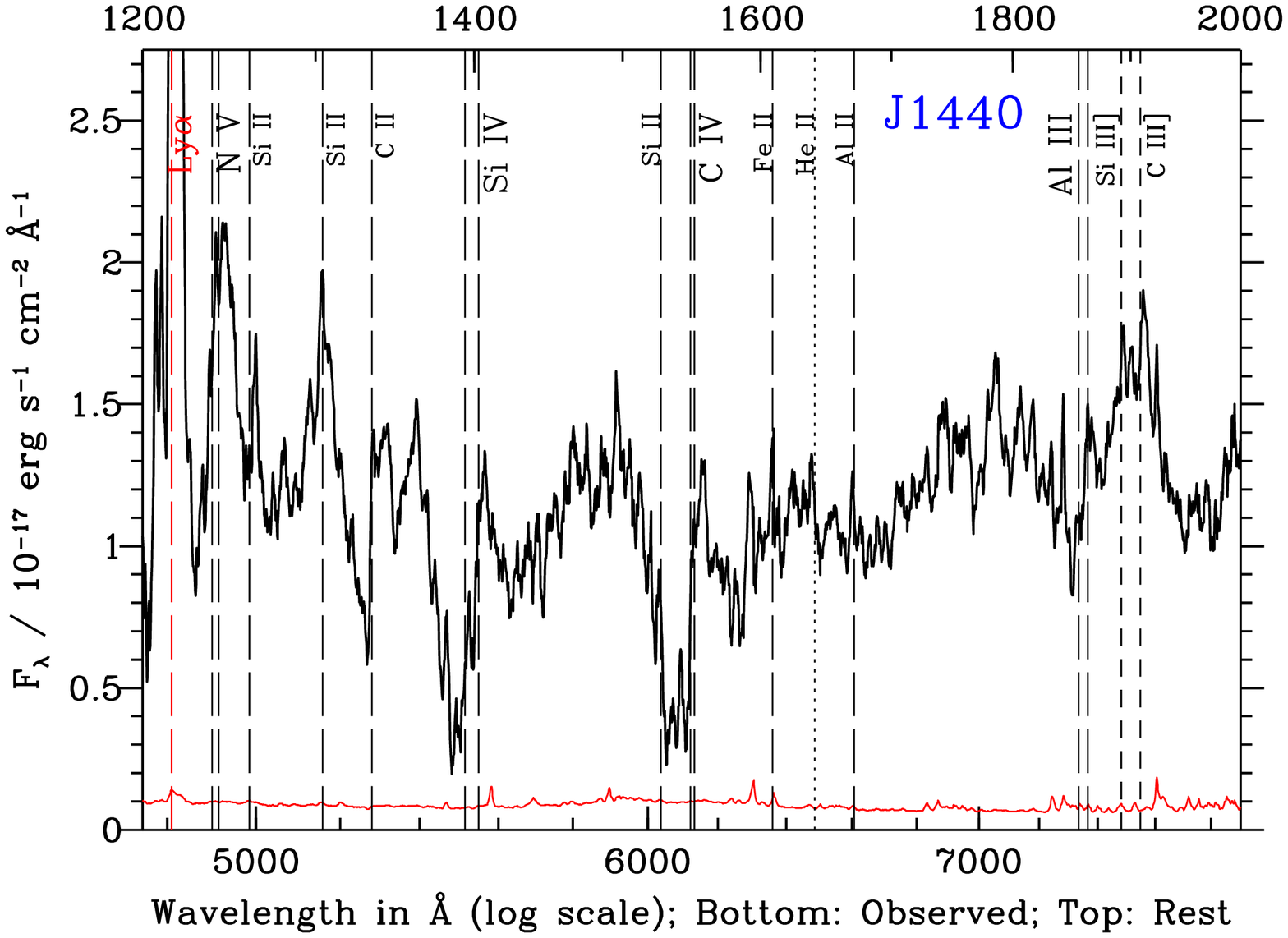} %35--105
\includegraphics[angle=0, width=0.497\textwidth]{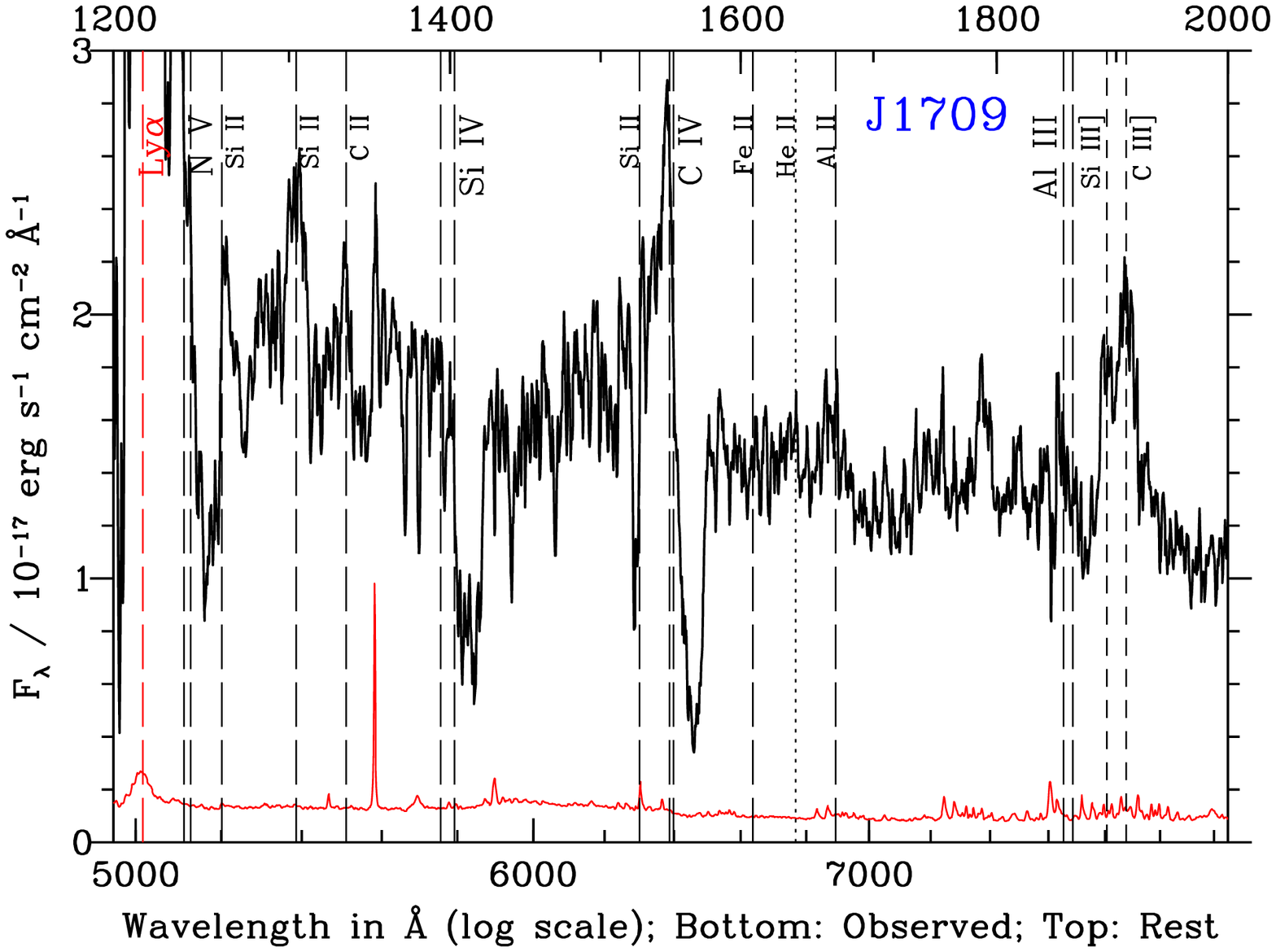}     %38--92
\includegraphics[angle=0, width=0.497\textwidth]{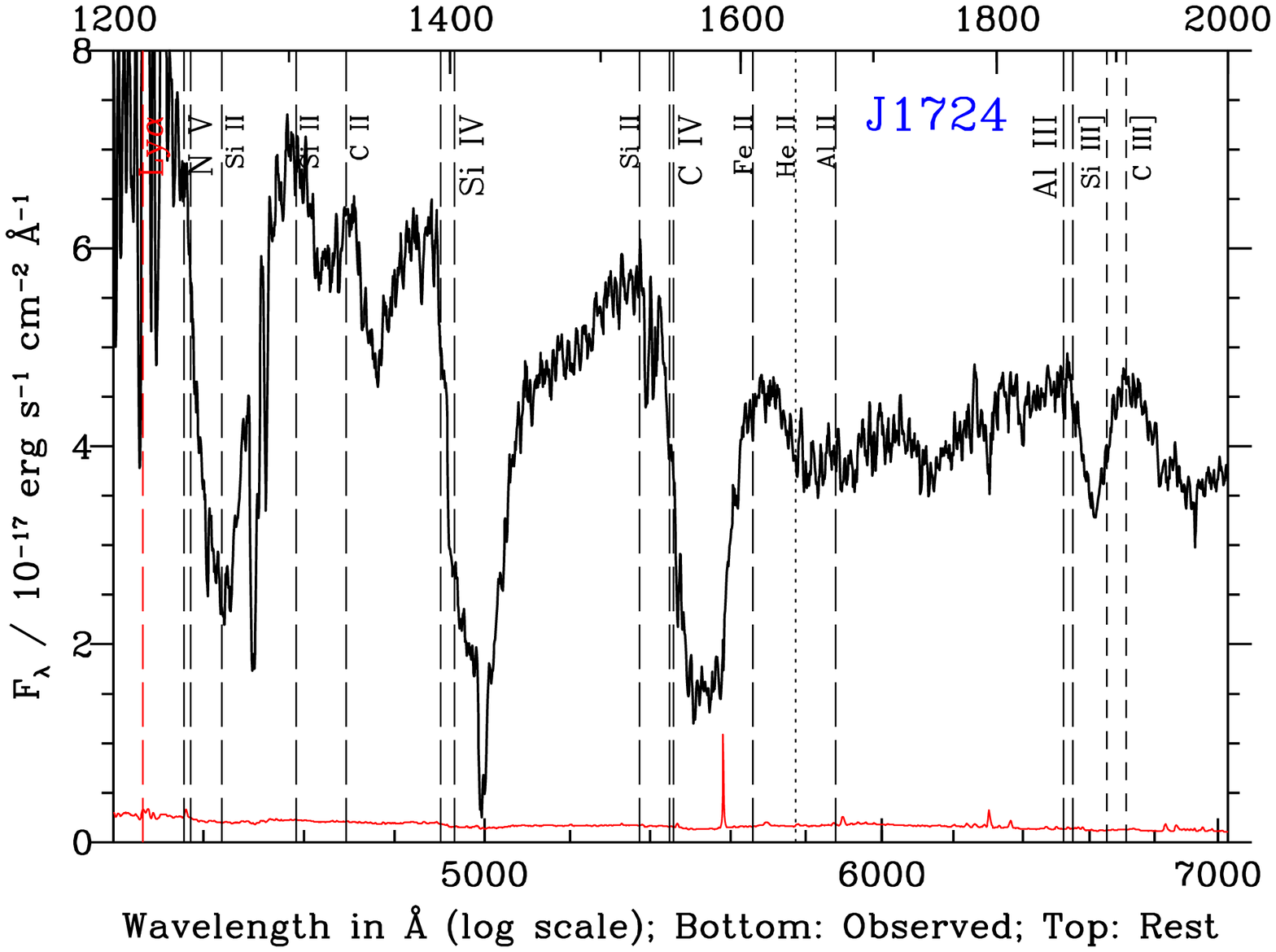} %35--105
\contcaption{} \end{figure*}

\begin{figure*} %%\vspace*{174pt} \makebox[\textwidth]{ % 0.497
\includegraphics[angle=0, width=0.497\textwidth]{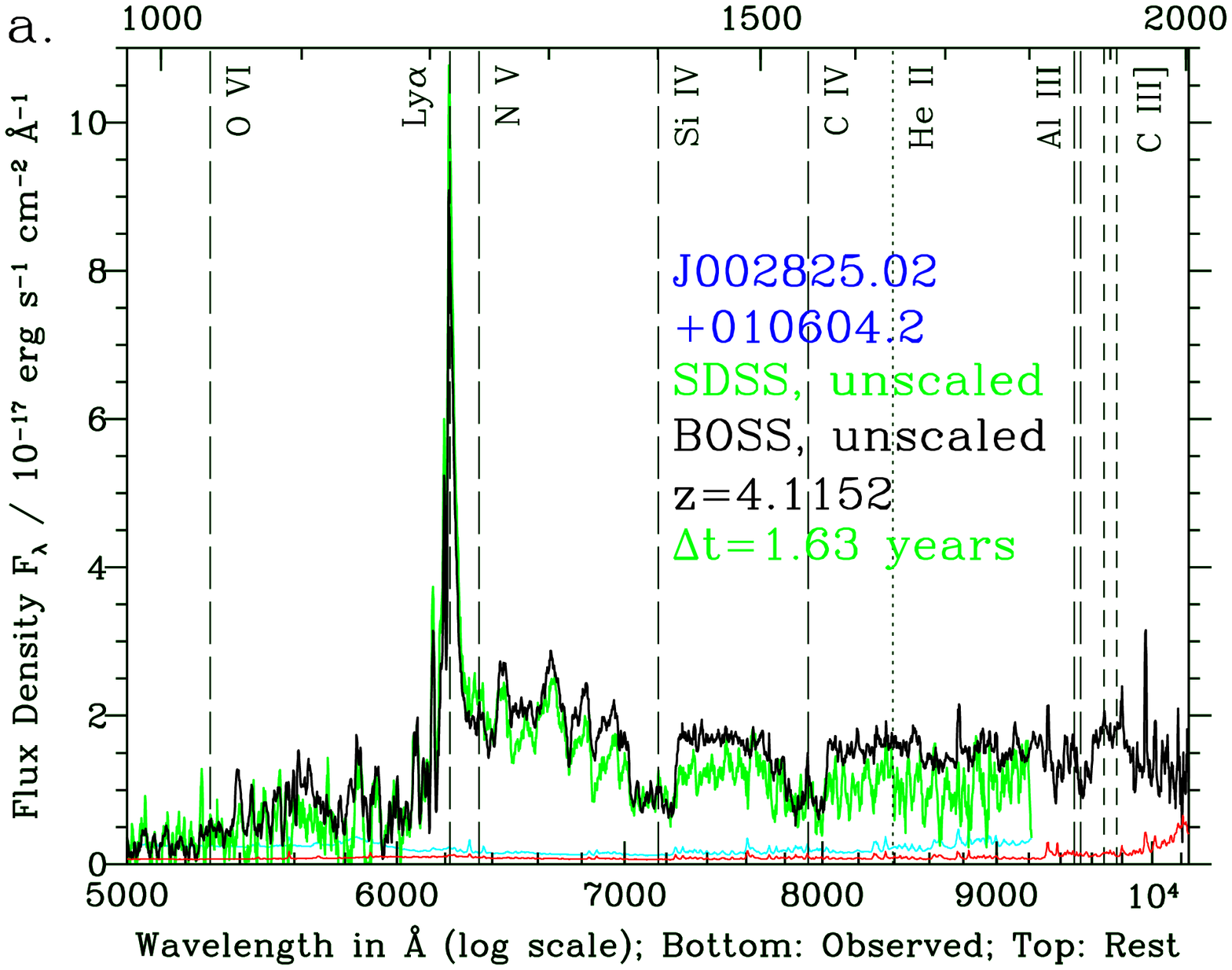}  %38--92
\includegraphics[angle=0, width=0.497\textwidth]{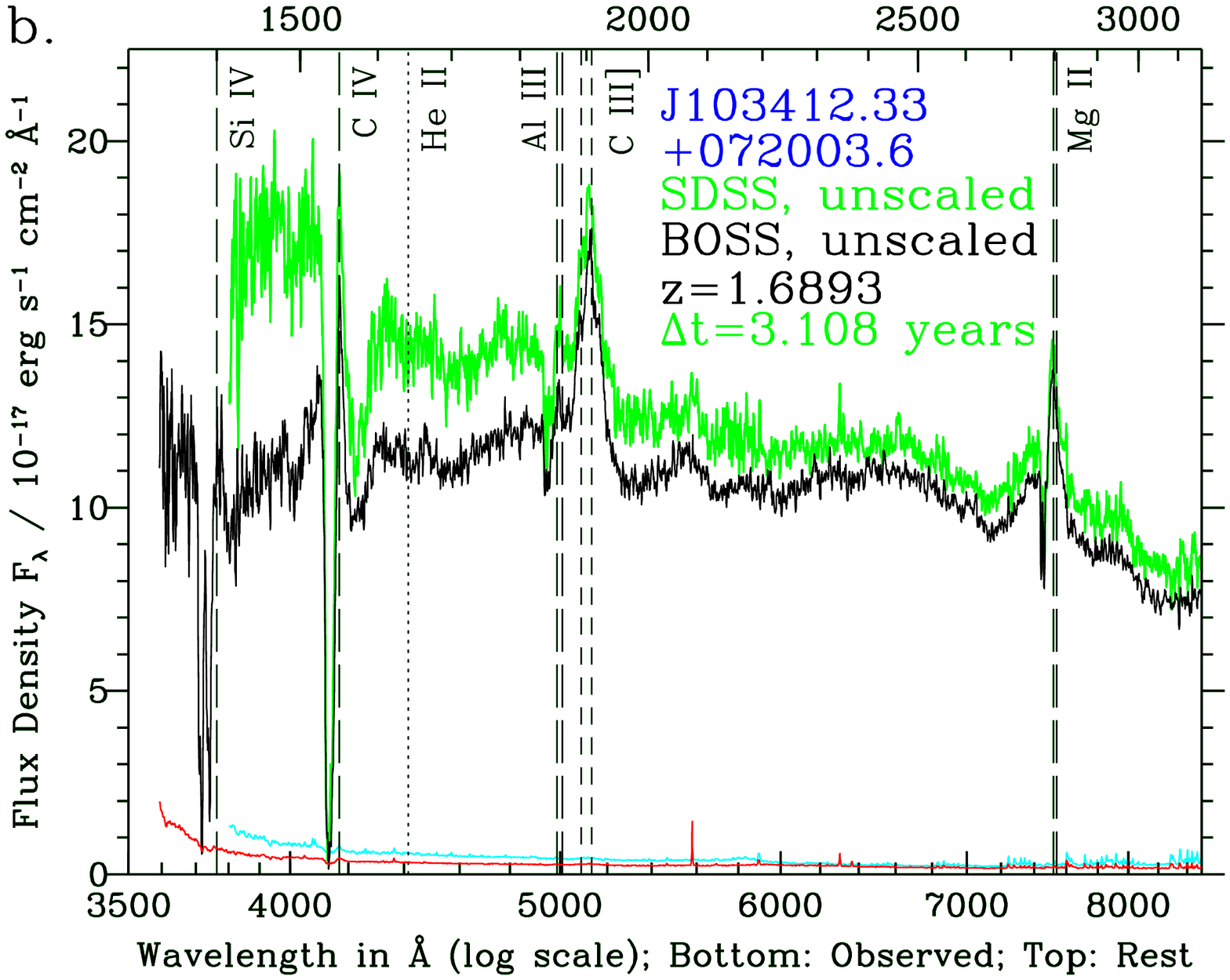} %35--85
\includegraphics[angle=0, width=0.497\textwidth]{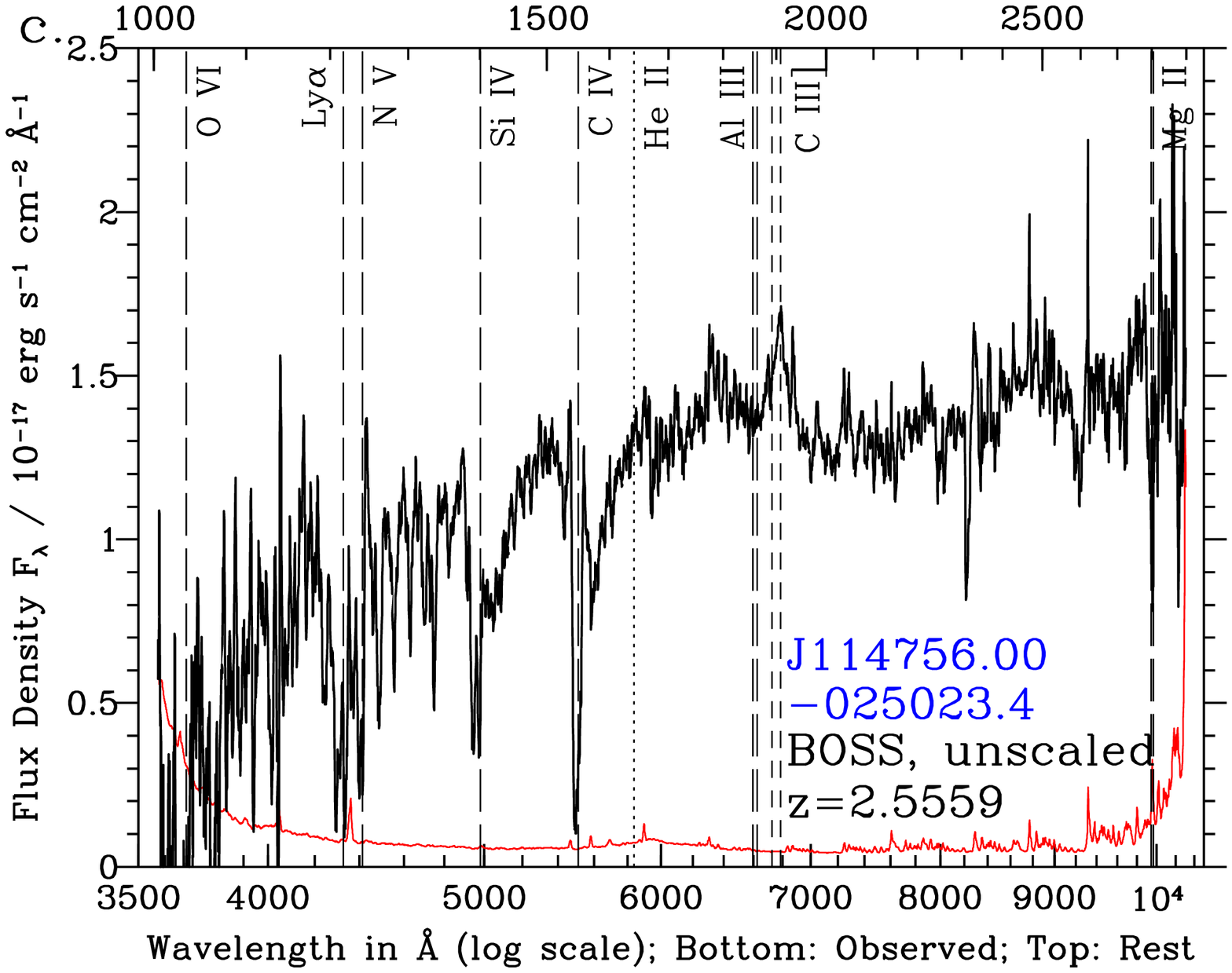}    
\includegraphics[angle=0, width=0.497\textwidth]{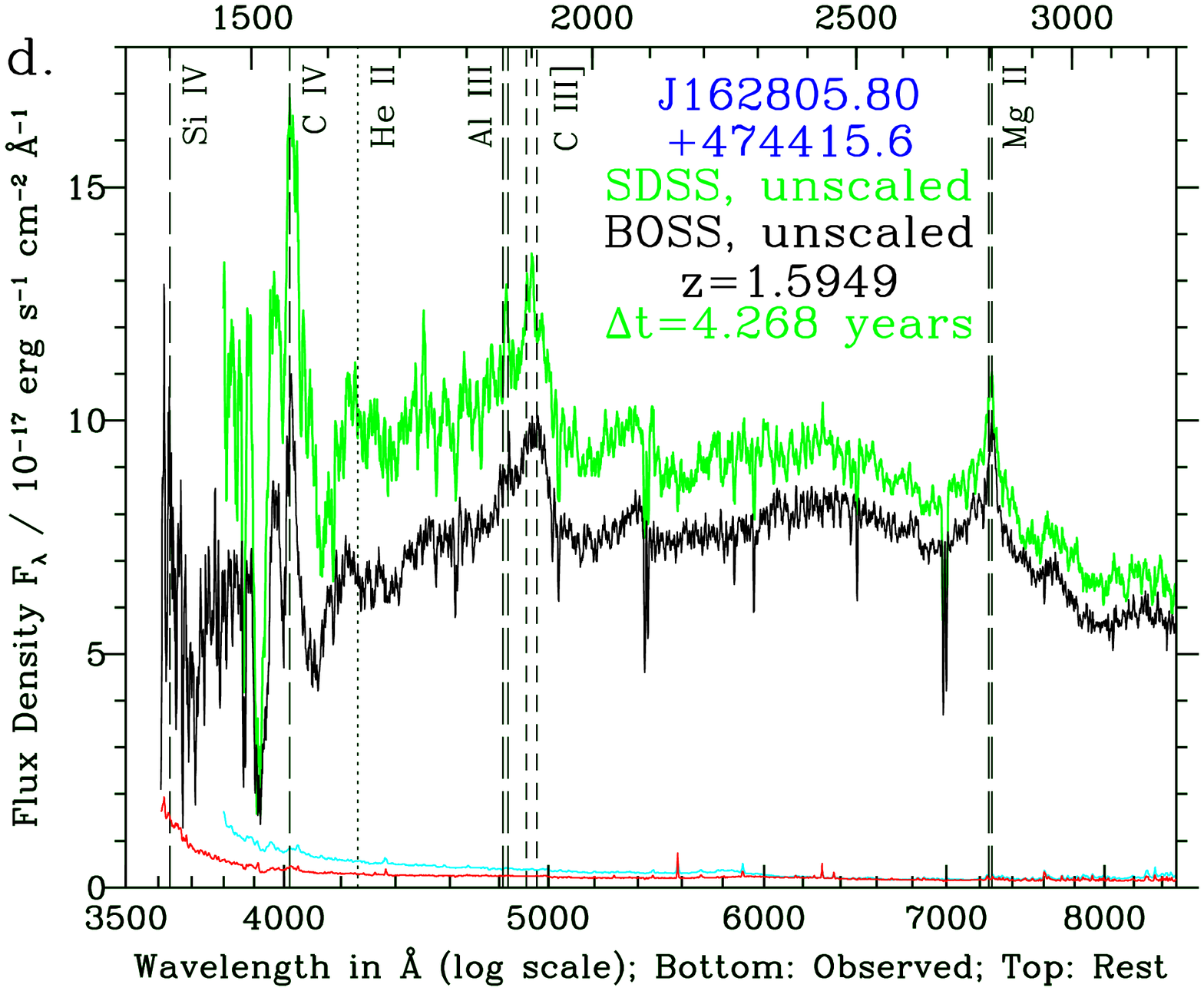}    %35--85
\caption{
Quasars which have redshifted absorption and SDSS spectra.  The smoothed 
SDSS spectra and error arrays are shown in green and cyan, respectively.
Smoothed second-epoch spectra and error arrays from BOSS 
are shown in black and red, respectively.
The separation of the spectra in time is given in rest-frame years.
For J1147, the SDSS spectrophotometry yields an erroneously 
steep spectrum and so we only show the BOSS spectrum.
}
\label{f_cands2} \end{figure*}

Over the course of the SDSS and now the BOSS, we have accumulated a sample of 
BAL quasars whose rest-frame spectra exhibit troughs with absorption at
wavelengths longer than the rest frame wavelength of one or more transitions.
These objects were found by visual inspection of over 100,000
quasar spectra by the authors,  %, primarily by PBH, PP and NFA.
including over 87,000 quasars in the Data Release Nine Quasar Catalog
\nocite{bossdr9q}({P{\^a}ris} {et~al.} 2012, hereafter DR9Q).

As with any qualitative selection criteria, 
while we cannot guarantee that our
inspections are complete, our numbers serve as lower limits to the
frequency of BAL quasars with redshifted troughs in BAL quasar samples.
(We can state that at an intermediate stage in the preparation of this paper,
nine out of the ten redshifted BAL quasars known at the time were recovered
by visual inspection of all visually identified BAL quasars in DR9Q.)
In Table 1 we give basic parameters for seventeen confirmed cases of redshifted
absorption in \civ, two previously known objects with redshifted absorption in
\mgii, and seven candidate cases of redshifted absorption.
Detailed notes on each confirmed case are given in Appendix \ref{notes}.  
We discuss our reservations about the candidate objects in Appendix \ref{maybe}
and present a number of rejected candidates in Appendix \ref{noway}.

The spectrum of the most striking object is shown in Fig. \ref{f_zoom}
and discussed below.
Spectra of all objects with redshifted high-ionization absorption
are shown in Fig. \ref{f_cands} (BOSS discoveries, with close-ups 
of the same spectra in Fig. \ref{f_closeups}) and Fig. \ref{f_cands2}
(SDSS discovery spectra in green, with second-epoch spectroscopy 
from the BOSS in black).  

SDSS J215704.26$-$002217.7 (J2157; Fig. \ref{f_zoom})
is the most dramatic example we have found of redshifted absorption.  
The quasar has a redshift $z=2.240\pm 0.002$, measured from its broad
\SiIII+\CIII\ emission as detailed in \S\,\ref{zsys}.\footnote{This redshift 
is conservative in that if redshifted \aliii\ absorption has significantly
affected the profile of this emission line, the true redshift will be lower and
the redshifted absorption will extend to even larger redshifted velocities.}
Because J2157 has a prominent blue wing of broad \civ\ emission, 
for comparison to it in Fig. \ref{f_zoom} we plot 
the composite spectrum of the quartile of quasars
with the largest \civ\ blueshifts in the study of \nocite{emshift}{Richards} {et~al.} (2002).  
Of their four composites, the one plotted
best matches the emission lines of J2157 in regions not affected by absorption.
The absorption troughs in J2157 in \NV, \SiIV\ and \CIV\ begin
at a small blueshift ($v$=$-1930$~\kms\ in \civ)
and extend smoothly to a large redshift ($v$=9050~\kms\ in \civ)
at more or less constantly decreasing depth.
There is also redshifted absorption in \PV\ (shaded in Fig. \ref{f_zoom})
and in \AlIII.
J2157 exhibits strong, narrow \lya\ emission 
accompanied by narrow \OI+\SIii$\lambda$1304, \HeIIsf,
\MgII, and possibly \civ\ emission.

The presence of redshifted absorption in multiple transitions makes the
identification of the troughs in J2157 and other such objects unambiguous.
The variety of the redshifted absorption troughs'
profiles, widths, and velocity ranges rule out an explanation of the troughs
as gaps between emission features (but see Appendix \ref{maybe} for additional
candidates where that explanation cannot be excluded).

\subsection{Systemic redshifts and redshifted trough extents} \label{zsys}

The identification of these troughs as redshifted depends on the systemic
redshift of these quasars being known to an uncertainty less than the trough
widths.  
The redshifted absorption extends to $v$$>$3000~\kms\ in all confirmed 
cases and up to $v$$>$10,000~\kms\ in a few cases (see \S~\ref{ai}).  
Such values are generally larger than the uncertainties in redshift 
measurements made using low-ionization broad emission lines. 
To estimate those uncertainties we use the results that, on average,
narrow \oiii\ in quasars is blueshifted by an average of $-$45$\pm$5~\kms\ from
the host galaxy redshift defined by \caii\ K absorption \nocite{hw10}({Hewett} \& {Wild} 2010),
\mgii\ is redshifted relative to \oiii\ by 97$\pm$269~\kms\ \nocite{emshift}({Richards} {et~al.} 2002), 
and the \ciii\ complex\footnote{Variations in the relative strengths of the 
transitions in the \ciii\ complex can introduce offsets of order a few 
100\,\kms\ into the determination of the systemic redshift from it.
This complex is a blend of \CIII, \SiIII, and \feiii\,UV34 \lalala 
1895,1914,1926, plus \AlIII\ if the line profiles are sufficiently broad.
At densities $\leq$10$^5$\,cm$^{-3}$ there will also be significant 
contributions from [\ciii\,$\lambda$1907 and [\Siiii\,$\lambda$1883 
\nocite{of06}({Osterbrock} \& {Ferland} 2006).} is blueshifted relative to \mgii\ 
by $-$827$\pm$604~\kms\ \nocite{2007AJ....133.2222S}({Shen} {et~al.} 2007).
Therefore, relative to the host galaxy redshift, 
\mgii\ redshifts should on average be redshifted by 52$\pm$269~\kms, and
\ciii\ redshifts should on average be blueshifted by $-$775$\pm$661~\kms.
We do not correct redshifts we obtain from those transitions for the above
offsets, but those offsets and scatter should be kept in mind as limits
on the current accuracy of our redshifts.

We examined the spectra of our objects to determine `inspection redshifts'
which were then compared to the various redshifts tabulated in DR9Q, and, 
for quasars in the SDSS Data Release Seven (DR7; \nocite{dr7}{Abazajian} {et~al.} 2009) Quasar Catalog
\nocite{dr7q}({Schneider} {et~al.} 2010), to the redshifts computed by \nocite{hw10}{Hewett} \& {Wild} (2010).
These objects have unusual spectra which affect the accuracy of redshifts
measured by fitting the overall spectrum with a Principal Component Analysis
(PCA) reconstruction with the redshift as a free parameter (the DR9Q full PCA
redshifts), but which rarely affect fits to the \ciii\ emission region.
Therefore, the default redshifts adopted for our objects are the redshifts 
measured from the peak of a PCA reconstruction of the \ciii\ emission line
(see DR9Q).
The available \ciii\ PCA redshifts agreed very well with our inspection redshifts
except in two cases, with an average 
$ \Delta v \equiv v_{C\,III]} - v_{insp} = (60 \pm 270)$~\kms\ excluding the
two special cases.  Those cases are the candidates J0050 ($\Delta v=-1400$~\kms)
and J1704 ($\Delta v=-2800$~\kms).  In the former case we believe the \ciii\ PCA
redshift to be incorrect and in the latter we believe the true redshift
is bracketed by our inspection redshift and the \ciii\ PCA redshift 
(see Appendix \ref{maybe}).
For objects without \ciii\ PCA redshifts available,
we use redshifts from HW10 whenever those redshifts were plausible.
The remaining cases are discussed individually in Appendix \ref{notes}.

All of our seventeen confirmed objects have maximum 
redshifted trough velocities of 3170~\kms\ or greater, equivalent to requiring
deviations from the mean of the \ciii\ redshift distribution of 3.6$\sigma$
or greater (we do not quote probabilities because we do not know how Gaussian 
the distribution is, especially in the tails).
Overall, our objects do not have maximum redshifted trough velocities
small enough that the redshifted absorption could be spurious,
arising from an outlying redshift measurement from the \ciii\ emission line.

Finally, a K-S test reveals no statistically significant
difference between the redshift distribution of our high-ionization objects
and that of the DR9Q BAL sample.

\subsection{Cases of particular interest} \label{cases}

We discuss here several objects of particular interest.  Notes on all objects 
with redshifted high-ionization absorption are given in Appendix \ref{notes}.

SDSS J094108.92$-$022944.7 (J0941; Fig. \ref{f_cands}e),
at $z$=3.446 from its DR9Q \ciii\ PCA redshift,
is morphologically classified as a galaxy in the SDSS.
Its image appears extended in the NW-SE direction.
These properties make J0941 a prime candidate for being a binary quasar system
where the BAL outflow of one quasar is silhouetted in front of a second quasar;
see \S\,\ref{binary}.
Note that J0941 was not targeted as a BOSS quasar candidate 
because all such candidates are required to be unresolved \nocite{bossQTS}({Ross} {et~al.} 2012).
It was targeted as part of the CMASS (``constant mass'') galaxy sample 
designed to select galaxies at $0.43 < z < 0.7$ \nocite{bossover}({Dawson} {et~al.} 2013).
It is unlikely to be a lensed quasar superimposed on a $z<0.7$ galaxy:
the rest-frame equivalent width %REW 
of broad \ciii\ in the observed spectrum is typical, so the observed flux 
must be dominated by the quasar and not a galaxy at $z<0.7$.
If J0941 is a lensed quasar, the lensing galaxy must be of sufficiently
high redshift or low optical luminosity (or both) to leave no signature
in the BOSS spectrum.

SDSS J101946.08$+$051523.7 (J1019; Fig. \ref{f_cands}f)
has \lya\ which peaks at $z$=2.4685,
but we adopt the DR9Q \ciii\ PCA redshift of $z$=2.452 as systemic.
J1019 has %systemic and 
mostly redshifted absorption in a wide range of transitions:
\feiii\ multiplets UV48, UV50 and UV34,
\aliii, \AlII, \civ, \SIiv, \CII, \SIii\,$\lambda$1304,
\SiII, \Nv, and apparently \CTHREE, \PV+\feiii$\lambda$1122, \SIV\ and
probably \OVI.  The last five transitions are in the \lya\ forest
and are thus less certain identifications.  If confirmed, \pv\ and 
\Siv\ would indicate a high column density absorber \nocite{borgaravPV}({Borguet} {et~al.} 2012).
Absorption from \feiii\,UV34 (EP 3.73\,eV) and \feiii\,UV48 (EP 5.08\,eV)
has been seen before in BAL quasars \nocite{sdss199}({Hall} \& {Hutsem{\'e}kers} 2003), but this is 
the first reported case of a trough from \feiii\,UV50 (EP 7.86\,eV).
Absorption from \feii\ is not clearly detected despite the detection 
of \SIii\ and the similarity of the two elements' abundances and
ionization potentials for ionization stages I--III.
We discuss this quasar further in \S\,\ref{dist}.

SDSS J162805.80$+$474415.6 (J1628 Fig. \ref{f_cands2}d) at $z$=1.5949
has broad \civ\ absorption blueshifted by up to $-$9810~\kms\ %$z$=1.514
and a redshifted \civ\ trough extending to 12400~\kms. %$z$=1.714 
We have previously outlined two alternative explanations for the latter trough
\nocite{unconventional}({Hall} {et~al.} 2004b).  
It could be a high
velocity ($-$46000~\kms) \aliii\ trough without accompanying \mgii. 
The required velocity is less than the maximum velocity known for \civ, but
considerably larger than the maximum velocity of 25000~\kms\ previously 
identified for \aliii\ \nocite{allenbal}(in SDSS J154303.24+264052.3; {Allen} {et~al.} 2011).
Furthermore, the lack of BAL quasars with \aliii\ seen between
25000~\kms\ and 45000~\kms\ makes this explanation unlikely.
The trough could also be \HeIIsf\ absorption associated with the 
blueshifted \civ\ outflow.
No unambiguous detection of \heii\ absorption in a BAL quasar has been reported,
but it has been seen in disc-wind outflows in Galactic objects \nocite{novalike}({Hartley} {et~al.} 2002).
Such absorption would indicate gas of high density 
and with a moderately high ionization parameter \nocite{wcp95}({Wampler}, {Chugai} \& {Petitjean} 1995).
Both of the above hypotheses remain technically viable for this object; %(see \S\,\ref{other});
however, the confirmed existence of redshifted \civ\ absorption
in other quasars makes these alternate explanations 
less likely.\footnote{Similarly, \nocite{maiolino+2004}{Maiolino} {et~al.} (2004) reported that
SDSS J104845.05$+$463718.4 (J1048) at $z=6.22$ has blueshifted
\SIiv\ and \civ\ absorption plus absorption which could be \civ\ redshifted by 
$v=3800$~\kms\ or \HeIIsf\ absorption in the blueshifted outflow.
The putative redshifted \civ\ absorption in J1048 appears less prominent in the
spectrum presented by \nocite{2010A&A...523A..85G}{Gallerani} {et~al.} (2010), however.  The trough may have
varied, but the significance of its detection is difficult to evaluate since no
error array accompanies either published spectrum.}

In addition to the new discoveries above, J1125 and J1128 from \nocite{sdss123}{Hall} {et~al.} (2002)
remain cases of redshifted \mgii\ absorption,\footnote{\nocite{sdss123}{Hall} {et~al.} (2002) 
also suggested that UN J1053$-$0058
and 3C 288.1 were candidates for redshifted absorption.
SDSS spectra obtained of those sources reveal that the absorption 
is not redshifted in either case.  It is at the systemic redshift as given by
\mgii\ for UN J1053$-$0058 and by numerous narrow lines for 3C 288.1.}
as discussed in \S\,\ref{intro}.  
In \S\,\ref{low} we discuss new BOSS spectra obtained for both objects.

%\clearpage

\subsection{Repeat spectra} \label{rptall}

\subsubsection{Repeat spectra of high-ionization cases} \label{rpt}

\begin{figure*} %%\vspace*{174pt} \makebox[\textwidth]{
\includegraphics[angle=0, width=0.490\textwidth]{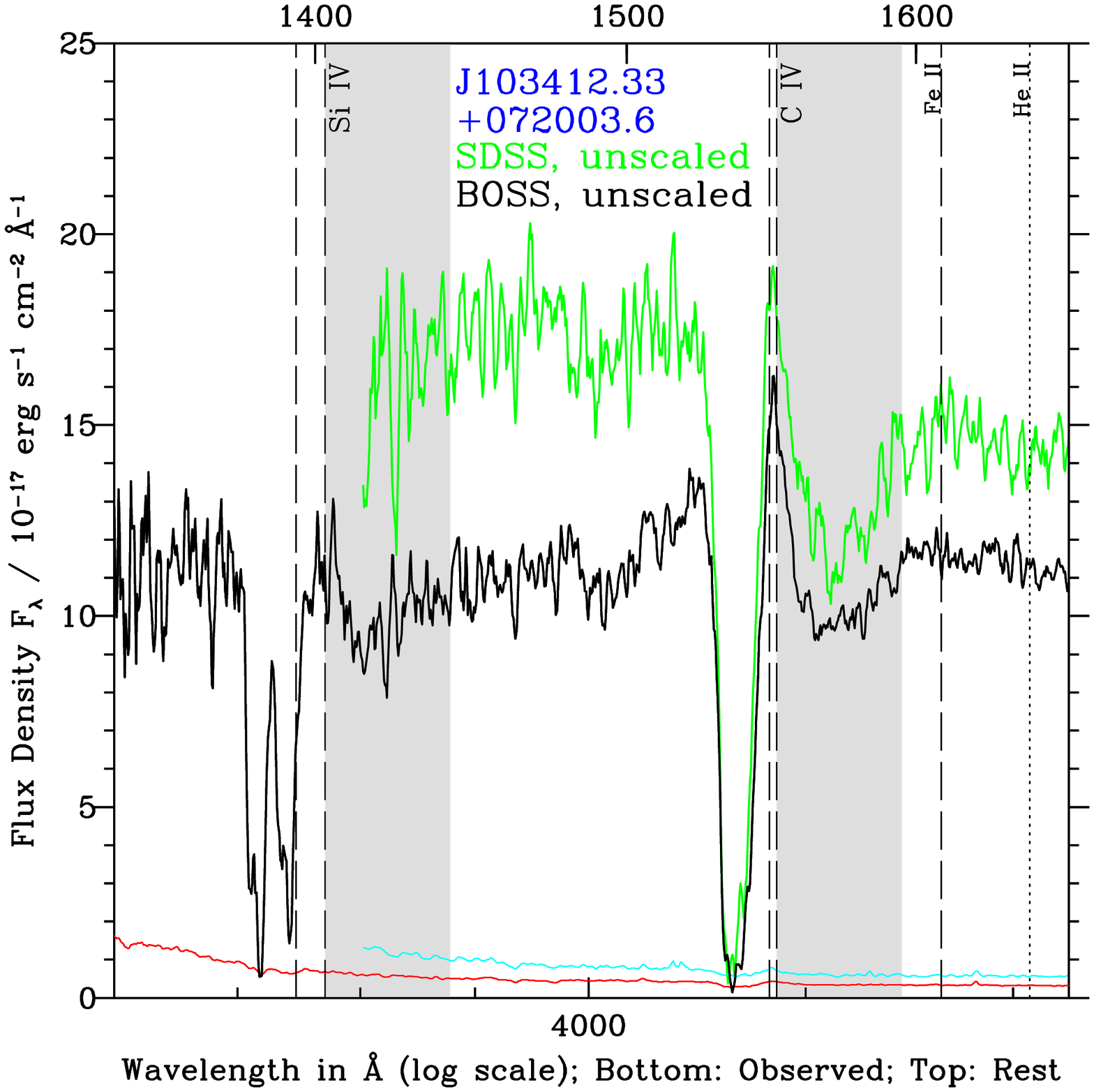} %}
\includegraphics[angle=0, width=0.490\textwidth]{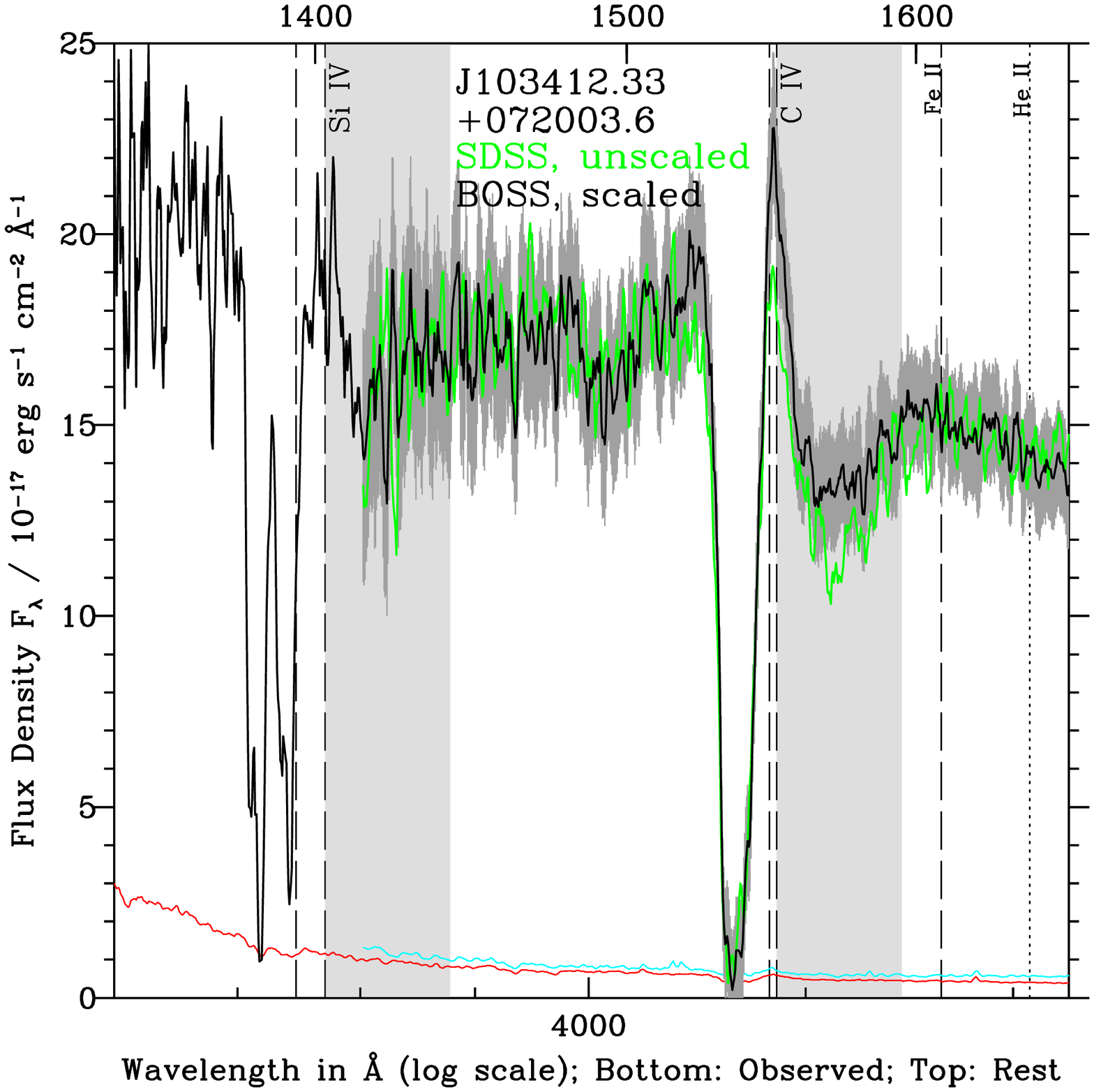} %}
\caption{
Spectra of J1034 in the \SiIV\ and \CIV\ region,
with dashed vertical lines showing the 
wavelengths of each component of each doublet at the adopted systemic redshift.
The vertical shaded regions show the velocity ranges where redshifted 
absorption is expected in the long-wavelength member of the doublets,
based on the relative velocities over which redshifted \civ\ absorption is
observed in the BOSS spectrum.
The first panel displays the SDSS spectrum (green, with error array in cyan)
and BOSS spectrum (black, with error array in red).
In the second panel, the BOSS spectrum has been multiplied by a constant
times a power-law to align the two spectra in continuum regions near \civ.
The grey error bars on the scaled BOSS spectrum show the $\pm 2\sigma$
statistical uncertainties ($\sigma^2=\sigma_{\rm SDSS}^2+\sigma_{\rm BOSS,scaled}^2$).
If the normalized absorption troughs in the two spectra were identical,
the grey region would include 95\% of the points in the green spectrum.
The redshifted \civ\ absorption has weakened in strength,
resulting in a stronger narrow emission peak for \civ,
while the blueshifted absorption has not changed significantly.
}
\label{fj1034} \end{figure*}

\begin{figure*} %%\vspace*{174pt} \makebox[\textwidth]{
\includegraphics[angle=0, width=0.490\textwidth]{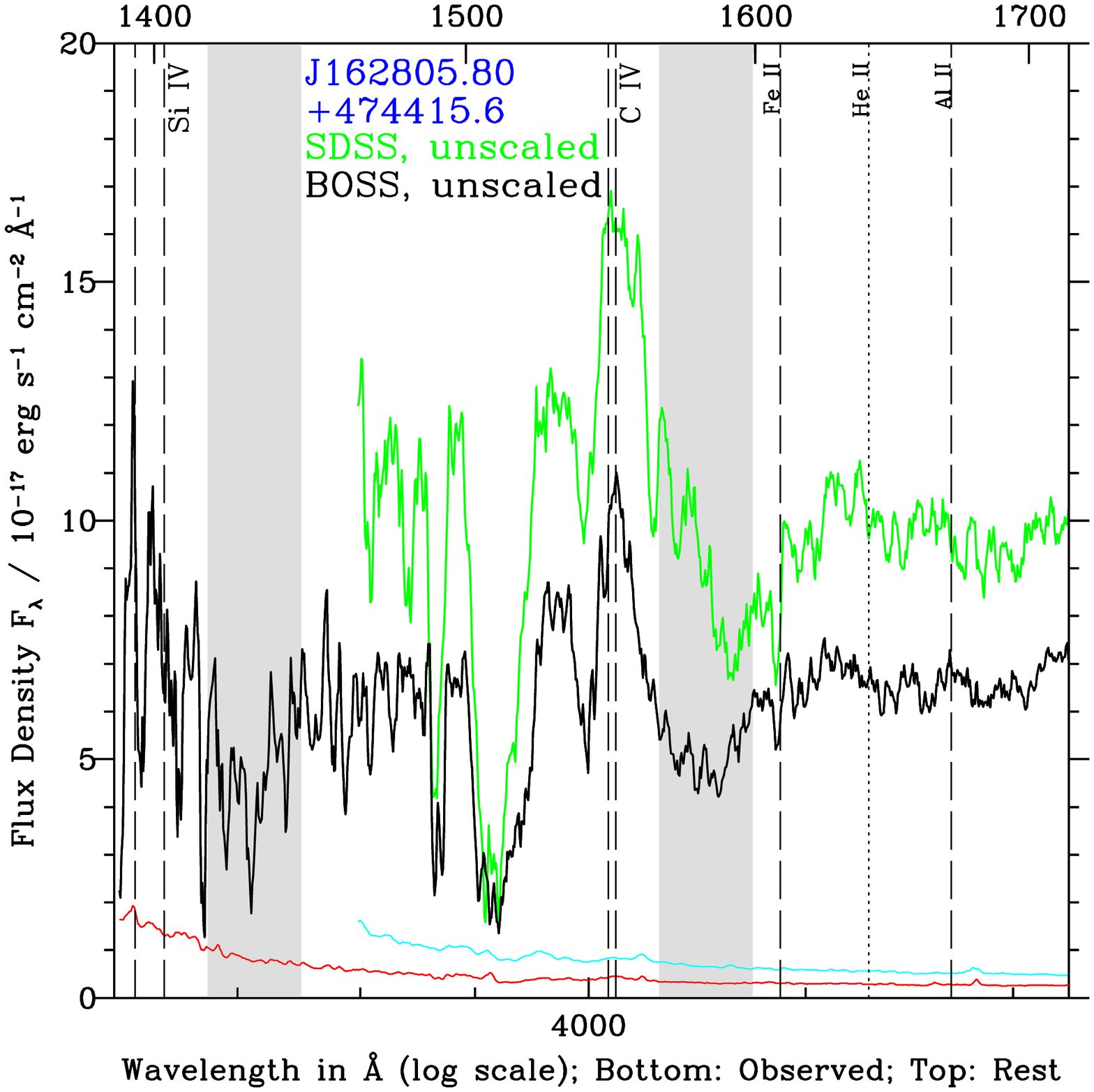} %}
\includegraphics[angle=0, width=0.490\textwidth]{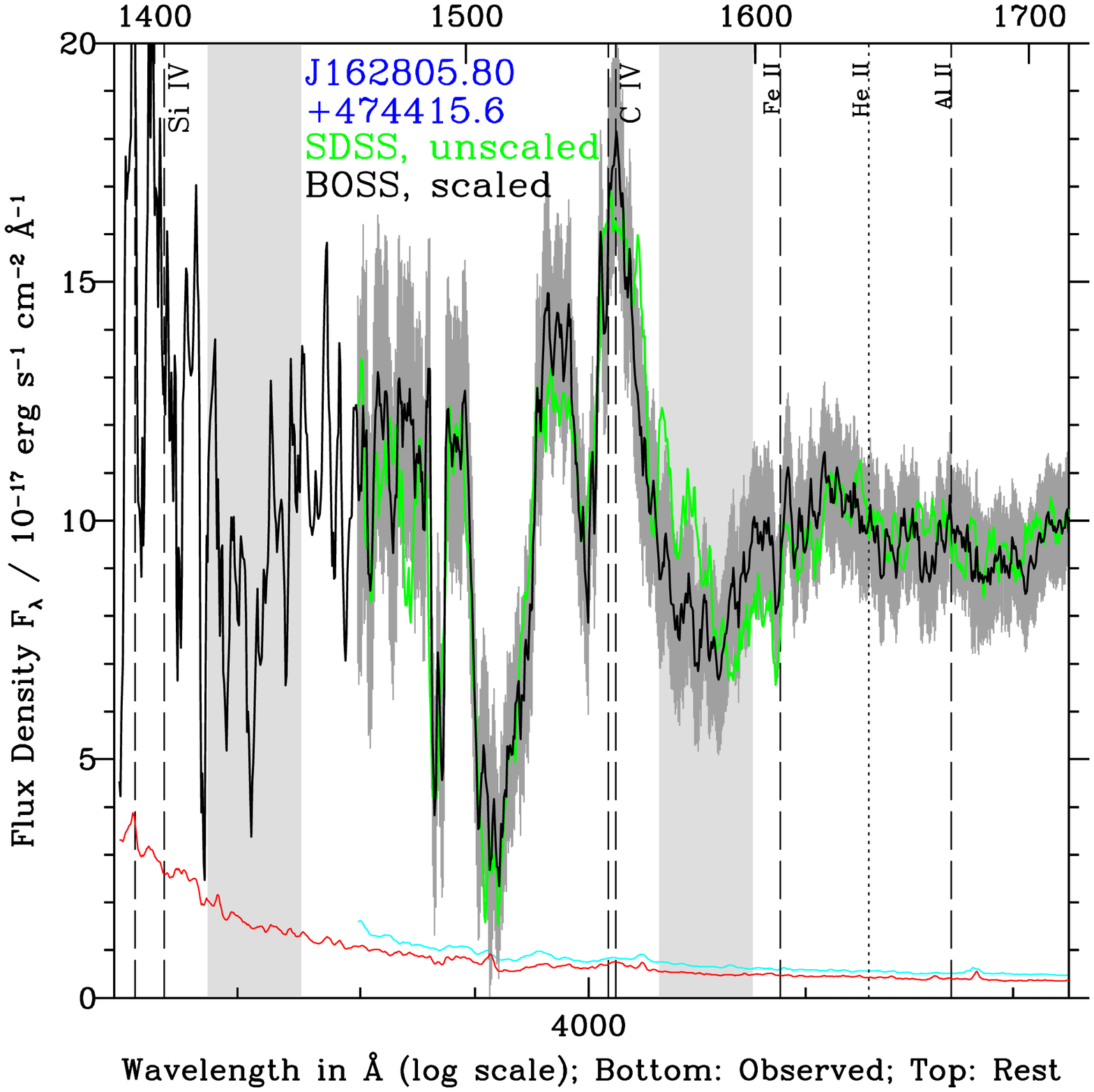} %}
\caption{
Spectra of J1628 in the \SiIV\ and \CIV\ region, displayed
in the same manner as in Figure \ref{fj1034}; see its caption for details.
The first panel shows the observed spectra,
while in the second panel the BOSS spectrum has been scaled to match
the SDSS spectrum in continuum windows near \civ.
Between the epochs of the SDSS spectrum (green) and the BOSS spectrum (black),
the redshifted \civ\ absorption has 
strengthened by $\sim 2\sigma$ per pixel at small redshifted velocities
and weakened by $\sim 2\sigma$ per pixel at large redshifted velocities.
}
\label{fj1628} \end{figure*}

\begin{figure*} %%\vspace*{174pt} \makebox[\textwidth]{
\includegraphics[angle=0, width=0.98\textwidth]{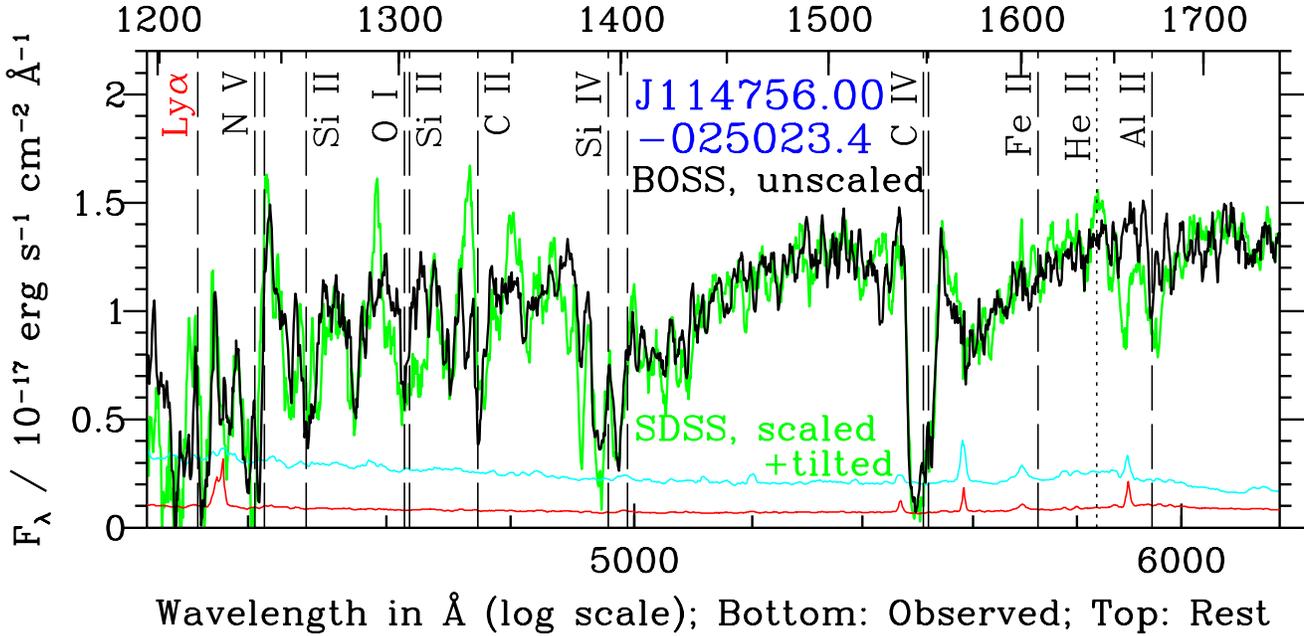} %}
\caption{For J1147, we show the BOSS spectrum (black, with error array in red) 
and the SDSS spectrum (green, with error array in cyan) which has been scaled 
by a constant times a power-law to match the BOSS spectrum in the wavelength
range shown.
}
  \label{fjothers}
\end{figure*}

\begin{figure*} %%\vspace*{174pt} \makebox[\textwidth]{
\includegraphics[angle=0, width=0.99\textwidth]{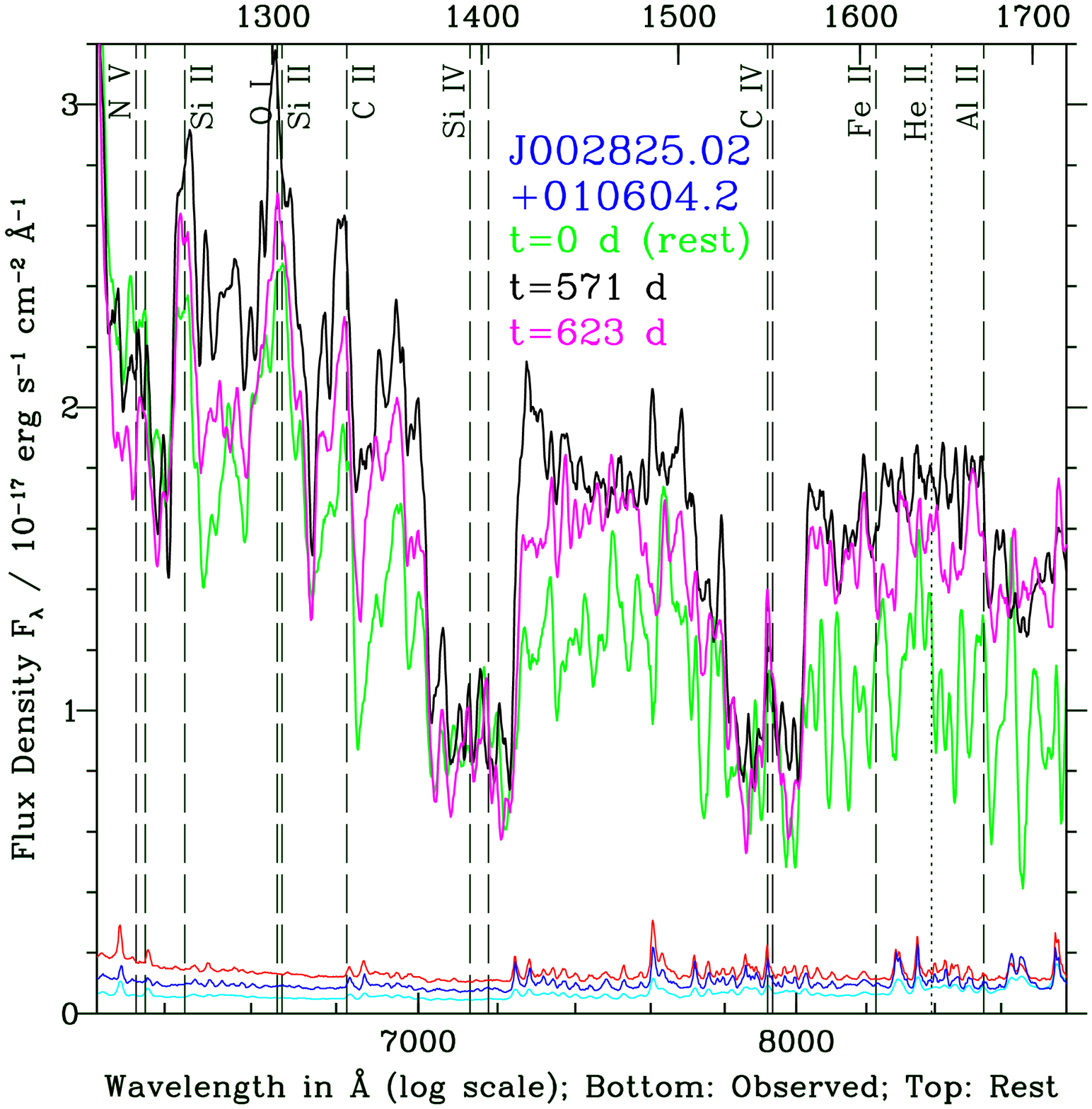} %}
\caption{For J0028, our one high-ionization object with three epochs of 
spectroscopy, we show the SDSS spectrum (green with cyan uncertainties),
the first-epoch BOSS spectrum (black with red uncertainties),
and the second-epoch BOSS spectrum (magenta with blue uncertainties).
All spectra have been heavily smoothed.
The times at which the spectra were taken are given in rest-frame days
starting from the epoch of the SDSS spectrum.
}
  \label{fj0028three}
\end{figure*}

Our sample includes four quasars with SDSS spectra of redshifted 
\civ\ absorption and two with SDSS spectra of redshifted \mgii\ absorption,
all of which now also have BOSS spectra.
Figure \ref{f_cands2} shows the unscaled SDSS and BOSS spectra for the
redshifted \civ\ absorption objects.
In the case of J1147, the SDSS spectrophotometry disagrees with both the SDSS
photometry and the BOSS spectrophotometry, and so we plot only the BOSS spectrum.
(There are also two quasars with candidate redshifted \civ\ absorption which
have SDSS spectra; see Figure \ref{f_cands3}.)

Between spectroscopic epochs separated by years in the rest frame,
quasars can vary in flux level and continuum slope as well as in absorption
properties (e.g., \nocite{fol87,2010ApJ...713..220G,2011MNRAS.411..247R}{Foltz} {et~al.} 1987; {Gibson} {et~al.} 2010; {Rodr{\'{\i}}guez Hidalgo},  {Hamann} \& {Hall} 2011).
To compare absorption troughs in these quasars' SDSS and BOSS spectra we
scale and tilt the BOSS spectra to match the SDSS spectra.\footnote{In only
one of our quasars with two spectra---namely, J1704---is the search for 
continuum variability affected by issues with the DR9 spectrophotometry of 
BOSS quasar survey targets taken on dates when washers were used to offset
those fibers to improve the spectral throughput in the blue \nocite{bossover}({Dawson} {et~al.} 2013).}
We scale by a constant factor to account for flux variability and tilt
through multiplication by a power-law to account for continuum slope differences
(apparent or real).  For display purposes, we normalize the spectra separately
for each figure rather than using a single global fit.

The results of this approach are shown for J1034 in Figure \ref{fj1034} 
and for J1628 in Figure \ref{fj1628}.
In each figure, the left- and right-hand panels show the unscaled and 
rescaled BOSS spectra (black), respectively, along with the unscaled 
SDSS spectrum (green).
In the right-hand panels, around each BOSS spectrum we show in grey 
the $\pm 2\sigma$ statistical uncertainty range,
where $\sigma$ accounts for both the SDSS and scaled BOSS uncertainties.
If the normalized absorption troughs in the two spectra were identical,
the grey region would include 95\% of the points in the green spectrum.

For J1034, in the 1135 rest-frame days between the SDSS and BOSS spectra the 
\civ\ absorption at the systemic redshift and at redshifted velocities weakened 
in relative strength, resulting in a stronger narrow emission peak for \civ.
The normalized blueshifted absorption did not change significantly.
The changes could be due to a strengthening of the underlying \civ\ emission
line plus a velocity-dependent broad-line region covering factor.
In any case, J1034 shows that redshifted and systemic absorption can vary in
normalized spectra independently of the blueshifted absorption. 

For J1628, in the 1559 rest-frame days between the SDSS and BOSS spectra 
the \civ\ emission line appears to have strengthened somewhat
(between 1520\,\AA\ and 1560\,\AA\ rest-frame).
The absorption at blueshifted velocities $v<-5000$~\kms\ did not change 
significantly.
The redshifted absorption trough has shifted to smaller redshifted velocities,
by about 2000~\kms\ on average.
Whether this is a case of deceleration or just covering factor variations
as a function of velocity cannot be determined without further observations.
If this is a case of gas undergoing decelerating infall 
due to outward radiative acceleration, 
the acceleration amounts to $|a| \simeq 1.5$~cm~s$^{-2}$.
For comparison, $|a|\simeq 100 \mbox{\ cm\ s}^{-2}$ is predicted in
the main acceleration region of a disc wind in the model of \nocite{mcgv}{Murray} {et~al.} (1995).
If the observed absorption arises in decelerating infalling gas,
and assuming constant ionizing luminosity, 
the deceleration should be seen to increase with time 
as the gas moves to smaller radii.

An additional result of this approach is shown in Fig. \ref{fjothers}.
No significant variations 
are seen in the absorption in J1147 over 886 rest-frame days.

Finally, in Figure \ref{fj0028three} we show one SDSS and two BOSS epochs
of spectroscopy for J0028, with no scaling.
In the 571 rest-frame days between the SDSS observation and the first
BOSS observation of J0028 (green to black spectra in Fig. \ref{fj0028three}),
the flux level of the unabsorbed continuum brightened by $\sim$50\%,
the flux levels within the two \SIii\ troughs and the \cii\ trough
increased by nearly 50\%, 
but the flux levels within the \SIiv, \civ\ and \Nv\ troughs,
both redshifted and blueshifted, did not change significantly 
except in the most highly blueshifted parts of the \SIiv\ and \civ\ troughs.
In the 52 rest-frame days between the two BOSS observations of J0028
(black to magenta spectra in Fig. \ref{fj0028three}),
the flux level across most of the spectrum decreased by $\sim$10\%.
Between both pairs of spectra, the trough variability showed no 
evidence for any change with velocity across the systemic redshift
dividing redshifted and blueshifted absorption.

\subsubsection{Repeat spectra of low-ionization cases} \label{low}

\begin{figure*} %%\vspace*{174pt} %\makebox[\textwidth]{
\includegraphics[angle=0, width=1.08\textwidth]{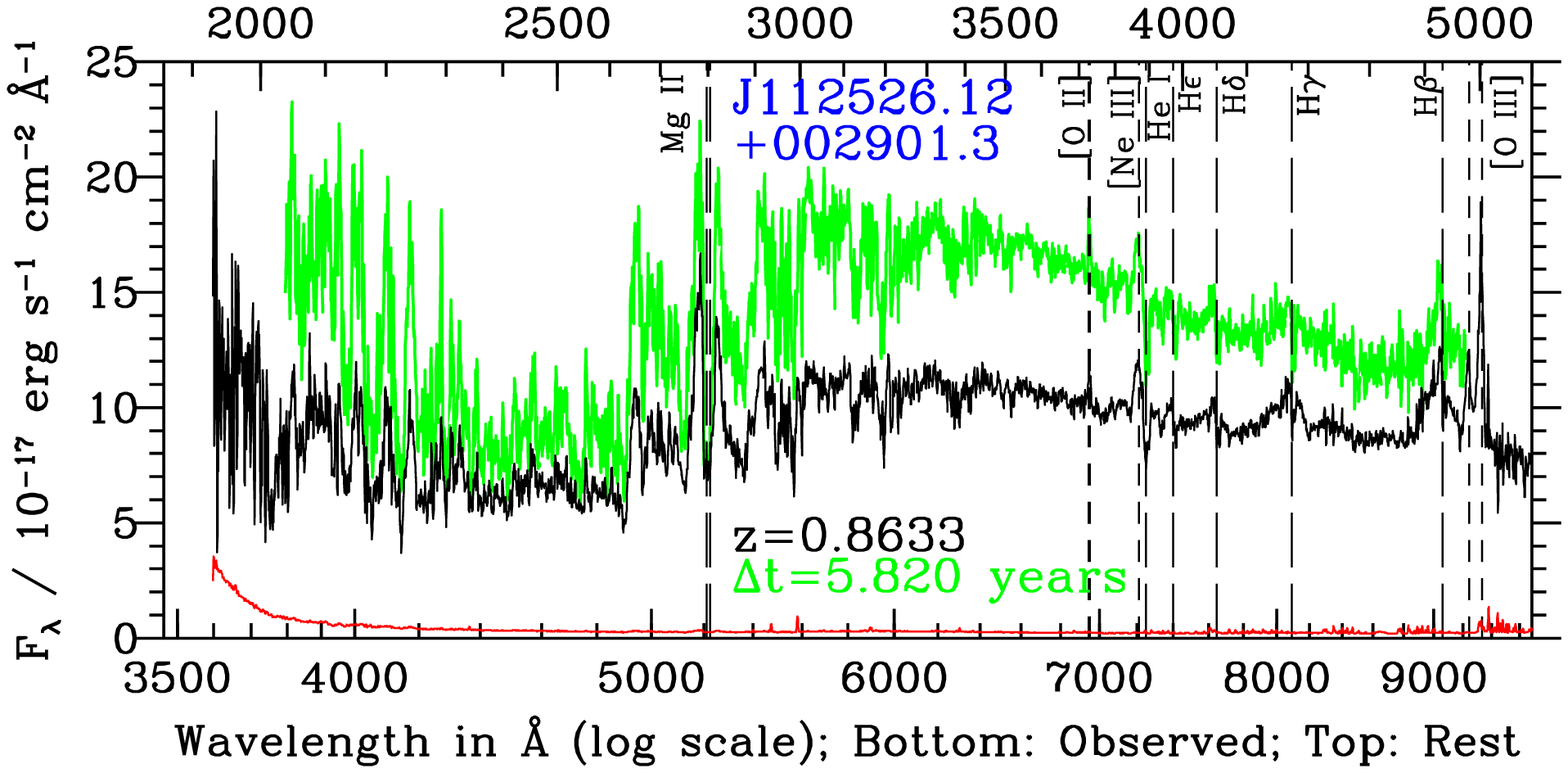}
\includegraphics[angle=0, width=1.08\textwidth]{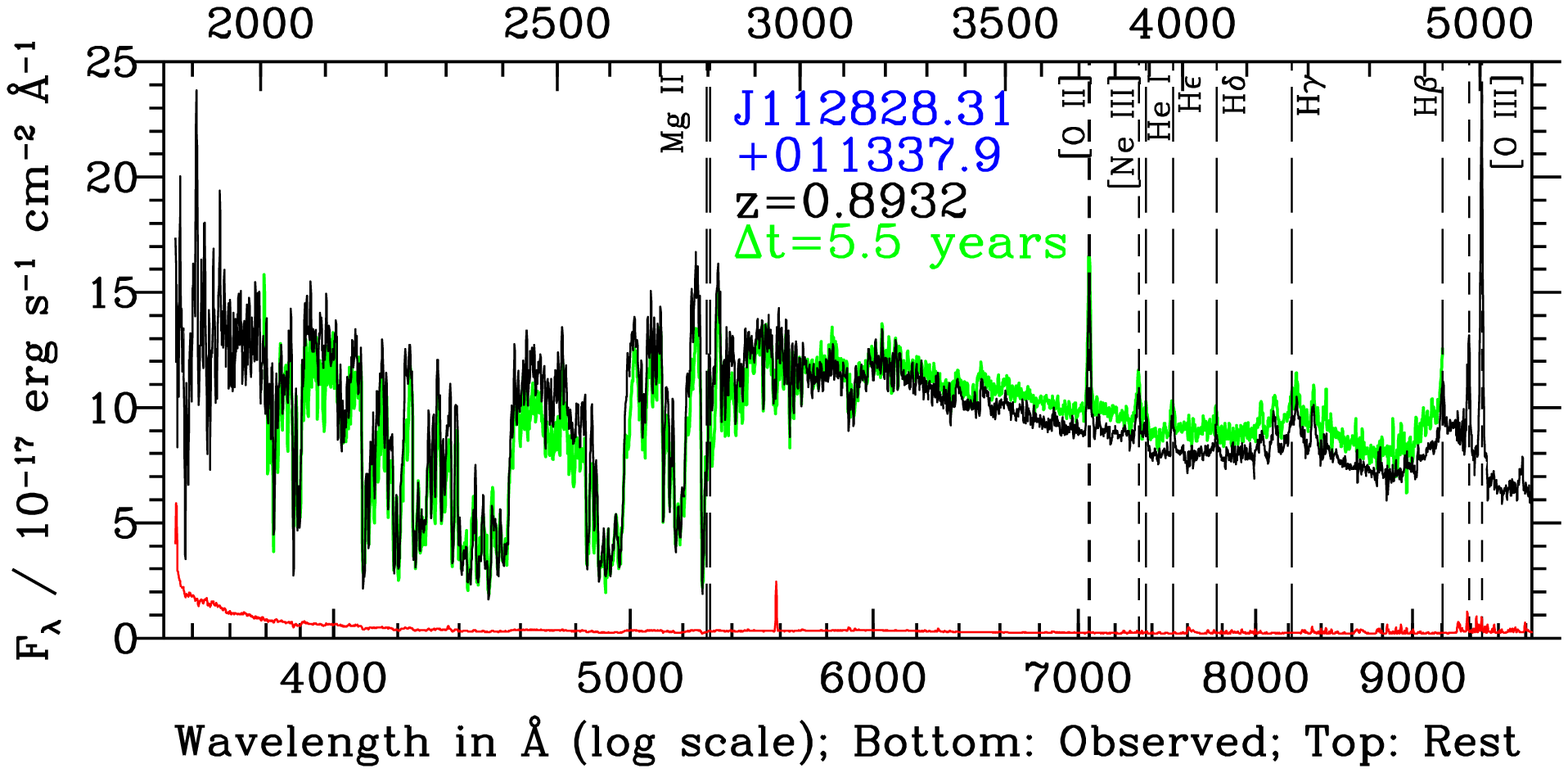}
\caption{SDSS (green) and BOSS (black) spectra of J1125 (top) and J1128 
(bottom), plotted as flux density per unit wavelength (in units of 10$^{-17}$
erg s$^{-1}$ cm$^{-2}$ \AA$^{-1}$) vs. wavelength in \AA\ on a log scale.
All spectra have been smoothed by a 3-pixel boxcar.  
The relatively narrow redshifted absorption present in both objects 
is difficult to see at this scale, but can be seen in Figure \ref{fj11s}.
There are two SDSS spectra available for J1128, which have been combined
into the single error-weighted average spectrum displayed.  The J1128 SDSS
spectra were taken 0.55 years apart in the quasar rest frame, which is why the
$\Delta t$ between the SDSS and BOSS spectra is less precise for J1128.
The red curves show the uncertainties on the smoothed BOSS spectra;
the uncertainties on the smoothed SDSS spectra are similar 
but for clarity are not shown.}\label{fj11}
\end{figure*}

\begin{figure*} %%\vspace*{174pt} %\makebox[\textwidth]{
\includegraphics[angle=0, width=1.08\textwidth]{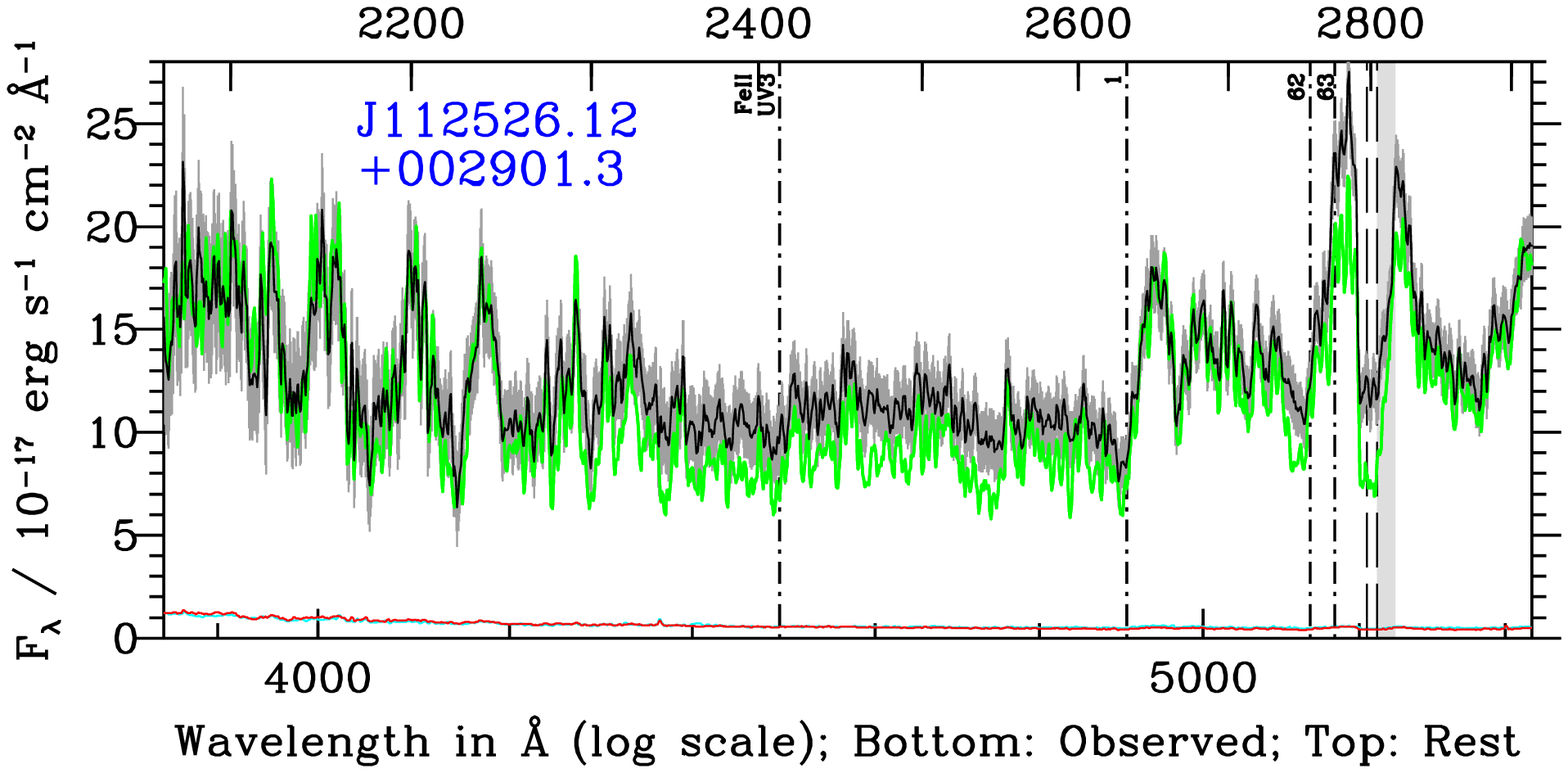}
\includegraphics[angle=0, width=1.08\textwidth]{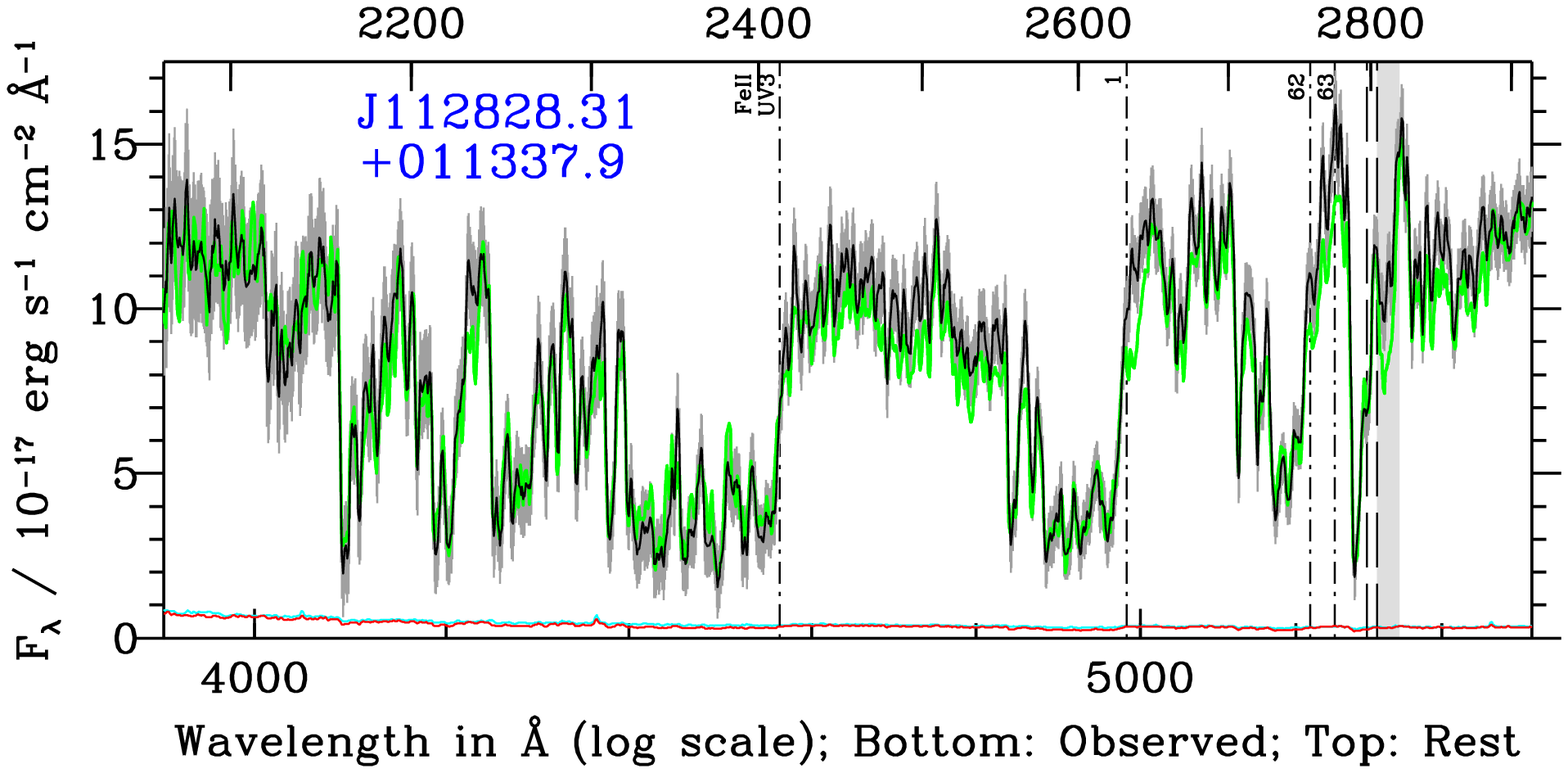}
\caption{SDSS (green) and scaled BOSS (black) spectra of J1125 (top) 
and J1128 (bottom) in a region near \MgII.  The BOSS spectra have been scaled
by a constant times a power-law to match the SDSS spectra 
in continuum regions near 2100\,\AA\ and 2910\,\AA\ rest frame.
The cyan and red curves show the uncertainties on the SDSS and scaled BOSS 
spectra, respectively.  
The grey region around each scaled BOSS spectrum (black) shows the 
$\pm 2\sigma$ statistical uncertainty range,
where $\sigma^2=\sigma_{\rm SDSS}^2+\sigma_{\rm BOSS,scaled}^2$.
If the normalized absorption troughs in the two spectra were identical,
the grey region would include 95\% of the points in the green spectrum.
The two dashed vertical lines show the wavelengths 
of the \mgii\ doublet at the systemic redshift.
The \mgii\ absorption is only 2200--3100~\kms\ wide in these objects,
roughly centered on the systemic redshift.  
The velocity range of redshifted absorption in each object is shown by 
the light grey shaded vertical bar which extends from 0\,\kms\ to
+1400\,\kms\ in J1125 and +1700\,\kms\ in J1128.
The dot-dashed vertical lines show the wavelengths 
of the longest members of four UV \feii* multiplets, 
denoted by their multiplet numbers from Moore (1950).}\label{fj11s}
\end{figure*}

Both quasars with previously known redshifted \mgii\ absorption, 
J1125 and J1128,
have also been observed by the BOSS (Figure \ref{fj11}).
We have adopted the redshifts from \nocite{hw10}{Hewett} \& {Wild} (2010) for both these objects.
The redshift for J1128 is almost identical to that used in \nocite{sdss123}{Hall} {et~al.} (2002).
The redshift for J1125 ($z=0.8633$, primarily from \oii) is 338~\kms\ smaller
than the $z=0.8654$ used in \nocite{sdss123}{Hall} {et~al.} (2002).  That latter redshift corresponds 
to narrow Balmer-line absorption which was recognized in
\nocite{sdss123}{Hall} {et~al.} (2002) and assumed to be host-galaxy absorption.  However, a small but
growing number of quasars are now known to exhibit Balmer-line absorption in
their outflows (\nocite{balmerbal}{Hall} 2007; \nocite{aoki10}{Aoki} 2010; and references therein).
Furthermore, \oii\ is offset by only $-$21$\pm$5~\kms\ on average from 
host-galaxy absorption in SDSS quasars \nocite{hw10}({Hewett} \& {Wild} 2010).  Therefore the Balmer-line
absorption is most likely part of the redshifted absorption in this object,
and we no longer adopt the redshift of that absorption as systemic.

Both quasars show broad Balmer-line emission, narrow \oii\ and \oiii\ emission,
and relatively narrow (few 1000~\kms\ wide) absorption from many ground-state 
(\feii) and excited-state (\feii*) transitions of singly ionized iron
\nocite{cem50}({Moore} 1950), as well as absorption in \mgii.
Superpositions of these narrow lines form the apparent broad troughs in the 
spectra.  There is also narrow \feii\ emission in the spectra; for example,
the narrow emission lines flanking H$\gamma$ in J1128 
are slightly redshifted \feii\ (\nocite{vjv04}{V{\'e}ron-Cetty}, {Joly} \&  {V{\'e}ron} 2004).

As in the previous section, to compare absorption properties at both epochs
we used a constant times a power law to rescale the BOSS spectra.  For these
objects we scaled to the SDSS spectra 
in continuum regions 
near 2100\,\AA\ and 2910\,\AA\ rest frame.
The resulting spectra are shown in Fig. \ref{fj11s}; again, the green spectrum
precedes the black spectrum.  
Around each BOSS spectrum (black), we show in grey the $\pm 2\sigma$ 
statistical uncertainty range. 

In both objects, the two normalized spectra show small but significant 
differences near \mgii\ which cannot be matched by smooth continuum
variations.  The \mgii\ absorption, emission, or both must have varied.

In J1125, the blueshifted \mgii\ absorption 
has weakened more than the redshifted absorption has.
The underlying \mgii\ emission may have strengthened 
relative to the continuum (Figure \ref{fj11} is consistent with 
constant emission-line flux and weakening continuum flux),
but different behaviour in blueshifted and redshifted absorption would still
be required if that were the case.
Stronger emission and apparently weakened blueshifted absorption 
are also seen in \hg\ and \hb\ (not shown).
Also, the \feii* absorption between 2300--2630\,\AA\ has apparently weakened,
with most of the scaled BOSS spectrum $>2\sigma$ above the SDSS spectrum.

In J1128, %J1128 has had less change in its absorption.
there have been $\sim 4\sigma$ decreases 
in the strengths of the 
redshifted absorption troughs in \mgii\ and \feii* multiplets UV1, UV62 and 
UV63, visible just to the right of the vertical lines marking those transitions
in the bottom panel of Figure \ref{fj11s}.  There has been no significant 
weakening in the blueshifted absorption in those transitions.

As with \civ\ absorption in the previous section,
it appears that redshifted \mgii\ absorption
can vary independently of blueshifted absorption.

%\clearpage
\section{Properties of BAL quasars with redshifted troughs} \label{props}

\subsection{Multiwavelength data} \label{multi}

We found no X-ray observations of useful depth of our objects in the
{\em Chandra}, {\em XMM-Newton}, {\em Suzaku}, {\em BeppoSAX}, {\em Swift XRT},
or pointed {\em ROSAT} archives, which we searched using HEASARC
Browse\footnote{http://heasarc.gsfc.nasa.gov/cgi-bin/W3Browse/w3browse.pl}.

We searched for counterparts of our objects within 2 arcseconds in the 
Wide-field Infrared Survey Explorer \nocite{wise}(WISE; {Wright} {et~al.} 2010) 
All-Sky Source Catalog.  Fourteen objects with confirmed \civ\ absorption 
were detected (all except J0805, J0828 and J1709),\footnote{The detection 
of J1439 is offset 1.7 arcseconds to the west, likely due to contamination
from a galaxy 4$\farcs$8 to the west which has a photometric redshift of
$z\sim 0.3$ \nocite{2008ApJ...674..768O}({Oyaizu} {et~al.} 2008).}
along with four of seven candidates (all except J1342 and J2133)
and all objects with confirmed or candidate redshifted \mgii\ absorption.  
Three of the five non-detections are the three faintest objects in our sample, 
all with $i>21$.  None of our objects exhibit obviously unusual $W1-W2$, 
$W2-W3$ or $i-W3$ colors as compared to other quasars in DR9Q.
A full multiwavelength study of these objects to determine how well their
spectral energy distributions match those of standard BAL and non-BAL quasars
will be very useful, but is beyond the scope of this work.

We note in passing that none of our confirmed objects and only two of our
candidates have radio detections in FIRST (Table 1), one with candidate
redshifted \civ\ absorption and one with candidate redshifted \mgii\ absorption.
The detection rate of DR9Q quasars at $z>1.57$ (where BOSS spectra cover
\SIiv\ to \civ) is 3.26\%, so between zero and two radio-detected objects in 
our redshifted \civ\ sample is statistically consistent with expectations.

%\clearpage

\subsection{Trough shapes} \label{uvw}

The absorption profiles of the redshifted troughs (or the redshifted parts of 
the troughs) in these objects resemble those of blueshifted troughs in many 
ways.  

When present, troughs from low-ionization species (ionization stages I to III)
are generally narrower and weaker than those from high-ionization species 
(stages IV and higher); e.g., \aliii\ in J2157 (Fig. \ref{f_zoom}), or
\SIii$\lambda$1304 and \CII\ in J1019 (Fig. \ref{f_cands}f and \ref{f_closeups}f).
The lack of objects with \civ\ absorption much stronger than absorption
in other transitions is unlikely to be due purely to the selection effect of 
typically requiring absorption in multiple troughs to
identify a quasar as having redshifted absorption.
We have found no cases where strong, redshifted \civ\ is
present but redshifted \SIiv\ is not; only J1324 (Appendix \ref{maybe})
comes close.

The residual intensities are similar in transitions with very different
transition probabilities from species of very different abundances,
indicative of partial covering of the emission regions at optical depth 
$\tau \gg 1$ \nocite{aea99b,2003ARA&A..41..117C}(e.g., {Arav} {et~al.} 1999; {Crenshaw}, {Kraemer} \&  {George} 2003).
Because the continuum emission region and the broad emission line region
(BELR) have very different sizes, the BAL gas may have different covering 
factors of each region.  Examination of these objects' spectra near 
\civ\ shows that both the continuum source and the BELR
can be fully covered by gas at systemic and blueshifted velocities.
However, in no case is the continuum source fully covered by redshifted 
absorbing gas.  Moreover, the redshifted gas need not cover the \civ\ BELR 
at all in any of these objects (though if it does not, the continuum source
would be close to fully covered in J0830, J1019 and J1709).
That possibility means that the redshifted absorbing gas could be located
interior to the BELR \nocite{aea99b}(e.g. {Arav} {et~al.} 1999), though it does not need to be
(we cannot rule out equal covering of the continuum source and the BELR).
The only exception to the above is the candidate J0050 (Appendix \ref{maybe}),
in which the absorption has nearly full coverage of both the continuum and 
the \civ\ BELR.

Broadly speaking, we can classify our quasars with high-ionization troughs 
into three categories based on the shapes of their absorption troughs:
\vshaped, $\mathsf W$-shaped, or $\mathsf U$-shaped.
See \S\,\ref{ai} for a discussion of trough shapes in terms
of quantitative absorption strengths.

Six of our quasars, including J2157, have \vshaped\  absorption
troughs which are strongest at their short-wavelength ends and
which extend smoothly to longer wavelengths at decreasing depths.
(In some cases the \civ\ trough is interrupted by narrow emission,
but the \SIiv\ trough shows the underlying \vshaped\  trough.)
These quasars are
J0148 (Fig. \ref{f_cands}a and \ref{f_closeups}a),
J0805 (Fig. \ref{f_cands}b and \ref{f_closeups}b),
J1146 (Fig. \ref{f_cands}g and \ref{f_closeups}g),
J1147 (a borderline case; see Fig. \ref{f_cands2}c and Fig. \ref{fjothers}),
J1439 (Fig. \ref{f_cands}i and \ref{f_closeups}i) and
J2157 (Fig. \ref{f_zoom})
plus, if confirmed, the candidates J1239, J1633, and J2133 (Appendix \ref{maybe}).
Our two low-redshift quasars with redshifted \mgii\ absorption (J1125 and J1128;
Fig. \ref{fj11s}) also have \vshaped\  troughs, albeit quite narrow ones.

Five of our quasars have what appear to be separate blueshifted and 
redshifted absorption troughs, with the same structure in \civ\ and \SIiv;
we refer to such troughs as $\mathsf W$-shaped.
These quasars are
J0828 (Fig. \ref{f_cands}c and \ref{f_closeups}c),
J0941 (Fig. \ref{f_cands}e and \ref{f_closeups}e),
J1034 (Fig. \ref{f_cands2}b),
J1440 (Fig. \ref{f_cands}j and \ref{f_closeups}j) and
J1628 (Fig. \ref{f_cands2}d)
plus, if confirmed, the candidates J1316 and J1342 (Appendix \ref{maybe}).
The blueshifted absorption is stronger than the redshifted absorption in all
cases, although the relative strengths are nearly equal in J0941 and J1324.
In the case of J1034, the weakness of the \civ\ emission line
between the two troughs means the absorption is likely to be continuous
over a span of $>$11000~\kms.  J1034 may have less absorption at the systemic
redshift or it may have a higher covering factor of the continuum than of the
emission-line region.

Six of our quasars have troughs which have relatively sharp edges
at both their short- and long-wavelength ends and which reach maximum depth
near the trough center (apart from occasional weak, narrow emission in \civ).
We refer to such troughs as $\mathsf U$-shaped.  These quasars are
J0028 (Fig. \ref{f_cands2}a),
J0830 (Fig. \ref{f_cands}d and \ref{f_closeups}d),
J1019 (Fig. \ref{f_cands}f and \ref{f_closeups}f),
J1323 (Fig. \ref{f_cands}h and \ref{f_closeups}h),
J1709 (Fig. \ref{f_cands}k and \ref{f_closeups}k) and
J1724 (Fig. \ref{f_cands}l and \ref{f_closeups}l)
plus, if confirmed, the candidates J0050 and J1704 (Appendix \ref{maybe}).

Note that four of our quasars have absorption only at redshifted velocities:
J1019, J1146, J1709 and J1724
plus, if confirmed, the candidate J0050 (Appendix \ref{maybe}).

\subsection{Low-ionization absorption} \label{lobals}

Among BAL quasars with redshifted \civ\ absorption, 
LoBAL quasars are greatly overrepresented.  Of our 17 confirmed 
cases of redshifted \civ\ absorption, 12 to 14 also have redshifted 
low-ionization absorption, and one more (J1034) has blueshifted
\aliii\ absorption but no clear redshifted \aliii\ absorption.  
The resulting LoBAL fraction is 56\%--92\% (90\% confidence range).
LoBAL quasars make up only $\sim$5\% of the population of all SDSS BAL quasars 
(and thus only 1\%--2\% of the entire SDSS quasar population; 
\nocite{trump06,allenbal}{Trump} {et~al.} 2006; {Allen} {et~al.} 2011).  
However, LoBAL quasars constitute a larger fraction of the BAL quasar 
population in near-infrared-selected quasar samples: a factor of two larger
comparing the studies of \nocite{daiss08}{Dai}, {Shankar} \& {Sivakoff} (2008) and \nocite{daiss12}{Dai}, {Shankar} \& {Sivakoff} (2012), and a remarkable
20 LoBAL quasars out of 21 BAL quasars in \nocite{2012ApJ...757...51G}{Glikman} {et~al.} (2012).
Our LoBAL classifications are based on the detection of either \CII\ or 
\AlIII\ troughs, and there are no plausible alternate identifications 
for those troughs.  

With only one FeLoBAL quasar in our sample, we can only say that 
they do not appear to be overrepresented 
among BAL quasars with redshifted \civ\ absorption 
to the same extent that LoBAL quasars are.

\SIiv\ absorption is also more common among BAL quasars with redshifted
\civ\ absorption than among all BAL quasars.
(This is unlikely to be a selection effect; see \S\,\ref{uvw}.)
\nocite{allenbal}{Allen} {et~al.} (2011) find \SIiv\ absorption in only 42.5\%$\pm$1.4\% of BAL quasars
with \civ\ absorption.
Redshifted \SIiv\ absorption is seen in all but one of our BAL quasars with 
redshifted \civ\ absorption and spectral coverage of \SIiv.  The exception is 
the candidate object J1342 (Appendix \ref{maybe}), and even it has
blueshifted \SIiv\ absorption.

\subsection{Absorption Index values} \label{ai}

\setcounter{table}{1}
\begin{table*}
 \centering
 \begin{minipage}{177mm}
 \caption{Trough Properties of Confirmed and Candidate BAL Quasars with Redshifted \civ\ Absorption}
  \begin{tabular}{@{}lllrrllrrr@{}}
\hline
Quasar & Plate-MJD-Fiber & Shape  & \multicolumn{2}{c}{\civ~blueshifted} & \multicolumn{2}{c}{\civ~redshifted} & \civ & \civ    & \civ\  \\
~      & ~               &        & \multicolumn{2}{c}{$v$~range,~\kms}  & \multicolumn{2}{c}{$v$~range,~\kms} & AI   & AI$^+$  & \ait\  \\
\hline
\multicolumn{10}{c}{Quasars with redshifted \civ\ absorption}\\
\hline			
J0028 & 0689-52262-363  & \ushape  &  $-8220$$\pm$780 &     $0$         &    0         &  3310$\pm$830 &  1382$\pm$658 &   843$\pm$270 &  2225$\pm$940 \\
...   & 3586-55181-0548 & \ushape  &  $-8080$$\pm$880 &     $0$         &    0         &  3510$\pm$120 &  2206$\pm$384 &  1381$\pm$92  &  3586$\pm$477 \\
...   & 4220-55447-0790 & \ushape  &  $-8840$$\pm$440 &     $0$         &    0         &  3510$\pm$250 &  2203$\pm$776 &  1483$\pm$255 &  3686$\pm$1029 \\
J0148 & 4273-55506-0334 & \vshape  &  $-3160$$\pm$160 & $-1440$$\pm$120 & 2020$\pm$210 & 6090$\pm$1080 &   538$\pm$159 &   664$\pm$383 &  1203$\pm$543 \\
J0805 & 4458-55536-0158 & \vshape  &  $-2000$$\pm$70  &     $0$$\pm$70  & 2140$\pm$120 &  3930$\pm$160 &  1040$\pm$145 &   327$\pm$145 &  1367$\pm$290 \\
J0828 & 3762-55507-0690 & \wshape  &  $-5290$$\pm$85  & $-1700$$\pm$210 & 1270$\pm$400 & 6130$\pm$1500 &   869$\pm$200 &   296$\pm$229 &  1166$\pm$513 \\
J0830 & 4489-55545-0364 & \ushape  &  $-2080$$\pm$70  &     $0$         &    0         &  3170$\pm$90  &   743$\pm$72  &  1326$\pm$102 &  2069$\pm$175 \\
J0941 & 3782-55244-0417 & \wshape & $-11600$$\pm$880 & $-2010$$\pm$1720 & 6370$\pm$2200 & 14780$\pm$340 & 3301$\pm$712 &  1718$\pm$793 &  5019$\pm$1505 \\
J1019 & 4802-55652-0629 & \ushape  &  ...             & ...             &  800$\pm$70  &  8300$\pm$160 &     0         &  2672$\pm$100 &  2672$\pm$100 \\
J1034 & 0999-52636-503  & \wshape  &  $-4340$$\pm$80  &  $-960$$\pm$70  & 1730$\pm$250 &  7600$\pm$70  &  2110$\pm$66  &   758$\pm$100 &  2868$\pm$165 \\
...   & 4852-55689-0592 & \wshape  &  $-4130$$\pm$70  & $-1030$$\pm$70  & 2360$\pm$70  &  6840$\pm$70  &  2035$\pm$7   &    81$\pm$28  &  2116$\pm$22  \\
J1146 & 4614-55604-0726 & \vshape  &  ...             & ...             & 1420$\pm$880 &  7980$\pm$540 &     0         &   176$\pm$159 &   176$\pm$159 \\
J1147 & 0329-52056-176  & \vshapeW &  $-2420$$\pm$80  &     $0$         &    0         &  8900$\pm$120 &  1706$\pm$352 &  1366$\pm$10  &  3071$\pm$356 \\
...   & 3790-55208-0290 & \vshapeW &  $-2210$$\pm$70  &     $0$         &    0         & 10560$\pm$70  &  1694$\pm$70  &  1780$\pm$335 &  3474$\pm$406 \\
J1323 & 4006-55328-0170 & \ushape  &  $-2900$$\pm$70  &     $0$         &    0         &  5600$\pm$70  &   849$\pm$26  &  1391$\pm$10  &  2240$\pm$26  \\
...   & 4050-55599-0624 & \ushape  &  $-3100$$\pm$160 &     $0$         &    0         & 5520$\pm$1080 &  1195$\pm$218 &  1338$\pm$605 &  2533$\pm$823 \\
J1439 & 4780-55682-0992 & \vshape  &   $-910$$\pm$70  &     $0$         &    0         & 9400$\pm$1800 &   306$\pm$33  &  1602$\pm$356 &  1908$\pm$389 \\
J1440 & 3868-55360-0886 & \wshape  & $-7100$$\pm$1070 &   $-90$$\pm$85 & 1800$\pm$1270 & 12600$\pm$400 &  3062$\pm$167 &  1307$\pm$509 &  4369$\pm$676 \\
J1628 & 0625-52145-121  & \wshape  &  $-9810$$\pm$240 & $-5710$$\pm$100 & 5610$\pm$590 & 12400$\pm$1000 & 2034$\pm$292 &   934$\pm$215 &  2967$\pm$506 \\
...   & 6322-56190-0599 & \wshape  &  $-9710$$\pm$80 & $-3500$$\pm$2000 & 2710$\pm$120 & 11100$\pm$1200 & 2058$\pm$141 &  1167$\pm$282 &  3225$\pm$422 \\
J1709 & 5014-55717-0763 & \ushape  &  ...             & ...             &  700$\pm$70  &  4430$\pm$70  &     0         &  1804$\pm$48  &  1804$\pm$48  \\
J1724 & 5002-55710-0600 & \ushape  &  ...             & ...             &    0$\pm$640 &  9350$\pm$540 &     0         &  3812$\pm$618 &  3812$\pm$618 \\
J2157 & 4197-55479-0030 & \vshape  &  $-1930$$\pm$85  &     $0$         &    0         &  9050$\pm$70  &  1329$\pm$34  &  2627$\pm$187 &  3956$\pm$221 \\
\hline
\multicolumn{10}{c}{Quasars with candidate redshifted \civ\ absorption}\\
\hline
J0050 & 4306-55584-0846 & \ushape  &  ...             & ...             &    0$\pm$200 &  1780$\pm$70  &     0         &  1397$\pm$10  &  1397$\pm$10  \\
J1239 & 0521-52326-221  & \vshape  &  $-5340$$\pm$160 & 0$\pm$200       & 2360$\pm$250 & 11670$\pm$590 &  1226$\pm$431 &  1152$\pm$155 &  2378$\pm$276 \\
...   & 4754-55649-0294 & \vshape  &  $-2340$$\pm$70  & 0$\pm$200       & 1770$\pm$160 & 11640$\pm$160 &  1139$\pm$55  &  1766$\pm$35  &  2905$\pm$21  \\
J1342 & 3987-55590-0124 & \wshape & $-10900$$\pm$1170 & $-4120$$\pm$120 &  920$\pm$70  &  6990$\pm$80  &  1385$\pm$129 &   793$\pm$14  &  2178$\pm$117 \\
J1633 & 4061-55362-0256 & \vshape  &  $-1700$$\pm$70  & $0$             &    0         &  7300$\pm$2800 &  930$\pm$14  &   601$\pm$121 &  1531$\pm$135 \\
J1704 & 1687-53260-150  & \ushape  &  $-9180$$\pm$70  & $0$             &    0         &  2410$\pm$240 &  4128$\pm$142 &   560$\pm$66  &  4688$\pm$209 \\
...   & 4177-55688-0220 & \ushape  &  $-9800$$\pm$350 & $0$             &    0         &  2550$\pm$120 &  4428$\pm$288 &   447$\pm$92  &  4875$\pm$380 \\
J2133 & 4088-55451-0426 & \vshape  &  $-1400$$\pm$100 & $0$             &    0         &  5200$\pm$80  &   891$\pm$260 &   854$\pm$190 &  1809$\pm$495 \\
\hline
\end{tabular}\\
The Plate-MJD-Fiber column gives the SDSS/BOSS spectroscopic plate number,
MJD of observation, and fiber number.
SDSS and BOSS observations can be distinguished because SDSS uses a three-digit fiber number
(001 to 600) while BOSS uses a four-digit fiber number (0001 to 1000).
Trough shapes are discussed in \S~\ref{uvw}.
The velocity ranges of absorption in \civ\ are measured relative to the redshift given in Table 1,
but do not include the systematic uncertainties on those systemic redshifts.
Entries with a 0 in both the blueshifted and redshifted columns
represent absorption troughs which are continuous across the systemic redshift.
Absorption indices are discussed in \S~\ref{ai}; the values here were calculated
using $v_{min}=450$~\kms.
\label{tab2}
\end{minipage}
\end{table*}

\begin{figure*} 
\includegraphics[angle=0, width=0.497\textwidth]{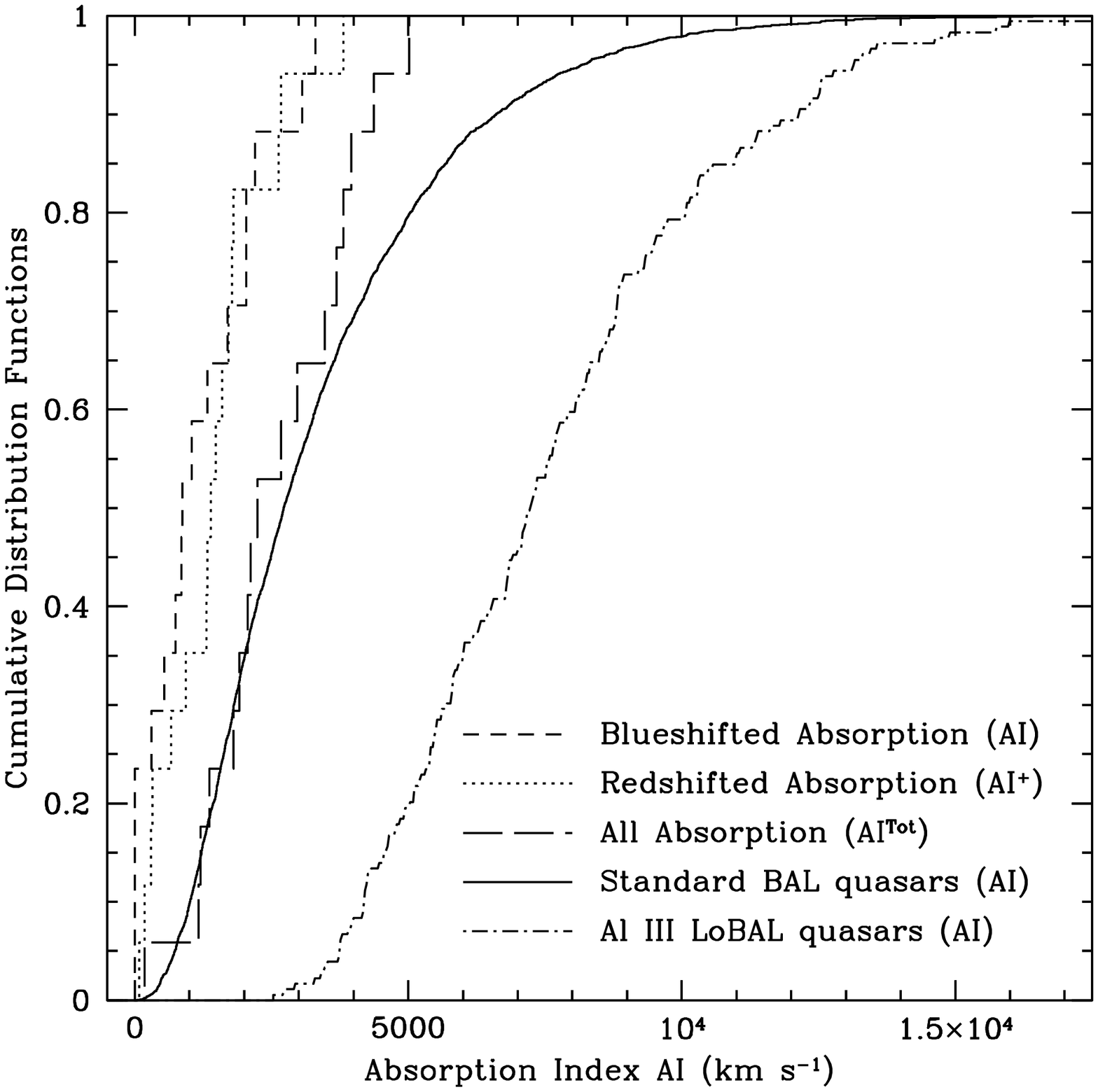} %}
\includegraphics[angle=0, width=0.497\textwidth]{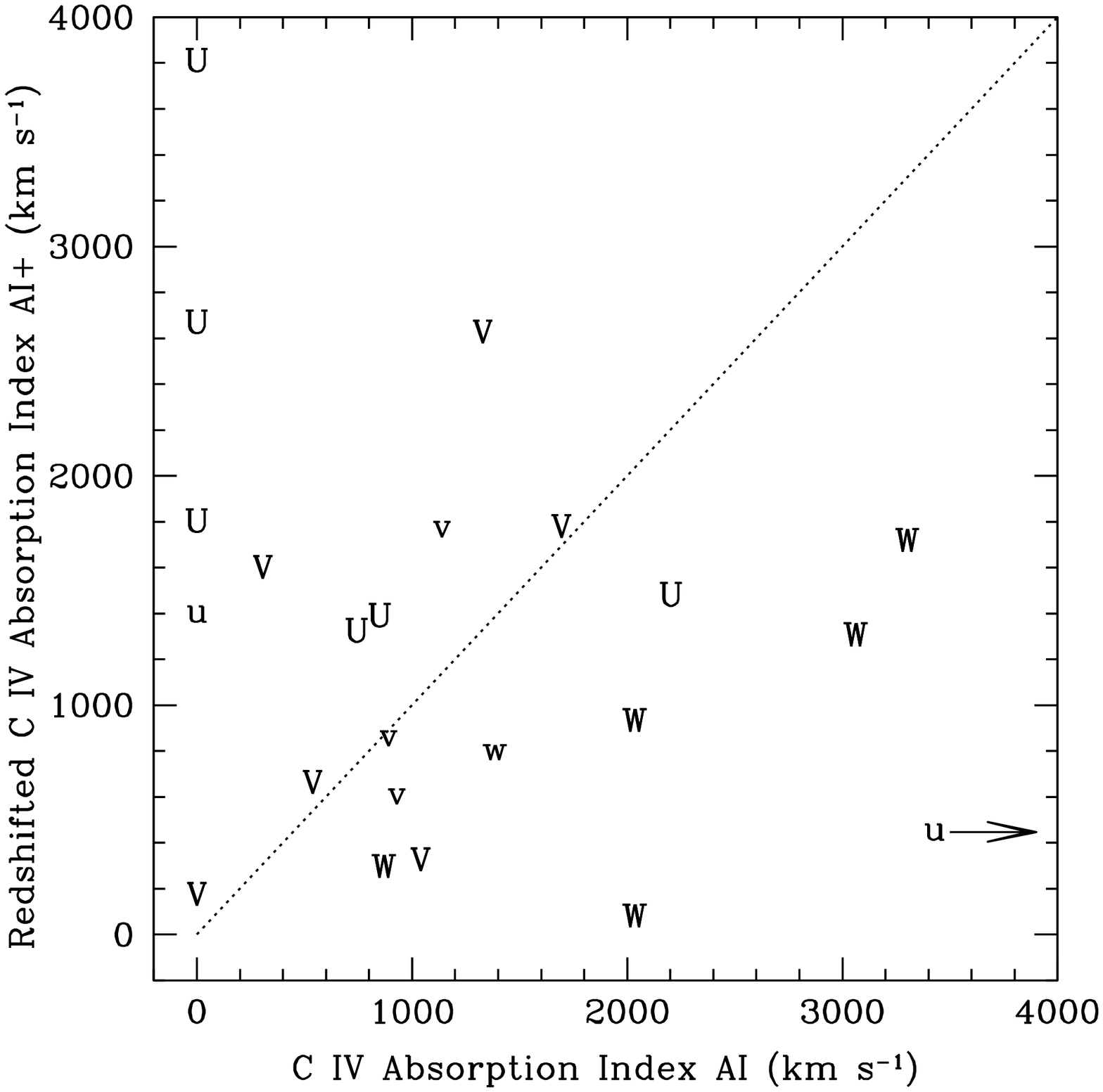} %}
\caption{{\bf Left:} 
The cumulative distribution functions of the \civ\ absorption index values 
(AI; see \S~\ref{ai}) for various quasar subsamples. 
From left to right, approximately:
redshifted-trough BAL quasars' AI values (short-dashed line),
AI$^+$ values (dotted line), and
AI$^{\rm Tot}$ values (long-dashed line);
visually identified BAL quasars from DR9Q (solid line);
DR9Q LoBAL quasars identified by \aliii\ absorption (dot-dashed line).
For each redshifted-trough object, we used the measurements from the 
spectrum with the highest signal-to-noise ratio (S/N) at 1325\,\AA--1625\,\AA.
{\bf Right:} AI$^+$ values (redshifted absorption strength) plotted 
vs. AI values (blueshifted absorption strength) for BAL quasars with
redshifted \civ\ absorption, both from the same highest-S/N spectrum.
The symbol plotted for each quasar is its trough type; see \S\,\ref{uvw}.
Uppercase letters are used for confirmed objects and
lowercase letters for candidate objects.
The dotted line shows 
the locus of equal redshifted and blueshifted absorption strength.
}
\label{f_rsai} \end{figure*}

The balnicity index (BI), which is essentially a modified equivalent width,
was introduced by \nocite{wea91}{Weymann} {et~al.} (1991) as a measure of 
the strength of a blueshifted absorption trough in a quasar spectrum.
The absorption index (AI; \nocite{sdss123,trump06}{Hall} {et~al.} 2002; {Trump} {et~al.} 2006) is a modification of
the BI which extends to narrower troughs and to blueshifted velocities 
closer to systemic.
We use the AI to measure the strength of blueshifted absorption, and we define
AI$^+$ as the corresponding measurement for redshifted absorption and 
\ait\ as the corresponding measurement for absorption at all velocities.
With $f(v)$ being the normalized rest-frame\footnote{For non-singlet
transitions, we use the rest frame of the longest-wavelength component of 
the transition.  Doing so yields a conservative measure of the AI$^+$.}
spectrum as a function of velocity from the systemic redshift, and with 
negative velocities representing blueshifted absorption, we have:
\begin{eqnarray} \label{eai}
{\rm AI}(v_{min}) = \int_{-c}^{0} [1-f(v)/f_{int}] ~C' ~dv \\
{\rm AI}^+(v_{min}) = \int_{0}^{c} [1-f(v)/f_{int}] ~C' ~dv \\
{\rm AI}^{\rm Tot}(v_{min}) = \int_{-c}^{c} [1-f(v)/f_{int}] ~C' ~dv %\\
\end{eqnarray} 
where $C'=1$ in contiguous intervals of width $v_{min}$ or greater 
within the integration limits
wherein the quantity in brackets is everywhere positive; otherwise, $C'=0$.
Although we will write AI instead of AI$(v_{min})$ in general,
in the definition we use AI$(v_{min})$ to emphasize that
AI values depend on the $v_{min}$ used.
In \nocite{sdss123}{Hall} {et~al.} (2002) and DR9Q, $v_{min}=450$~\kms\ and $f_{int}=0.9$ have 
been used.  For ease of comparison with DR9Q, we adopt those values
also.\footnote{\nocite{trump06}{Trump} {et~al.} (2006) adopted $v_{min}=1000$~\kms\ and used
$f_{int}=0.9$ to calculate $C'$ but $f_{int}=1$ to calculate AI values,
so that the AI would be a true equivalent width measured in \kms.}
We have given the limits to the integrals as the speed of light to emphasize
that the ideal AI would measure the absorption in a given transition at all
outflow velocities, even though in practice limits on the integrals will 
be set by confusion with absorption from other transitions.

We present \civ\ absorption indices for our quasars in Table 2.
AI values were calculated automatically in the DR9Q only for 
five confirmed and three candidate objects, due to low signal-to-noise 
ratios or poor continuum fits in other objects.
We therefore calculated our own AI, AI$^+$ and \ait\ values for all our 
quasars to ensure internally consistent measurements.  We first performed 
weighted-average smoothing on each spectrum, using a smoothing box size of 
the minimum odd number of pixels required to yield an average signal-to-noise 
ratio $\geq$10 in the 1325\,\AA--1625\,\AA\ wavelength range; typically,
this was 7 pixels (each BOSS pixel spans 69\,\kms\ in the observed frame).
We then normalized each smoothed spectrum by a fifth-order polynomial
continuum fit to wavelengths usually free from strong emission or absorption
(see Appendix A of \nocite{sdss123}{Hall} {et~al.} 2002), adjusted as necessary for each object 
individually.
The AI values calculated from these normalized continua are
expected to be underestimated relative to AI values from DR9Q because DR9Q
accounts for emission lines in the normalizing continuum and we do not.
We find this to be the case for seven out of eight objects where a comparison
is possible.
It is of course preferable to account for emission lines, but to do so 
accurately for these unusual objects will require study of their continuum
and emission-line properties which is beyond the scope of this work.
We calculated \ait\ between $-$15000~\kms\ and 15000~\kms\ in general, 
but adjusted the limits for individual quasars as needed
to avoid obvious intervening narrow absorption and 
to include all intrinsic absorption from the transition under consideration.
We do not use the traditional blueshifted limit of $-$25000~\kms\ to avoid
counting redshifted \SIiv\ as blueshifted \civ.
We estimated the uncertainties for the velocity limits 
and AI, AI$^+$ and \ait\ values by repeating our measurements 
on spectra normalized using a different method 
--- that of \nocite{2012ApJ...757..114F}{Filiz Ak} {et~al.} (2012) ---
and then calculating
the dispersion between those two measurements of each quantity.
The average uncertainty on the AI values is $\sim$20\%.
For the velocity uncertainties, we added in quadrature 
a base uncertainty of 69\,\kms, equivalent to one spectral pixel.
The average velocity uncertainty is $\sim$450\,\kms; note that this
is in addition to the systematic uncertainty on the quasar redshift
(the velocity zeropoint).

In the left panel of Figure \ref{f_rsai} 
we compare the cumulative distribution functions of 
non-zero
\civ\ AI values for visually identified BAL quasars from DR9Q (solid line)
with those of redshifted-trough BAL quasars' AI values (short-dashed line),
AI$^+$ values (dotted line), and AI$^{\rm Tot}$ values (long-dashed line).
Although the distribution of AI$^{\rm Tot}$ 
in our small sample of BAL quasars with redshifted troughs appears to lack
the tail to high values seen in the AI distribution of all DR9Q 
visual BAL quasars, the difference is not statistically significant:
there is a 52\% chance of identical intrinsic distributions.
(We calculate all probabilities here using 
a two-sample Kuiper's test \nocite{nr3}({Press} {et~al.} 2007), 
a variant of the K-S test which is equally sensitive
to deviations at all values of the distribution.)
The differences between the AI distribution in all visual BAL quasars 
and the AI or AI$^+$ distributions of BAL quasars with redshifted troughs 
are only marginally statistically significant, due to our small sample size
(3.1\% or 0.12\% chances of identical intrinsic distributions, respectively,
equivalent to $2.1\,\sigma$ or $3\,\sigma$).
The AI and AI$^+$ distributions of BAL quasars with redshifted troughs 
are themselves statistically indistinguishable 
(67\% chance of identical intrinsic distributions).

Note, however, that despite commonly having low-ionization absorption
(\S~\ref{lobals}), redshifted-trough BAL quasars are completely inconsistent
with being a random sample of LoBAL quasars,
as traced by DR9Q quasars with \aliii\ absorption (dot-dashed line).
LoBAL troughs are preferentially seen in standard BAL quasars when the
\civ\ absorption is strong (see, e.g., Figures 5 and 6 of \nocite{allenbal}{Allen} {et~al.} 2011),
but the \civ\ AI distribution of redshifted-trough BAL quasars lacks the tail
to large AI values seen in standard BAL quasars, let alone in LoBAL quasars.

In the right panel of Figure \ref{f_rsai} we plot the redshifted absorption
index AI$^+$ versus the (blueshifted) absorption index AI for confirmed and
candidate quasars with redshifted absorption.  
There is no overall correlation between AI$^+$ and AI.

There is a tendency for differently shaped troughs to inhabit different regions
of the diagram.
In $\mathsf W$-shaped troughs, the redshifted absorption is 
always weaker than the blueshifted absorption.
The opposite is true in most \vshaped\  
and especially in most $\mathsf U$-shaped troughs.
Selection effects may help explain the latter: $\mathsf U$-shaped and 
\vshaped\  troughs with only a small redshifted component might have
been missed in the assembly of our sample, given redshift uncertainties 
and the difficulty of detecting weak absorption atop broad emission lines.
However, selection effects cannot explain the lack of $\mathsf W$-shaped 
troughs with stronger redshifted than blueshifted absorption.

More objects and better systemic redshift determinations will be useful in
interpreting this diagram.  
If the separation of different trough shapes in this diagram persists
with more objects added and after more study of selection effects,
that may suggest that different mechanisms are responsible for different
trough shapes.

To summarize, in our objects the AI and AI$^+$ values are uncorrelated overall,
but the \ait\ values
are a statistical match to
the AI distribution for standard BAL quasars.
This result is somewhat surprising, given that our objects
are predominantly LoBALs (see \S~\ref{lobals}) and that LoBALs have stronger
average \civ\ absorption than HiBALs.
This contradiction indicates that there is likely 
a different origin for at least the low-ionization
absorption in our objects as compared to standard BAL quasars.

\subsection{How rare are these objects?} \label{frac}

For an approximate estimate of how common BAL quasars with redshifted troughs
are, we begin with the SDSS DR5 BAL catalog of \nocite{gibsonbal}{Gibson} {et~al.} (2009)
and the SDSS DR6 BAL catalog of \nocite{allenbal}{Allen} {et~al.} (2011).
The strength of a BAL trough can be measured by its balnicity index 
(\S~\ref{ai}).
\nocite{gibsonbal}{Gibson} {et~al.} (2009) quote values for both the traditional balnicity index (BI),
measured between blueshifts of $-$3000~\kms\ and $-$25000~\kms,
and a modified balnicity index BI$_0$ measured between blueshifts 
of 0~\kms\ and $-$25000~\kms.  \nocite{allenbal}{Allen} {et~al.} (2011) only quote BI values.

We have found four SDSS quasars with redshifted \civ\ absorption 
(Figure \ref{f_cands2}) and two SDSS quasars with redshifted \mgii\ absorption 
(Figure \ref{fj11}) which we can compare to SDSS BAL quasar samples.

Of 4664 quasars with \civ\ BI$_0$$>$0 in \nocite{gibsonbal}{Gibson} {et~al.} (2009), 
3 have redshifted absorption
(J0028 does not make that cut).
Of 3880 quasars with \civ\ BI$>$0 in \nocite{gibsonbal}{Gibson} {et~al.} (2009), 
only J1628 has redshifted absorption.
Of 3317 quasars with \civ\ BI$>$0 in \nocite{allenbal}{Allen} {et~al.} (2011), 
only J1034 has redshifted absorption.
Thus, in the SDSS at most about 1 in 1600 BAL quasars 
with blueshifted \civ\ absorption also has redshifted \civ\ absorption 
(0.064\%, with a 90\% confidence range of 0.028\%$-$0.124\%,
based on the BI$_0$ statistics and determined following \nocite{geh86}{Gehrels} 1986).

The quasars we have discovered with redshifted \civ\ absorption 
in the BOSS DR9Q are only marginallly consistent with that result.
Of our seventeen quasars with redshifted absorption, two were only included in 
the BOSS as ancillary targets because of that absorption (J1034 and J1628),
one was targeted as a galaxy (J0941), and three others have
only redshifted absorption.  The remaining eleven % including J0028,J1147
have a parent sample of 7228 visually identified BAL quasars at $z>1.57$
(where BOSS spectra cover \SIiv\ to \civ) in the DR9Q \nocite{bossdr9q}({P{\^a}ris} {et~al.} 2012).
Thus, approximately 1 in 660 DR9Q quasars with blueshifted absorption also has
redshifted absorption (0.15\%, with 90\% confidence range 0.11\%$-$0.21\%).
Note that the overall incidence of quasars with redshifted BAL troughs among
BOSS DR9Q quasar targets at $z>1.57$ is approximately 1 in 5000 (14/69674).

In summary, the frequency of BAL quasars with both blueshifted and redshifted
\civ\ absorption differs by a factor of two between the SDSS and the BOSS 
surveys, although the 90\% confidence ranges overlap due to small-number 
statistics.  If the difference in frequency is confirmed, it might mean that 
the incidence of redshifted absorption troughs decreases with increasing quasar 
UV luminosity, as the mean luminosity of BOSS quasars is lower than that of 
SDSS quasars by about a factor of four.
However, a K-S test between the $M_i$ distributions 
of our sample and of other BOSS BAL quasars reveals no statistically 
significant discrepancy (22.5\% chance of identical intrinsic distributions).

\begin{figure*}
\includegraphics[angle=0, width=1.000\textwidth]{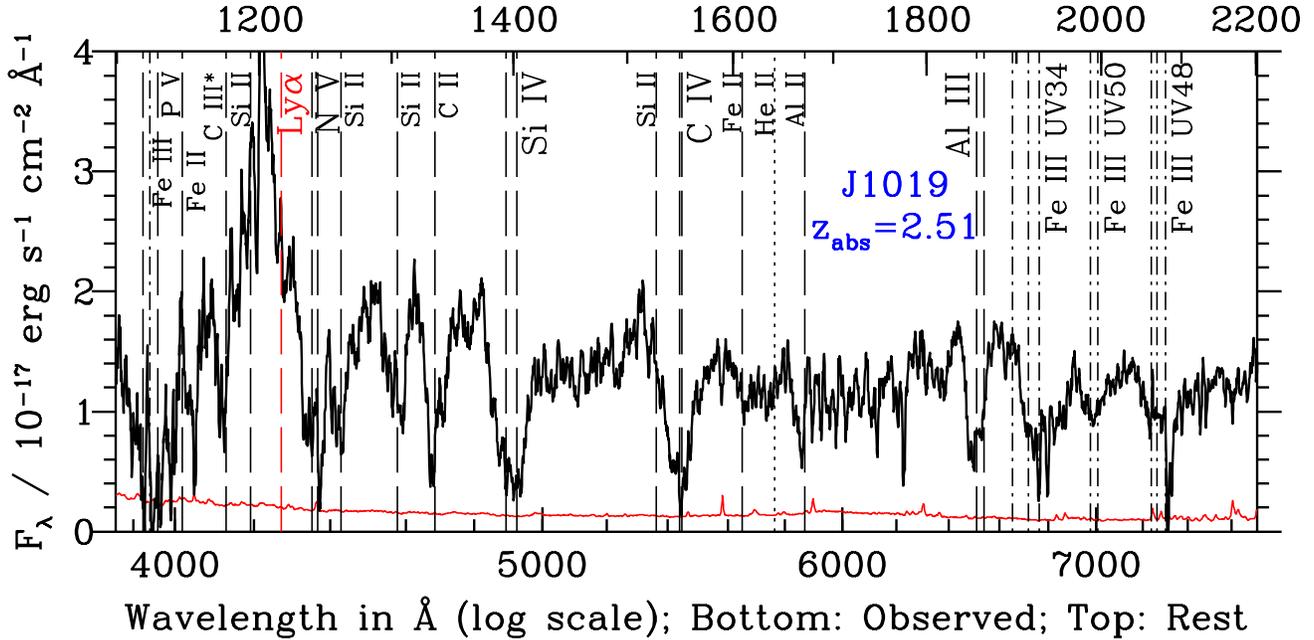} %}
\caption{Detailed view of the spectrum of J1019.  The lower axis shows observed
wavelengths and the upper axis shows wavelengths in the rest frame of maximum
apparent absorption at $z$=2.51 (redshifted by 5000\,\kms\ from the systemic
redshift of $z$=2.452).  Dashed vertical lines identify the wavelengths of notable
transitions, most of which are detected.  Dot-dashed lines indicate \feiii\ 
transitions, and the dotted line shows the wavelength at which absorption in
\HeIIsf\ would be seen.}
\label{f_j1019zoom} \end{figure*}

The statistics are somewhat different for redshifted \mgii\ absorption.
Of 332 quasars with \mgii\ BI$_0$$>$0 in \nocite{gibsonbal}{Gibson} {et~al.} (2009),
two have redshifted absorption
(or three, if the candidate J1316 is confirmed; see Appendix \ref{maybe}).
Of 244 quasars with \mgii\ BI$>$0 in \nocite{gibsonbal}{Gibson} {et~al.} (2009),
two have redshifted absorption.
Of 222 quasars with \mgii\ BI$>$0 in \nocite{allenbal}{Allen} {et~al.} (2011),
two have redshifted absorption.
Thus, approximately 1 in 130 BAL quasars with blueshifted \mgii\ absorption
also has redshifted \mgii\ absorption (0.60\%, with a 90\% confidence range of 
0.19\% to 1.35\%, based on the BI$_0$ statistics of the two confirmed objects).
We are unwilling to place much emphasis on this higher abundance of redshifted
\mgii\ troughs based on two objects, especially since the 90\% confidence
ranges overlap, but it is clearly worthy of further investigation.

\subsection{A distance constraint on the absorber in SDSS J101946.08+051523.7} \label{dist}

In Fig. \ref{f_j1019zoom} we show a detailed view of the spectrum of
J1019 (\S\,\ref{cases}).
J1019 may have absorption in \Cthree, a set of transitions arising from
levels with an excitation potential (EP) 6.5 eV above the ground state.  
\Cthree\ can be a useful density diagnostic due to the range of critical 
densities of its different levels \nocite{2005ApJ...631..741G}(section 4 of {Gabel} {et~al.} 2005).
J1019 also has absorption in
\feiii\,UV34 \lalala 1895,1914,1926
(EP 3.73\,eV, $n_{crit}\simeq 10^5$\,cm$^{-3}$),
\feiii\,UV48 \lalala 2062,2068,2079
(EP 5.08\,eV, $n_{crit}\simeq 10^{8.5}$\,cm$^{-3}$) and
\feiii\,UV50 (EP 7.86\,eV, $n_{crit}\simeq 10^{9.5}$\,cm$^{-3}$),
a multiplet spanning 1987--1996\,\AA.
The \feiii\ troughs appear to extend to redshifted velocities as large as
seen in \civ\ and larger than seen in \aliii.  
The short-wavelength end of the \feiii\ UV34 trough appears to be filled in
by emission from a combination of \ciii, \Siiii, and \feiii\ UV34 itself.

Strong absorption from \feiii\ without strong accompanying \feii~absorption
is rare in part because it occurs only for absorbers with column density
too low to form a hydrogen ionization front (outside of which \feii\ will form)
and with electron densities sufficiently high ($\geq 10^{10.5}$\,cm$^{-3}$) 
to increase the recombination of \feiv\ to 
\feiii\ \nocite{2011NewA...16..128R}({Rogerson} {et~al.} 2011, section 4).  Such densities have 
also been seen in outflowing gas in NGC~4051 \nocite{2012ApJ...746....2K}({King}, {Miller} \&  {Raymond} 2012).
J1019 and possibly J1316 (Appendix \ref{maybe}) represent the first reported
detections of \feiii\ UV50 absorption in quasars.
Its rarity is explained by the relative populations 
of the ground states of the UV34, UV48 and UV50 multiplets 
at $n_e\geq 10^{10.5}$\,cm$^{-3}$ and $T=10^4$~K; 
namely, 1:0.16:0.016 \nocite{1998ApJ...492..650B}({Bautista} \& {Pradhan} 1998).

Thus, the density in the medium causing the absorption at redshifted velocities
of thousands of \kms\ in J1019 is higher than the gas-phase densities found
everywhere else in a galaxy, except in accretion discs 
\nocite{hq10}(e.g., Fig. 1 of {Hopkins} \& {Quataert} 2010).  
Even protostellar cores in molecular clouds only reach gas-phase densities of
$\simeq 10^7$\,cm$^{-3}$ \nocite{2007prpl.conf...17D}({di Francesco} {et~al.} 2007).  If this dense gas
is in fact accreting onto the quasar, it must have been compressed
at some stage in the accretion process.

This lower limit on the density $n_e$ enables us to set an upper limit on the
distance $d$ of the absorbing gas from the ionizing continuum source by
rearranging the definition of the ionization parameter $U$:
$d=\sqrt{Q/4\pi cn_eU}$ where $Q$ is the number of hydrogen-ionizing photons
emitted per second.  J1019 is almost exactly an order of magnitude less luminous
than the quasar studied by \nocite{2009ApJ...706..525M}{Moe} {et~al.} (2009), so its value of $Q$
should also be an order of magnitude less: $\log Q = 56.5$.
With \feiii\ most abundant at $\log U \simeq -2\pm 0.5$ 
\nocite{dek02b}({de Kool} {et~al.} 2002),\footnote{It is possible to see \feiii\ absorption at
ionization parameters $U\leq 2.5$, but the relative strengths of absorption 
from \feiii\ and from \feii\ in the object cannot be reproduced
at such parameters.  To see \feiii\ UV50 at $\tau\geq 0.5$ over the width of
the trough in J1019 requires $N_H\geq 10^{21.5}$\,cm$^{-3}$, which is 
sufficient to generate a hydrogen ionization front and accompanying
\feii\ absorption unless $\log U \geq -1.5$ (Figure 6 of \nocite{dek02b}{de Kool} {et~al.} 2002).
We adopt a slightly lower value of $\log U=-2$ to place a conservative
upper limit on the distance of the absorbing gas.}
we conservatively estimate that the absorbing gas in this object is located at
$d<0.5$~pc, with an uncertainty in this upper limit of about a factor of two.  
High resolution spectroscopy of J1019 (and possibly J1316) will be 
extremely useful for measuring or setting limits on the column densities in the 
\feiii\ lines to obtain more accurate value of $n_e$ and $U$ and thus of $d$.

\section{Possible explanations} %for Absorption at Redshifted Velocities} 
\label{poss}

Having established the reality and incidence of redshifted absorption in
BAL quasars, we now consider possible explanations for redshifted
absorption, more than one of which may be at work in our sample.
We consider gravitational redshifts, infall, rotating outflows, 
binary quasars, and the relativistic Doppler shift.  

Where relevant, as a fiducial example we consider
a 10$^8$ $M_\odot$ black hole surrounded by a standard
\nocite{ss73}{Shakura} \& {Sunyaev} (1973, hereafter SS73) disc with $\alpha=0.1$
accreting at a fraction $\dot{m}=0.5$ of the Eddington rate.
The disc's surface temperature profile is approximately $T\propto r^{-3/4}$
(SS73 Eq. 3.5, but see \nocite{2011ApJ...729...34B}{Blackburne} {et~al.} 2011),
so the physical diameter of the region of an SS73 accretion disc which 
emits significant continuum at wavelength $\lambda$ can be written as
\begin{equation} \label{e_rcon}
2\rcon(\lambda) = ({\rm 2.65~light{\rm -}days})~
\dot{m}^{1/3} M_8^{2/3} \left(\frac{\lambda}{1550\,{\rm \AA}}\right)^{4/3}
\end{equation}
for a black hole of mass $M_8 \times 10^8 M_\odot$.
The crossing time of the 1550\,\AA\ continuum region of this fiducial
BH and accretion disc for gas moving at $x$ percent of the speed of light
is 210/$x$ days.  For example,
gas moving at 300~\kms\ across the line of sight,
a plausible speed in the core of a massive galaxy,
would have a crossing time of 5.75 years.

\subsection{Gravitational redshifts} \label{grav}

\nocite{2009MNRAS.393.1433D,2010MNRAS.406.1060D}{Dorodnitsyn} (2009, 2010) have shown that various wind
velocity laws plus gravitational redshifting can lead to a variety of
redshifted and blueshifted troughs being produced by spherically symmetric 
winds, provided they are launched within $\sim 100~R_{Sch}$ of the black hole.
However, we can rule out gravitational redshifts as 
the sole explanation for redshifted absorption in our objects.
The gravitational redshift of a transition seen in absorption or 
emission at radius $r$ around a Schwarzschild black hole is given by 
\begin{equation}
z_g = \left(1-{R_{Sch}\over r}\right)^{-1/2}-1.
\end{equation}
A purely gravitational redshift explanation for our objects
would require absorption from gas at $50 > r/R_{Sch} > 11$ to explain 
redshifted velocities between 3000 and 14,700~\kms.
Such radii are comparable to 
the ultraviolet continuum emission region sizes of quasars.
Unshielded gas at such small radii would be too highly ionized 
by that continuum (or too dense, in the accretion disc itself)
for ions such as \civ\ to exist \nocite{mc98}(e.g., {Murray} \& {Chiang} 1998).
However, gravitational redshifting may contribute to the observed redshifted
velocities of absorption from a rotating, shielded wind launched from 
within or just outside the continuum emission region; see \S~\ref{rotvel}.

\subsection{Infall} \label{infall}

Simple infall of gas on large scales 
is a reasonable explanation for absorption at redshifted
velocities characteristic of dark matter halos; i.e. up to a few 100~\kms\ in
galaxies\footnote{Silhouetted absorption from an expanding structure within 
the galaxy hosting the quasar is a related possibility.  For example, supernova
remants (SNRs) are larger than quasar continuum emission regions,
can show absorption from a wide range of ionization stages, and 
have asymmetries which can vary the ratios of blueshifted and redshifted
absorption (though probably not enough to explain $V$-shaped troughs).
Some SNRs are likely to be created at relatively small 
distances from quasars as a result of star formation in the outer regions 
of quasar accretion discs \nocite{1999A&A...344..433C}({Collin} \& {Zahn} 1999), which would increase
the covering factor of SNRs around quasars.
However, the density in such discs is likely to be high enough to greatly
suppress the expansion of such SNRs \nocite{2005ApJ...630..167T}(\S\,2.1 of {Thompson}, {Quataert} \&  {Murray} 2005).
SNRs can have expansion velocities of 10,000~\kms\ or more,
but the gas moving at such velocities is typically too highly ionized
to absorb at UV wavelengths \nocite{2010ApJ...725..894H}({Hayato} {et~al.} 2010).  The swept-up gas
which can show absorption in ions such as \ovi\ typically does so over 
a velocity range of only hundreds of \kms\ \nocite{1976ApJS...32..681J}({Jenkins}, {Wallerstein} \&  {Silk} 1976).
Furthermore, absorption from cool supernova ejecta free-expanding at thousands
of \kms\ in the interior of a SNR is seen only in elements such as 
Ca, Si and Fe in ionization stages I to IV \nocite{2005ApJ...624..189W}({Winkler} {et~al.} 2005),
which can produce spectra very different from those of our objects or of
standard BAL quasars \nocite{1997ApJ...477L..53W,1999ApJ...514..195F}({Wu} {et~al.} 1997; {Fesen} {et~al.} 1999).
Thus, a silhouetted SNR explanation seems unlikely for these objects, 
although it should be kept in mind.}
and up to a few 1000~\kms\ for galaxies in galaxy clusters.
However, the {\em widths} of the absorption troughs in our objects rule out 
direct infall explanations except in the narrowest cases ($<$2000~\kms\ or so).
Absorption from an infalling galaxy (or the quasar host galaxy) could 
conceivably produce such linewidths %of up to 2000~\kms\ 
if its gas is affected by a merger or a starburst-driven wind.
\nocite{2007ApJ...659..283I}{Iono} {et~al.} (2007) find gas motion spanning a range of
1200~\kms\ in the central kpc of the merging binary AGN NGC~6240.
\nocite{2010ApJ...717..289S}{Steidel} {et~al.} (2010) find that \OI\ absorption in
Lyman-break galaxies can span a velocity range of at least 700~\kms\ infall
to $-$1000~\kms\ outflow (their Figures 6, 9 and 10).
In our sample, only J0050 (see Appendix \ref{maybe}), J1125, and perhaps J1128
have such narrow troughs.
Absorption from a galaxy is a possible explanation for J0050.
However, the \feii\ absorption blanketing the spectrum of J1125 and J1128 is
much stronger than is ever seen in star-forming galaxies \nocite{feulirg}({Farrah} {et~al.} 2005).
Also, neither J1125 nor J1128 host luminous starbursts detectable in
the far-IR \nocite{2012ApJ...745..178F}({Farrah} {et~al.} 2012).

To generate absorption at infall velocities $>$3000~\kms\ toward the UV
continuum sources in these quasars requires gas infall along our line of sight
(and others, if the infall is not purely radial)
down to radii small enough to generate such velocities.
To a good approximation, 
the maximum velocity gain achievable through infall to radius
$r$ is the escape velocity from that radius ($\sqrt{2GM_{BH}/r}$),
so if redshifted absorption at velocity $v=\beta c$ is caused by gas radially
infalling towards the quasar, then $r=R_{Sch}/\beta^2=2GM_{BH}/v^2$.  
The maximum observed $\beta \sim 0.049$ ($v\sim14700$~\kms)
in our sample requires infall to $r\sim 420\,R_{Sch}$.
This calculation agrees within a factor of two with the 3-D simulations of
\nocite{kp09}{Kurosawa} \& {Proga} (2009), in which infall velocities of 4500~\kms\ to 
7000~\kms\ can be reached at $\sim 2000\,R_{Sch}$ (their Figures 7 and 8),
and of \nocite{2012MNRAS.424..728B}{Barai}, {Proga} \&  {Nagamine} (2012), in which infall velocities of 1500~\kms\ to
2000~\kms\ can be reached at $\sim 10,500 R_{Sch}$ (their Figure 3).
Such velocities are reached in their models if the infall is weakly rotating 
and is viewed close to perpendicular to its rotation vector.
The corresponding 2-D simulations of \nocite{progaOK08}{Proga}, {Ostriker} \& {Kurosawa} (2008)
yield very similar maximum infall velocities (Proga, personal communication).
Similarly, the 2-D simulations of \nocite{los12}{Li}, {Ostriker} \& {Sunyaev} (2012) show that free-falling inflows
can reach $\sim 910 R_{Sch}$.
However, it remains to be seen whether or not these simulations can reproduce
the large optical depths observed at these redshifted velocities in our objects.

Simulations of gas inflow on larger scales in galaxy nuclei have been performed
by, e.g., \nocite{lev08}{Levine} {et~al.} (2008), \nocite{hq10}{Hopkins} \& {Quataert} (2010),
\nocite{2011MNRAS.413.2633H}{Hobbs} {et~al.} (2011) and \nocite{gaspari2013}{Gaspari}, {Ruszkowski} \&  {Oh} (2013).  
\nocite{lev08}{Levine} {et~al.} (2008) find that a flattened but thick circumnuclear disc ($h/R\sim 0.1$) 
is formed, containing large-scale structure such as spiral waves and bars.
However, the average inflow velocity is only 10\% of the rotational velocity 
at each radius in their simulation domain ($r>0.1$\,pc).
\nocite{hq10}{Hopkins} \& {Quataert} (2010) find inflow velocities of 0.01--0.3 times the rotational velocity 
in their simulations (their \S\,4).  At scales of $\sim$1--10\,pc,
they find that eccentric discs or single-armed spirals can form and drive gas
inwards.  However, even in the extreme case of plunging hyperbolic orbits for
such gas, the maximum inflow velocity achieved will be only $\sqrt{2}$ times
the Keplerian velocity at the innermost radius of the flow plus the turbulent
velocity with which the gas began its infall ($<$1000~\kms).
\nocite{2011MNRAS.413.2633H}{Hobbs} {et~al.} (2011) argue that supersonic turbulence can lead to the 
formation of dense filaments which fall ballistically towards the black hole.
Their model may be a promising way to explain our objects, but their study
is confined to scales $>$1\,pc, making direct comparison difficult.
\nocite{gaspari2013}{Gaspari} {et~al.} (2013) make a similar argument using simulations that extend
down to sub-parsec scales, but do not report gas infall velocities.

The absorbing gas in our objects has avoided overionization and is
experiencing a net outward force from scattered UV radiation, so why 
is it infalling?  One possible answer is that the absorbing gas is
unusually dense, which might also help explain the
high fraction of LoBALs in our sample (\S\,\ref{lobals}).
Denser gas will have a lower ionization parameter, all else being equal,
and is plausibly more likely to form optically thick structures.
Such structures are less likely to be driven into outflow by a quasar's 
radiation pressure (in the absence of velocity gradients, only the inner face
of an opticallly thick gas clump will scatter UV lines to provide an outward 
force, but the entire clump mass will feel the inward force of gravity).
Thus, dense gas clumps formed by swept-up lower-density gas 
may end up in infall \nocite{NayakshinZubovas2012}({Nayakshin} \& {Zubovas} 2012).

In this scenario, more variability should occur at larger redshifted velocities
than at smaller ones, because gas with larger infall velocities 
is likely to have larger transverse velocities as well.
(Note that standard BAL quasars exhibit greater trough variability at higher 
trough outflow velocities, which is likely due to higher transverse velocities
in such troughs; see
\nocite{2011MNRAS.413..908C}{Capellupo} {et~al.} 2011 and \nocite{2012ApJ...757..114F}{Filiz Ak} {et~al.} 2012.)
For troughs broader than can reasonably be produced outside the nucleus of a 
galaxy ($>$1000~\kms\ or so), the longer the troughs are seen {\sl not} 
to vary, the more ordered the infall has to be,
and the less likely infall becomes as a hypothesis.

A key test of the infall hypothesis will be to compute absorption profiles
for the simulations to determine if they can match the apparent optical
depths of the ionization stages seen in these objects \nocite{sim12}(e.g., {Sim} {et~al.} 2012).
In observed cases where the absorption profiles are smooth over 
a large velocity range, the infall motions must be ordered so that 
our line of sight intercepts most of the infall trajectory.
In that case the gas must have lost considerable angular momentum 
and have avoided shocking to unobservably high temperatures
and ionization stages during the accretion process.
Gas with such properties will be rare, 
but redshifted BAL troughs are also rare.  
Therefore, an explanation postulating infall velocities
of $>$3000~\kms\ in some of these objects remains potentially viable.

However, in this scenario,
infalling redshifted troughs and outflowing blueshifted troughs should have
nothing to do with each other, so there is no reason to expect their absorption 
index distributions to be statistically indistinguishable, as observed
(\S\,\ref{ai}),
nor to expect the distribution of their summed absorption indices (\ait) 
to be a statistical match to the AI distribution for standard BAL quasars.

%\subsubsection{Covering factor changes in infalling gas} \label{adiabat}

Furthermore, we can argue against infalling gas in some objects following 
\S\,4.2 of \nocite{2009ApJ...703.1394M}{Martin} \& {Bouch{\'e}} (2009), who point out that the covering factor of
optically thick structures in a wind changes in a predictable way with radius.
In BAL quasars where \civ\ and \SIiv\ are of comparable depth, the
absorbing gas is almost certainly optically thick, with the depth of
the troughs determined by the covering factor of the absorbing structures
\nocite{2003ARA&A..41..117C,2008ApJ...681..954A}({Crenshaw} {et~al.} 2003; {Arav} {et~al.} 2008).  
Regardless of whether these structures are outflowing or infalling, 
if they are located in
a gaseous halo with pressure profile $P\propto r^{-b}$ then their volume 
will adjust at different radii so as to maintain pressure equilibrium
($PV^\gamma={\rm constant}$).  We can thus write $V\propto r^{b/\gamma}$ 
where we assume $\gamma=5/3$, appropriate for a monatomic, ideal gas
undergoing adiabatic compression.
For clouds that keep the same shape and orientation as their volume changes,
the covering factor $C(r)$ is given by the area of the clouds ($\propto V^{2/3}$)
relative to the area of the sphere at that radius: $C\propto r^{(-2+2b/3\gamma)}$.  
For an isothermal sphere $b=2$, yielding $C\propto r^{-6/5}$.
For the covering factor to decrease with decreasing radius would require 
an implausibly steep pressure profile with $b>5$.

For radially infalling gas, we have $v\propto r^{-0.5}$,
which yields $C\propto v^{2.4}$ for the isothermal case 
and $C\propto v^{(4-4b/3\gamma)}$ in general.
{\em Thus, unless $b>5$, troughs from infalling gas 
undergoing adiabatic compression
should have covering factors which increase with redshifted velocity}
until the gas leaves our line of sight or the ion in question vanishes
due to increasing ionization (from shocks, photoionization of lower-density
fragments created from instabilities in the main clump, or both).
Conceptually, gas clumps free-falling toward a quasar 
do not shrink as fast as the surface area around the quasar shrinks 
(unless the pressure increases implausibly fast with decreasing radius), 
leading to an increasing covering factor with increasing redshifted velocity.
This effect can be seen in the simulated clumpy accretion flows of
\nocite{2012MNRAS.424..728B}{Barai} {et~al.} (2012, their Figure 2 or Figure 7).

Note that the ionization parameter $U$ depends on the ionizing luminosity $L$, 
the distance $r$, and the gas density $n\propto 1/V$:
$U\propto L/nr^2 \propto r^{-2+b/\gamma} = r^{-2+3b/5}$.  Thus, under our 
assumptions the ionization parameter will increase with decreasing radius 
for $b<\frac{10}{3}$, following $U\propto r^{-0.8}$ in the isothermal case.
With $v\propto r^{-0.5}$, we have $U\propto v^{0.4}$ in the isothermal case 
and $U\propto v^{1-3b/10}$ in general.
There is no clear sign in our objects for an overall trend of increasing 
ionization (as traced by the \aliii/\civ\ ratio) with increasing redshifted 
velocity. 
It can be is difficult to determine the extent of redshifted troughs 
of \aliii\ due to confusion with the broad emission complex at 1900\,\AA.
Nonetheless, in some objects (e.g., J0830 and probably J1709) the troughs of
\aliii, \Nv\ and \civ\ all appear to have the same redshifted extent, while in
others (e.g., J1019 and J1724), \aliii\ is limited to less redshifted velocities.

In summary: for the case of adiabatic compression,
infalling gas is ruled out as an explanation for
redshifted troughs with \vshaped\ profiles, but infall remains
a possibility for those $\mathsf U$ and $\mathsf W$-shaped profiles wherein
the covering factor initially increases with increasing redshifted velocity.
Such an increase is found in seven out of the eleven objects 
with $\mathsf U$ or $\mathsf W$-shaped profiles in our sample
(J0828, J0941, J1019, J1146, J1440, J1709 and J1724) and in the candidates 
J0050 and J1342, and may be found in other objects (e.g., J0830)
depending on the exact values of their systemic redshifts.
The above conclusion does depend on our assumption of constant cloud shape.
Abandoning that assumption, reconciling the case of adiabatic compression 
with our observations would require $b$ between 2 and $\frac{10}{3}$ 
(to match the behaviour of $U$ with velocity) 
and clouds which radially elongate in those respective cases 
by a factor of $>$10$^{6/5}$=16 to $>$10$^{2/3}$=5 
per factor of ten decrease in radius (to match the behaviour of $C$ with velocity).
Detailed simulations involving heating and cooling may also show infall to still
be a possible explanation for our objects.  

\subsection{Rotating accretion disc winds} \label{rotation}

As mentioned in \S\,\ref{intro},
\nocite{gan01}{Ganguly} {et~al.} (2001) and \nocite{sdss123}{Hall} {et~al.} (2002) have
noted that redshifted absorption can occur when 
an extended emission source is seen through a rotating outflow.  
Specifically, the outflow must have a rotational velocity which dominates 
the component along our line of sight of its poloidal (nonazimuthal) velocity.
To yield redshifted velocities of $\geq$1000~\kms,
the inner radius of the outflow should also be no more than $\sim$10
times larger than the radial extent of the continuum source $r_{\rm con}$
(\nocite{sdss123}{Hall} {et~al.} 2002, \S\,6.5.2 and thereafter).
In a rotating disc-wind model, discs viewed at high inclination (close to edge-on)
will always be more likely to produce redshifted absorption.  
Unobscured high inclination sightlines will be rare in
models where the obscuring torus (or equivalent) is coplanar with the disc, 
but are possible with tilted but not twisted discs as obscurers 
\nocite{le10}(e.g., {Lawrence} \& {Elvis} 2010).

The rotating wind scenario is consistent with the result that
the \ait\ values for these objects have the same statistical
distribution as in standard BAL quasars.  Conceptually, one can imagine
gas which is distributed over a range of blueshifted velocities in
standard BAL quasars being seen near its launch radius,
where it is rotating over a range of blue- and red-shifted velocities.

The high fraction of LoBALs in our sample (\S\,\ref{lobals}) might be
understandable in this scenario if the absorbing gas is of relatively high density, 
resulting in outflows with lower terminal velocities.
That would enable redshifted velocities due to rotation to be dominant over a
larger range of outflow distances in these objects than in standard BAL quasars.

\subsubsection{Velocity profiles of rotating disc wind absorption} \label{rotvel}

To estimate the relevant properties of a rotating disc-wind, we consider
our fiducial 10$^8$ $M_\odot$ black hole and SS73 disc with
$r_{\rm con}(1550\,{\rm \AA}) = 2.7\times 10^{15}$~cm 
= 92\,$R_{Sch}$.
We adopt an inner wind launching radius of 
$r_{\rm wind}=3.7 \times 10^{16}$~cm = 1255\,$R_{Sch}$
\nocite{elv00,mc98}({Elvis} 2000; {Murray} \& {Chiang} 1998), at which $v_{\rm circ}=6000$~\kms.
For a wind terminal velocity greater than 8500~\kms\ (the escape velocity 
from $r_{\rm wind}$), we find that redshifted velocities of up to 
600~\kms\ in \civ\ can be reached in either the Elvis or Murray et~al. model
(assumed launching angles of 27$^\circ$ and 6$^\circ$, respectively).

Larger redshifted velocities can be produced if 
the inner wind launching radius is closer to the continuum emission region.
That could arise if the continuum region is larger than predicted by SS73,  
which observationally does seem to be the case: 
quasar half-light radii at rest-frame 1736\,\AA\ are larger than predicted
by a factor of $\sim$5 \nocite{2012ApJ...751..106J}({Jim{\'e}nez-Vicente} {et~al.} 2012)
to $\sim$15 \nocite{2011ApJ...729...34B}({Blackburne} {et~al.} 2011), on average.
Increasing the continuum source $r_{\rm con}$ by a factor of $\sim$5
makes redshifted velocities of up to 2150~\kms\ possible.
A factor of $\sim$15 increase, so that $\rcon=\rwind$, makes 
a redshifted velocity of up to $v_{\rm circ}(\rwind)=6000$~\kms\ possible.
Such a velocity is within a factor of two or three of the largest
redshifted velocities seen in our sample.

Equating the highest redshifted velocities in our sample with 
$v_{\rm circ}(\rwind)$ implies a wind launch radius as small as 
$\rwind = 200\,R_{Sch}$ in this model.
Such an $\rwind$ would be within the observationally determined UV continuum
emission region.  The simulations of \nocite{pk04}{Proga} \& {Kallman} (2004) indicate that winds can be
launched from radii as small as $40\,R_{Sch}$, but how much \civ\ absorption
is likely to arise at those radii is an open question.

The redshifted velocities from a rotating disc wind might also increase with
$M_{BH}$.  
For an SS73 disc at fixed Eddington ratio,
$\rcon\propto M_{BH}^{2/3}$ (Equation \ref{e_rcon}); 
if $r_{\rm wind}$ scales the same way then 
$v_{\rm circ}(r_{\rm wind})\propto \sqrt{M_{BH}/r_{\rm wind}} \propto M_{BH}^{1/6}$, increasing the maximum
redshifted velocities by a factor of 2.15 for the most massive black holes
with $M_{BH}=10^{10}$ $M_\odot$.  
However, if $r_{\rm wind}$ scales linearly
with $M_{BH}$ then the maximum redshifted velocity is independent of $M_{BH}$.

Two properties of our objects' absorption profiles are relevant to the
rotating wind model.

All our objects except J0028 (and the candidates J1342 and J1704)
have \civ\ absorption which extends farther in the
redshifted velocity direction than in the blueshifted velocity direction.
Azimuthal asymmetries in rotating winds can in principle explain this result;
such asymmetries will eventually disappear due to rotation, and objects 
with winds hosting such asymmetries will still over the long term have larger 
blueshifted than redshifted velocities.  

Also, 
rotating winds launched at large radii cannot have redshifted trough depths
greater than 50\%, as seen in J0830, J1019, J1709, J1724 and probably J2157.
Because at most half of the outflow is rotating away from the absorber,
only half of the continuum source will be seen through redshifted absorption.

Both of these properties may be explainable with gravitational redshifting.
If the absorption arises at small enough radii, gravitational redshifting 
will shift the absorption profile to more redshifted velocities,
making it possible for more than half of the continuum source to be seen
through redshifted absorption.  
The 3800\,\kms\ average difference between maximum absolute blueshifted and
redshifted velocities would require some absorption to arise at an average
$r = 40\,R_{Sch}$.  However, the lack of cases of absorption extending 
farther in the blueshifted direction than in the redshifted direction,
and the lack of any blueshifted absorption in some cases, might require
a failed wind in these objects.  In other words, in this model 
the redshifted absorption might have to arise at such small radii
that the observed velocities are due solely to rotation and gravitational
redshifting, with no actual outflowing velocity component to extend the
profiles farther in the blueshifted direction than in the redshifted direction.

\subsubsection{Numerical simulations of rotating disc winds} \label{rotnum}

Some numerical disc-wind outflow models have predicted absorption troughs 
spanning both blueshifted and redshifted velocities, at least
for some parameter choices.  
For examples, see Figures 7
and 9 of \nocite{kwd95}{Knigge}, {Woods} \& {Drew} (1995), Figure 1 of \nocite{pea02}{Proga} {et~al.} (2002), Figure 1 of \nocite{proga2003}{Proga} (2003),
and Figure 3 of \nocite{pk04}{Proga} \& {Kallman} (2004).\footnote{Absorption troughs spanning
roughly $\pm$2000~\kms\ around a systemic redshift have been seen in nova-like
variables in our Galaxy (e.g., Figures 1c and 2c of \nocite{novalike}{Hartley} {et~al.} 2002).
Those authors attributed the absorption to rotationally broadened
photospheric features from an accretion disc, which is not a viable explanation
for quasars because the broadening will be $\gg$0.1$c$ in the ultraviolet
continuum emitting region of quasars (few tens of $R_{Sch}$).}

Furthermore, some profiles in \nocite{pea02}{Proga} {et~al.} (2002) show distinct blueshifted and 
redshifted absorption troughs (their Figure 1, panels c, d, h and i), 
similar to those seen in J1034 and J1628.  These troughs arise
at relatively large inclination angles ($i>45^\circ$) when considering only 
absorption from wind radii $<$10 times the continuum source size.
When the same models are extended to follow the wind to larger radii
\nocite{proga2003}({Proga} 2003), additional absorption at small line-of-sight velocities 
is seen between the blueshifted and redshifted absorption, resulting in 
an absorption profile spanning both blueshifted and redshifted velocities.

Therefore, one possibility for producing distinct blueshifted and redshifted
absorption troughs is a rotating wind more ionized at large radii than in
the models of \nocite{proga2003}{Proga} (2003).  Such a situation could occur if the density in
the wind were to drop off more rapidly with distance from the black hole
than assumed in those models (due to more efficient acceleration, for example),
or perhaps if the overall density in the wind were lower.
Such density profiles being uncommon could help explain the rarity 
of BAL quasars with redshifted absorption.  
Further work on simulated line profiles including gravitational redshifting
and resolved continuum emission regions 
\nocite{pprhh}(e.g., {Proga}, {Rodriguez-Hidalgo} \&  {Hamann} 2012)
will be useful in untangling how redshifted absorption troughs arise.

\subsubsection{Absorption variability in rotating disc winds} \label{rotabs}

Absorption at only systemic and redshifted velocities
can be produced by an azimuthally asymmetric 
outflow with absorbing gas present in the receding part of the silhouetted
rotating outflow but not the approaching part.  That scenario makes a
specific prediction that the parcels of redshifted absorbing gas will each move
to larger redshifted velocities as the flow rotates, so that the low-velocity 
end of the trough moves to higher velocities until the trough disappears.
Any new absorption components rotating into view should
appear as blueshifted absorption.
Note that the component of the poloidal velocity perpendicular to our line of 
sight can complicate the above picture by limiting the range of line-of-sight 
velocities through which a gas element flows as it crosses the continuum source.

Azimuthal asymmetries can also produce variability in objects in which
rotating disc winds produce both blueshifted and redshifted troughs.
The variability of the absorption in J1125 (Fig. \ref{fj11s})
is potentially consistent with the model: the blueshifted absorption weakened
more than the redshifted, consistent with rotating absorbing gas moving from
blueshifted to redshifted velocities.
However, the variability of the blueshifted and redshifted absorption in 
J1034 (Fig. \ref{fj1034}),
J1128 (Fig. \ref{fj11s}),
and J1628 (Fig. \ref{fj1628})
does not follow the above pattern, arguing against the rotating disc-wind model.
In the first two cases the redshifted absorption weakened while the blueshifted
did not (although that behaviour might be explained by poloidal velocity,
as mentioned above).  
In the case of J1628, the redshifted absorption trough moved toward
zero velocity, consistent with a decelerating inflow but not a rotating disc.

We can write the time required for gas with an azimuthal velocity equal to the
Keplerian velocity at its radius of $r\geq r_{\rm con}$ to pass in front 
of the continuum-emitting region of the accretion disc ($r<r_{\rm con}$) 
in units of the orbital timescale at $r_{\rm con}$ as
\begin{equation} \label{e_tcross}
\frac{t_{cross}(r)}{t_{orb}(r_{\rm con})}=\frac{0.5-[\arccos(\rcon/r)]/\pi}{({r_{\rm con}/r})^{3/2}}.
\end{equation}
As $r$ increases, this ratio declines from 0.5 at $\rcon$
to a minimum value of 0.412 at 1.2$\rcon$
before increasing to $\sim$0.5 at 2.3$\rcon$ and $\sim$1 at 10$\rcon$.
(Note that the above is a lower limit to the true crossing time, as angular 
momentum conservation lowers the rotational velocity of outflowing gas 
far from its launch radius below the Keplerian velocity at that radius.
However, such gas is unlikely to produce redshifted absorption.)

For the 1550\,\AA\ continuum-emitting region
around our fiducial $10^8$\,$M_\odot$ black hole,
$t_{orb}(\rcon)=0.24$ years and 
$t_{cross}(\rwind)=0.28$ years,
where $\rwind=3.7\times 10^{16}~{\rm cm}$ is the inner edge of our fiducial wind.
For $\rcon$ five times larger,
$t_{orb}(\rcon)=2.71$ years and 
$t_{cross}(\rwind)=1.46$ years.
Because $\rcon \propto M_{BH}^{2/3}$ (Equation \ref{e_rcon})
and $v_{orb}(\rcon)\propto (M_{BH}/\rcon)^{1/2}$,
$t_{orb}(\rcon)\sim \rcon/v_{orb}(\rcon) \propto M_{BH}^{1/2}$.
Assuming $\rwind\propto \rcon$, 
both $t_{cross}(\rwind)$ and $t_{orb}(\rcon)$ are a factor of ten longer 
for the most massive black holes ($10^{10}$\,$M_\odot$).

Thus, variability on a timescale dependent 
on the continuum source size and black hole mass is expected in this model,
with the most common variability pattern being absorption appearing 
at blueshifted velocities and moving to redshifted velocities.
Note that variability can be avoided
if the winds have sufficiently small azimuthal variations in their covering
factors and/or transmitted flux over more than 
one orbital time.
(and large azimuthal variations in $\tau$ can occur without causing
variability as long as $\tau \gg 1$ is always satisfied.)
Such high levels of symmetry would only be possible for $\mathsf U$-shaped and
$\mathsf W$-shaped troughs, as azimuthal asymmetry is required to explain
\vshaped\  troughs in this model.
Whether such high levels of symmetry are possible is an open question, 
but some BAL outflows do appear to have substantial azimuthal symmetry
\nocite{2010ApJ...713..220G}(e.g., {Gibson} {et~al.} 2010).

\subsubsection{Summary: rotating disc winds} \label{rotsum}

A rotating disc wind model can in principle produce the redshifted
{\em velocities} observed in these objects, but such a model has 
difficulty explaining the observed variability in several of our objects.
Gravitational redshifting may help explain
the prevalence of absorption that extends farther 
in the redshifted velocity direction than the blueshifted,
as well as the four or five cases of redshifted troughs with depths greater
than 50\% of the continuum, but overall, 
a simple rotating disc wind model does not cleanly explain all objects in
our sample.  
Determining how often,
if ever, rotating outflows can exhibit the parameters required to match these 
redshifted troughs will be a useful goal for analytic and numerical simulations.

\subsection{Silhouetted outflows in binary quasars} \label{binary}

Quasars inhabit massive dark matter halos, which are highly clustered
\nocite{tp98,2007AJ....133.2222S}(e.g., {Tegmark} \& {Peebles} 1998; {Shen} {et~al.} 2007).
Therefore, close binary quasar systems should exist, 
and in fact numerous examples are known (e.g., \nocite{joebinaryq}{Hennawi} {et~al.} 2006).
A binary quasar system can produce a quasar with a redshifted BAL trough
if the following conditions are met:\\
1) The quasar physically farthest from us is at a redshift
given by the emission lines seen in the SDSS spectrum.\\
2) The quasar physically closest to us:\\
$\bullet$~emits many fewer photons along our sight-line than does 
the background quasar (so that its broad emission lines are not obvious
in the combined spectrum),\\
$\bullet$~produces a BAL outflow,\\
$\bullet$~and is oriented such that the background quasar backlights at least
part of the outflow with a relative radial velocity vector that produces
redshifted absorption.\footnote{It is possible 
for the wind from the background quasar to be oriented so as to pass between us
and the foreground quasar, but such an orientation can only produce redshifted
absorption through the relativistic Doppler effect (\S\,\ref{Doppler}).}
This scenario is viable in cases where the quasars are far enough
apart ($\geq 100$~pc or so) that they do not dramatically disrupt
each other's accretion discs, fueling, or BAL outflows.

A large projected area of the flow on the sky increases the chance of 
backlighting, and some BAL outflows are known to extend to kpc scales 
\nocite{edarav11}(e.g., {Edmonds} {et~al.} 2011, and references therein).
Different parts of the outflow along the same sightline
could have different redshifts or blueshifts.
The binary scenario may be favored for \vshaped\  troughs, which 
have the appearance a standard BAL outflow might have
if it were pointed away from the observer and then backlit.

We inspected available images of our objects for evidence
supporting a binary quasar scenario for their spectra. 
The SDSS imaging pipeline classifies all objects as unresolved except for
J0941, which is morphologically extended (\S\,\ref{cases}).\footnote{J1034 
has an unresolved, red object 4$\farcs$62
away which was not selected as a quasar candidate in the SDSS, 
and therefore is likely to be a late-type star.}
Near-IR imaging from UKIDSS \nocite{ukidss1}({Warren} {et~al.} 2007)
confirms that J0941 is not a pure point source, but yields no evidence
for binarity in J0805, J1019, J1439 or J2157.
Also, deep coadds of SDSS imaging data in Stripe 82 \nocite{s82coadd}({Annis} {et~al.} 2011)
show no evidence for binarity in J0028 or J2157.
However, the angular separation of any putative binary quasar could be much
smaller than the value needed to make SDSS or UKIDSS images elongated, 
and the foreground quasar could also be obscured. 
To confirm or reject the binary scenario for individual objects
would require high-resolution imaging, multiwavelength photometry,
or near-IR spectroscopy to search for narrow line emission at two redshifts.
Note that there is no evidence for two narrow-line redshifts in our two
low-redshift quasars with redshifted \mgii\ absorption (Figure \ref{fj11}).

The conditions outlined in this section are
very similar to those proposed by \nocite{cosmosBH}{Civano} {et~al.} (2010) as one scenario to explain
the $z=0.359$ AGN CXOC J100043.1$+$020637 (hereafter CXOC J1000).
That object hosts two compact optical sources in the same galaxy and has a
broad-line redshift which is higher by $\sim$1200~\kms\ than the narrow-line redshift.
It also exhibits a strong X-ray iron emission line accompanied by redshifted
iron absorption.  In contrast to our objects, the absorption in CXOC J1000
is deepest at the
longest wavelengths (greatest redshifted velocities).
\nocite{cosmosBH,cosmosBH2}{Civano} {et~al.} (2010, 2012) suggest that the foreground object in that system 
could be a heavily obscured Type 2 (narrow-line) AGN 
with an outflowing, high-velocity, highly ionized wind 
backlit by the X-ray continuum of a background Type 1 (broad-line) AGN.
The velocity vector of the backlit portion of the wind 
must be oriented so as to produce a redshift.
The large velocity difference between the redshfts of the two AGN
could arise from a gravitational interaction.

Another effect may be at work in this model, for both
CXOC J1000 and any binary quasar in our sample.  If the wind from the
foreground object has a sufficiently large velocity, the relativistic Doppler
shift can produce significantly redshifted absorption even when the backlit
wind is directed near the plane of the sky.  
We discuss the relativistic Doppler shift in \S\,\ref{Doppler}.

\subsubsection{Expected number of silhouetted outflows in the BOSS DR9Q} \label{binfrac}

We can estimate whether or not the abundance of BAL quasars with
redshifted troughs is consistent with a scenario explaining them as
silhouetted outflows in binary quasar systems.

We start with 69674 $z>1.57$ BOSS quasars examined for redshifted troughs.
At $z>1.57$, 5~kpc is a separation small enough for the system to appear 
unresolved in SDSS imaging.
Approximately 0.04$\pm$0.01\% of luminous SDSS quasars have binary companions 
at projected radii between $\sim$5 and $\sim$100~kpc 
\nocite{2007ApJ...658...99M,2008ApJ...678..635M,2010ApJ...719.1672H}({Myers} {et~al.} 2007, 2008; {Hennawi} {et~al.} 2010).
The binary fraction at projected radii $\leq$5~kpc is likely to be a factor
of $\sim$10 lower, based on how long merging or interacting galaxies
spend at a given projected separation \nocite{2011ApJ...737..101L}({Liu} {et~al.} 2011).
Thus, we expect only $\sim$3 binary quasars at $\leq$5~kpc separation in the
69674 BOSS DR9Q quasars.  
Even if we make the extreme assumption that all quasars in such close binaries
host BAL outflows with covering factor $C$=100\% out to a radius of 5~kpc,
only half of such binaries will have an optically faint or obscured quasar
as the foreground object and an optically brighter quasar 
as the background object, as required in our scenario.

Thus, while a small number of silhouetted outflows may be present in our
sample, to explain our entire sample with them would require
binary quasars at projected radii $\leq$5~kpc to be a factor of $\sim$10
more abundant than current studies indicate.
Furthermore, their outflows would have to have a geometry and distribution
of covering factors able to explain why both blueshifted and redshifted 
absorption are seen in 10 out of 13 objects targeted by the BOSS as $z>2.15$ 
quasar candidates.  That fraction is much larger 
than expected if blueshifted and redshifted absorption are uncorrelated.

\subsection{Relativistic Doppler shifts} \label{Doppler}

A final possibility for producing redshifted absorption is the relativistic
Doppler shift \nocite{1905AnP...322..891E}({Einstein} 1905).  In its own rest frame, an ion
absorbs at the rest wavelength of a given transition.  In a frame in which
the ion is moving (in any direction) then --- in addition to the normal
Doppler shift --- time dilation means that the ion will absorb photons of a 
lower frequency (longer wavelength).  Normally the time dilation factor is 
negligible, but it has been observed in the Galactic microquasar SS~433 in 
emission lines which arise from two oppositely directed jets moving at
0.26$c$ near to the plane of the sky \nocite{ss433}({Abell} \& {Margon} 1979).
If there is absorbing gas in some or all of these BAL quasars moving at a
sufficiently high speed over a range of angles close to the plane of the sky,
the resulting absorption will be seen at a range of velocities which can
span both blueshifts and redshifts relative to the quasar rest frame.

In Appendix \ref{rds} we discuss some underappreciated aspects of the 
relativistic Doppler shift relevant to line {\em absorption} (for example, 
absorption at a given blueshifted wavelength does not uniquely determine
the velocity of the outflow even if its angle to the line of sight is known).
Few papers on the effects of the relativistic Doppler shift on absorption have
appeared in the astronomical literature 
\nocite{1968JRASC..62..105S,1998PhyU...41..941H,2011AnP...523..239W}({Sher} 1968; {Hovsepyan} 1998; {Wang} 2011),
in contrast to the numerous papers applying it to line {\em emission} 
in quasars (beginning with \nocite{1982ApJ...258..425M}{Mathews} 1982),
In this section we consider only the potential application 
of the relativistic Doppler shift to the absorption in our objects.
We use $\psi$ to denote the angle relative to our line of sight,
with $\psi=0^\circ$ pointing away from us and $\psi=90^\circ$
denoting the plane of the sky.

As an example,
for a flow reaching the highest speeds seen in quasar outflows 
($\beta\approx 0.2$), a range of angles centered slightly toward
the line of sight ($\psi\approx 92.5^\circ\pm 5^\circ$) would be
sufficient to explain the trough in J2157 
(approximately $-$2000~\kms\ to 10,000~\kms).
For a flow reaching a top speed of only $\beta\approx 0.1$, reproducing 
the same trough would require twice as wide a range of angles, centered
slightly away from the line of sight ($\psi\approx 87.5^\circ\pm 10^\circ$).
Similarly, gas at a range of speeds from $0<\beta<0.2$ at $\psi=90^\circ$
would produce redshifted troughs spanning 0 to +6000~\kms,
but an identical trough could be produced by gas at lower velocities 
directed slightly away from the observer ($\psi<90^\circ$).

Producing redshifted troughs via the relativistic Doppler effect
does not require large gas inflow velocities, but does require 
even larger velocities nearly transverse to our line of sight
to a quasar's continuum source.  
One scenario which could generate large transverse velocities is
a nearly vertical outflow launched from small radii, where the initial
rotational velocity and radiative acceleration are both large,
and viewed at a large inclination angle.

The relativistic Doppler effect also could be a factor
if any object is involved in a binary quasar scenario (\S\,\ref{binary}).  
In that scenario, the outflow velocity 
as well as the rotational velocity can contribute to the transverse velocity.

In either case, large transverse velocities should result in variability in 
the redshifted troughs on the crossing timescale of the continuum source
by the gas.
Gas moving at $0.1c$ crosses the 1550\,\AA\ continuum region of our fiducial
$10^8$ $M_\odot$ BH + SS73 disc in 21 days.
The crossing time would be at most 2.2 times longer at 2800\,\AA\ than at 
1550\,\AA\ for an SS73 disc,
but could be 5 to 15 times longer at both wavelengths (105 to 315 days)
given the observationally inferred sizes of quasar continuum source regions
(\nocite{2011ApJ...729...34B}{Blackburne} {et~al.} 2011 and \S\,\ref{rotvel}).
For the most massive black holes (10$^{10}$ $M_\odot$), assuming a continuum
source 5 to 15 times larger than a SS73 disc that nonetheless scales as
$\rcon \propto M_{BH}^{2/3}$ (Equation \ref{e_rcon}),
the continuum source crossing time at $0.1c$ would be 6 to 18 years.

\section{Discussion} \label{end}

We have presented observations of seventeen quasars with redshifted broad 
absorption in \civ\ and \SIiv\ (and often in numerous other transitions) and new
observations of two previously known quasars with redshifted broad absorption
in \mgii.  

Roughly 1 in 1000 BAL quasars has redshifted \civ\ absorption (\S\,\ref{frac}),
although such absorption appears to be more common in the SDSS-III/BOSS DR9 
quasar catalog (1 in 660) than in the SDSS-I/II DR7 quasar catalog (1 in 1600).
Why these objects are so rare, given that infalling matter is
ultimately the power source for all quasars, is not clear.

The absorption in these objects can extend to velocities of $>$10,000~\kms\ and 
appears just as saturated as absorption in standard, blueshifted BAL troughs does.
The trough shapes can be classified as either
$\mathsf U$-shaped, \vshaped\  or $\mathsf W$-shaped (\S\,\ref{uvw}).
Although absorption at the systemic redshift in these objects sometimes reaches
zero flux within the noise, none of their redshifted absorption troughs do so,
meaning that the redshifted absorption does not fully cover the continuum 
source.  
The redshifted absorption does not need to cover the broad emission line region 
to explain the observations (with one exception), but we
cannot rule out equal covering of both the continuum and emission ine regions.

We have used the absorption index (\S\,\ref{ai}) to measure the strengths 
of the blueshifted absorption (AI), the redshifted absorption (\aip), and 
the total absorption (\ait) in these objects.
The AI and \aip\ values are uncorrelated overall, but their distributions 
are statistically indistinguishable.
The \ait\ distribution is statistically consistent with the AI distribution
of standard BAL quasars, 
although the \ait\ distribution lacks a tail to very strong troughs.
In other words, the redshifted velocities of these troughs are unusual, 
but their overall strengths are not.

Based on our sample, at least 50\% of quasars with redshifted \civ\ absorption
are LoBAL quasars, as compared to 5\% of standard quasars being LoBAL quasars
(\S\,\ref{lobals}).
However, the \ait\ distribution of these objects is not at all consistent 
with the AI distribution of standard LoBAL quasars (\S\,\ref{ai}).
The low-ionization gas in these objects appears to have a different origin
than in standard LoBAL quasars.

In one or two cases where high-excitation \feiii\ absorption is seen 
(\S\,\ref{dist}),
we infer that the absorber has a very high density ($n_e\geq 10^{10.5}$\,cm$^{-3}$)
and a small distance from the ionizing continuum source ($d<0.5$~pc).

We have discussed possible explanations for redshifted troughs including
gravitational redshifts (\S\,\ref{grav}), 
infall (\S\,\ref{infall}), 
rotating disc-wind outflows (\S\,\ref{rotation}), outflows from one member of
a binary quasar pair seen in absorption against the other (\S\ref{binary}), 
and the relativistic Doppler effect in
outflows moving close to transverse to our line of sight (\S\,\ref{Doppler}).
None of these explanations by themselves
seems likely to explain all our objects.

The binary quasar scenario can explain the observed
range of trough shapes, velocities and absorption indices.
However, a factor of $\sim$10 excess in the numbers of close binary quasars 
over the predictions of current studies would be needed to explain our sample
entirely as binary quasars; this scenario cannot be the dominant explanation.

As alluded to in the title of our paper, we are left with two main scenarios
to explain these objects.  It is possible that some of our objects will be 
explained by one scenario and some by the other.

Infalling gas undergoing adiabatic compression
is expected to have an increasing covering factor with infall
velocity, making it a plausible explanation only for some 
$\mathsf U$-shaped and $\mathsf W$-shaped troughs.
The infall velocities observed in our objects have been seen in simulations of
gas infall in quasars, but whether the gas at those velocities can match the
large optical depths observed in \civ\ and other species is not clear.
In this scenario, the high fraction of low-ionization redshifted troughs 
would be due to preferential infall of high-density gas.
However, the match between the distributions of \ait\ in BAL quasars with
redshifted absorption and AI in standard BAL quasars is not explained 
by this model.

Rotation-dominated disc winds can explain the distributions of absorption
indices in these objects, at least for
$\mathsf U$-shaped and $\mathsf W$-shaped troughs.
Rotation plus gravitational redshifting can in principle explain
the range of trough shapes and velocity extents in this scenario.
Low-ionization absorption might be preferentially seen in this model due to
winds with relatively high-density gas having relatively low terminal
velocities (perhaps not even exceeding the escape velocity), enabling 
redshifted velocities due to rotation to dominate the absorption profile.
However, the variability properties available for a subset of our objects
(see below) do not match the predictions of this scenario.

\subsection{Variability in different scenarios} \label{var}

The observed variability of BAL troughs is believed 
to be due primarily to tranverse motions of absorbing gas 
across the continuum source \nocite{2010ApJ...713..220G,fbqs1408}(e.g., {Gibson} {et~al.} 2010; {Hall} {et~al.} 2011),
although cases are known where the variability in narrower quasar outflows are
more consistent with changes in the ionization of the absorbing gas 
\nocite{2007ApJ...660..152M,2011MNRAS.410.1957H}({Misawa} {et~al.} 2007; {Hamann} {et~al.} 2011).
In this section we discuss what each scenario considered herein for the origin of
redshifted BAL troughs might predict for the variability of those troughs.
Future spectroscopic observations of these objects will be required to test these predictions.
In all cases, what matters for variability is the transverse motion of the
absorber and its size relative to the continuum source, not the distance of 
the absorber from the continuum source; as both are at the same cosmological 
distance from us, only their relative sizes matter.

In each of our suggested
models for redshifted troughs, strong variability on timescales of less
than a year is possible because large transverse motions are possible.  
Not observing rapid variability in any of these objects over time
would constrain the rotating wind model to ever greater azimuthal symmetry,
the infall scenario to more ordered infall,
the binary quasar scenario to smaller transverse velocities,
and would rule out a substantial contribution from a relativistic Doppler shift.

In the infall scenario, 
more variability should occur at larger redshifted velocities,
because gas with larger infall velocities is likely to have
larger transverse velocities as well.
In the quasar J1628, a shift of the absorption to smaller average redshifted 
velocities has been seen.
If the absorption is due to gas undergoing decelerating infall due to outward
radiative acceleration, future observations should show a continued shift 
to smaller average redshifted velocities and increasing deceleration.

Variability in the rotation-dominated wind scenario has been discussed in 
\S\,\ref{rotabs}.  The pattern expected for the simplest variability scenario 
(changes in absorption strength propagating from blueshifted to redshifted 
velocities) has been seen in only one of the four cases where two epochs of 
high S/N spectroscopy are available.  
A lack of variability in \vshaped\  outflows on rest-frame timescales
of two years or more would rule out this scenario for such objects.

In the binary quasar scenario, variability at blueshifted and redshifted
velocities will be uncorrelated.  The level of variability in this scenario
depends on the transverse velocity of the silhouetted outflow relative to the
background quasar.  The variability could be comparable to that in standard 
BAL quasars, but might be lower if BAL outflows decrease substantially in 
velocity at large outflow distances due to the sweeping up of ambient gas
\nocite{2011arXiv1108.0413F}({Faucher-Giguere}, {Quataert} \&  {Murray} 2011).

In the case of absorption redshifts arising from a relativistic Doppler shift, 
the variability of such absorption is likely to be more substantial than the
variability of a standard BAL trough, as the large transverse velocity of the 
gas causes it to traverse the continuum source quickly.

\subsection{Ly-alpha emission} \label{lya}

A final unusual aspect of these quasars with redshifted broad absorption
is that most of them have strong and relatively narrow \lya\ emission 
(nine of fifteen objects with \lya\ coverage have such emission).
Although it may be possible to produce apparent narrow \lya\ emission via
\lya\ and \Nv\ absorption of underlying broad emission \nocite{nlya}(e.g., {Hall} {et~al.} 2004c), 
absorption in those lines is not always strong in these objects.  
A full comparison of the \lya\ properties of these objects and of DR9Q quasars
is beyond the scope of this paper,
but the prevalence of strong, narrow \lya\ emission argues against a
rotating accretion-disc scenario for redshifted \civ\ absorption, as there is 
no obvious explanation for such \lya\ emission in such scenarios.

Our objects' \lya\ FWHMs are near the upper end of the 
500--2000~\kms\ range of the FWHMs of \lya\ emission from 
the regions of neutral gas near high-redshift galaxies known as \lya\ blobs
(\nocite{2006A&A...452L..23N,2007MNRAS.378L..49S}{Nilsson} {et~al.} 2006; {Smith} \& {Jarvis} 2007, and references therein).
The fluxes of our objects' narrow \lya\ lines are about an order of magnitude 
larger than in the most luminous \lya\ blobs known \nocite{2000ApJ...532..170S}({Steidel} {et~al.} 2000),
so the lines must arise predominantly from illumination by the quasar.
The strong, relatively narrow \lya\ emission in our quasars might arise due to 
larger than usual amounts of neutral gas in the quasar host galaxy and 
surrounding environs.  An excess of such gas would be consistent with 
the infall explanation for these objects.
Narrow-band imaging or longslit spectroscopy of these objects to ascertain 
the spatial extent of their \lya\ emission might 
constrain the origin of the gas responsible.

\subsection{Future work} \label{future}

A variety of further observations of these objects would be useful
to determine their nature.
Near-infrared spectroscopy could provide accurate redshifts and possible
evidence for binarity via narrow emission lines in the rest-frame optical.
Continued monitoring of these objects' spectra at observed optical wavelengths
will establish the variability properties of the full sample.
High-resolution UV/optical spectroscopy of some objects
may better constrain the column densities of the absorbing gas 
and thereby its physical properties and location as well.

X-ray observations could help discriminate between the different models
we have put forward.
Quasars observed to have the same strong X-ray absorption seen in standard
BAL quasars with the same trough strength and ionization range distributions
as our sample are candidates for having rotating disc winds.
In the infall scenario, however, the redshifted gas observed in front of these
quasars' UV emitting regions may not cover the smaller X-ray emitting region,
or may have the same covering factor for both regions (if the absorbing gas
consists of many small clouds, for example).  
Thus, in this scenario quasars with only redshifted absorption may have 
X-ray properties similar to those of normal, non-BAL quasars  
(i.e., no significant X-ray absorption, normal X-ray spectral shape, and
nominal optical to X-ray flux ratio).

Larger samples of quasars with redshifted troughs would better establish 
how common they are and how their trough 
properties compare to those of standard, blueshifted-trough BAL quasars. 
The goal of the BOSS is to obtain at least 150,000 spectra of $z>2.15$ quasars,
and the DR9Q catalog contains only $\sim$62,000 such objects; therefore, an 
increase in the sample size a factor $\sim$2.5 can be expected in the BOSS alone.
Follow-up spectroscopy of candidate cases and a dedicated search for redshifted
\mgii\ absorption extending to lower redshifts (especially where \oii\ emission
at the systemic redshift can be seen) could increase this number still further.
Furthermore, the existence of redshifted BAL troughs raises the possibility that
the processes producing them could also create narrower redshifted troughs.
That possibility may complicate the interpretation of the velocity distribution
of narrow absorbers around quasar redshifts solely in terms of infall and 
clustering (e.g., \nocite{2008MNRAS.388..227W}{Wild} {et~al.} 2008).

Regardless of the origin of redshifted BAL troughs, they provide
a rare but useful probe of quasars and their environs.
Any which are due to infall of gas offer a test of models of gas dynamics
and AGN fueling and feedback in galaxies during epochs of AGN activity.
Any which are due to rotating accretion disc winds 
offer a test of models of such winds.
Any which are due to binary quasars offer a new view of BAL outflows; namely,
along lines of sight which do not intersect the quasar which launched them.
Such cases offer probes of BAL outflow properties at known projected distances 
from their launch site, enabling new tests of models of BAL dynamics.

\section*{Acknowledgments}
We thank D. Proga,
M. Bautista, D. Edmonds and N. Murray for discussions,
and the referee for a careful review.
PBH thanks NSERC for its research support, the Institute of Astronomy at
the University of Cambridge for hosting his sabbatical,
and the Aspen Center for Physics (NSF Grant \#1066293) for its hospitality.
WNB and NFA are supported by NSF grant AST-1108604.
This research has made
extensive use of NASA's Astrophysics Data System Bibliographic Services.
and of the Atomic Line List at http://www.pa.uky.edu/$\sim$peter/atomic/.

Funding for SDSS-III has been provided by the Alfred P. Sloan Foundation, the Participating Institutions, the National Science Foundation, and the U.S. Department of Energy Office of Science. The SDSS-III web site is http://www.sdss3.org/.  SDSS-III is managed by the Astrophysical Research Consortium for the Participating Institutions of the SDSS-III Collaboration including the University of Arizona, the Brazilian Participation Group, Brookhaven National Laboratory, University of Cambridge, Carnegie Mellon University, University of Florida, the French Participation Group, the German Participation Group, Harvard University, the Instituto de Astrofisica de Canarias, the Michigan State/Notre Dame/JINA Participation Group, Johns Hopkins University, Lawrence Berkeley National Laboratory, Max Planck Institute for Astrophysics, Max Planck Institute for Extraterrestrial Physics, New Mexico State University, New York University, Ohio State University, Pennsylvania State University, University of Portsmouth, Princeton University, the Spanish Participation Group, University of Tokyo, University of Utah, Vanderbilt University, University of Virginia, University of Washington, and Yale University.

%\bibliographystyle{mn}
%\bibliography{}

%% --------------------------------------------------------------------
%% Tue Jun 11 19:07:59 2013
%%   This file was generated automagically from the files
%%   rsub3.bbl and rsub3.tex using
%%     ./nat2jour.pl
%%   This file should accompany rsub3-aas.tex.
%% --------------------------------------------------------------------

\appendix

\section{Notes on Individual Objects} \label{notes}

{\bf SDSS J002825.02$+$010604.2} (J0028; Fig. \ref{f_cands2})
has a redshift of $z$=4.1152$\pm$0.0102 \nocite{hw10}({Hewett} \& {Wild} 2010),
from emission features of \lya, \Siiii, and \ciii.
(Only a visual redshift is available in DR9Q.)
J0028 has absorption troughs spanning the systemic redshift in
\civ, \SIiv, \Nv, and \pv.  
The \SIiv\ trough is continuous, but the \civ\ trough is interrupted
by weak, narrow \civ\ emission at the systemic redshift.
There is also emission and absorption in \SIii, \oi, and \cii.
The three epochs of spectroscopy on this quasar are shown in
Figure \ref{fj0028three} and are discussed in \S\,\ref{rpt}.

{\bf SDSS J014829.81$+$013015.0} (J0148; Fig. \ref{f_cands}) 
has a DR9Q \ciii\ PCA redshift of $z$=3.061$\pm$0.008.
It has strong, narrow \lya\ emission and accompanying narrow \civ\ emission
which appears to interrupt a BAL trough spanning both blueshifted and
redshifted velocities, as seen in \SIiv.  
Similar troughs interrupted by narrow emission are seen in \ovi\ and \Nv.
There may be some \feii* absorption longward of \civ, but \feii* cannot 
explain the redshifted absorption seen in \ovi, \Nv\ and \SIiv. 
At the blue end of the \SIiv\ trough where the absorption is strongest,
absorption is also seen in \pv, \Siv, possibly \ciii*, \cii, and \aliii.
There also appears to be strong, blueshifted emission of \civ\ and \SIiv.

{\bf SDSS J080544.99$+$264102.9} (J0805; Fig. \ref{f_cands}) 
has a DR9Q \ciii\ PCA redshift of $z$=2.703$\pm$0.008.
Its spectrum is remarkably similar to that of J0148, except that
the \cii\ absorption is weaker and only weak blueshifted emission
is discernible in \civ\ and none at all in \SIiv.

{\bf SDSS J082818.81$+$362758.6} (J0828; Fig. \ref{f_cands}) 
has a DR9Q full PCA redshift of $z$=2.366$\pm$0.005,
which is a better match to the spectrum than the DR9Q \ciii\ PCA $z$=2.370.
The true systemic redshift may be even smaller, 
as J0828 has strong \lya\ emission with an apparent peak at $z$=2.361,
but we adopt a systemic $z$=2.366 to be conservative.
It has strong, blueshifted absorption in \Nv, \cii, \SIiv, \civ, 
\aliii, and \mgii. 
It has redshifted absorption in \SIiv\ and \civ,
and probably in \Nv, \cii, and \aliii.

{\bf SDSS J083030.26$+$165444.7}
(J0830; Fig. \ref{f_cands}) has narrow \lya\ emission at $z$=2.4345$\pm$0.0005;
there is also narrow emission at the same redshift in \Nv, \SIiv, and \civ.  
The DR9Q full PCA redshift is $z$=2.423$\pm$0.005 
(no \ciii\ PCA redshift is available).
We adopt the narrow emission redshift as systemic to be conservative
about the extent of the redshifted absorption.
There is redshifted absorption in \Cthree, \Nv, \cii, \SIiv, \civ, \aliii, and
possibly \lya, extending to at least $z$=2.465 in all troughs
from at least $z$=2.417 in \Cthree, \Nv, \SIiv\ and 
\civ\ ($-$1530 to +2650 \kms, $\Delta v=4180$~\kms)
and from $z$=2.43 in \lya, \cii\ and \aliii\ ($-$390 to +2650 \kms, $\Delta v=3040$~\kms).
\lya\ absorption is confused with broad and narrow \lya\ emission, making the
absorption identification uncertain; however, the redshifted velocity of the
end of the trough for other ions matches that of the putative \lya\ trough.

{\bf SDSS J094108.92$-$022944.7} (J0941; Fig. \ref{f_cands}) 
has $z$=3.446$\pm$0.002 from its DR9Q \ciii\ PCA redshift.
It has a red continuum and striking, narrow emission lines of 
\ovi, \lya, \Nv, \SIiv\ (very weak), and \civ.
J0941 has blueshifted absorption in \ovi, \pv, \Nv, \SIiv, and \civ.
It has redshifted \civ\ absorption which is roughly symmetric 
around the emission redshift with the blueshifted \civ\ absorption.
There appears to be accompanying redshifted \SIiv\ absorption,
and possibly redshifted \pv\ and \Nv\ absorption,
but there is no sign of redshifted \ovi\ absorption.

SDSS J0941 is unique in our sample in that 
it is morphologically classified as a galaxy; see \S\,\ref{cases}.

{\bf SDSS J101946.08$+$051523.7} 
(J1019; Fig. \ref{f_cands}) has a DR9Q \ciii\ PCA redshift of $z$=2.452.
It has broad emission in \Nv, \aliii, \ciii,
\feiii\ UV34, UV50 and UV48, and possibly \mgii.
Blue wings of \SIiv\ and \civ\ emission also appear to be present.
There is redshifted absorption 
(from $z$=2.46 to $z$=2.54, or $v$=+800~\kms\ to $v$=+7500~\kms)
in \feiii, \aliii, \alii, \civ, \SIiv, \cii,
\SIii\ and \Nv.  Shortward of \lya, there is absorption
consistent with redshifted \Cthree, \pv+\feiii, \Siv\ and probably \ovi, but
given the resolution of the BOSS spectrum,
\lya\ forest contamination makes those identifications less certain.
The \civ\ and \feiii\ absorption extend to $z$=2.55 ($v$=+8300 \kms),
and may extend to $z$=2.57 ($v$=+10100 \kms).
The \feiii\ absorption is seen in multiplets UV34, UV48, and UV50.
As discussed in \S\,\ref{cands},
BAL troughs in UV34 and UV48 have been seen before,
but this is the first reported case of UV50 absorption.
An intervening absorber at $z$=1.6135
is responsible for narrow absorption at 4050, 4200 and 4365\,\AA.

{\bf SDSS J103412.33$+$072003.6} 
(J1034; Fig. \ref{f_cands2}) at $z=1.6893\pm 0.0018$ \nocite{hw10}({Hewett} \& {Wild} 2010)
has \civ\ absorption both blueshifted ($\sim$3100~\kms\ wide up to $v\simeq -4200$~\kms)
and redshifted ($\sim$5200~\kms\ wide, up to $v\simeq 7250$~\kms)
in its SDSS and BOSS spectra, plus blueshifted \aliii\ and \mgii.  
In addition, its BOSS spectrum shows both blueshifted and redshifted \SIiv.
In both \civ\ and \SIiv,
the blueshifted absorption is much stronger than the redshifted absorption.
There is no sign of \OII\ emission in its BOSS spectrum.

{\bf SDSS J114655.05$+$330750.1} (J1146; Fig. \ref{f_cands}) 
has a DR9Q \ciii\ PCA redshift of $z$=2.780.
It has weak, redshifted absorption in \Nv, \SIiv, \civ,
and probably \aliii\ and \ovi.
J1146 has the shallowest absorption and the smallest \ait\ value in our sample.

{\bf SDSS J114756.00$-$025023.4} (J1147; Fig. \ref{f_cands2}) % nmf
has $z$=2.5559$\pm$0.0056 \nocite{hw10}({Hewett} \& {Wild} 2010) from \Siiii+\ciii\ emission.
(The DR9Q \ciii\ PCA redshift is $z$=2.5600, but despite the small difference
our inspection shows that the HW10 redshift provides the better fit.)
However, the continuum slope of the SDSS spectrum of this object is in error.
The $g-r$ and $r-i$ colors synthesized from the spectrum are much redder than
the corresponding colors from the SDSS imaging and the BOSS spectrum.
Therefore, we only plot the BOSS spectrum in Figure \ref{f_cands2}.
J1147 shows absorption in \SIiv\ and \civ\ which is strongest
at $z$=2.540 ($-$1340~\kms) 
and extends to redshifted velocities, 
generally weakening with increasing redshifted velocity.
There is no clear redshifted \Nv\ trough, but it may be concealed by narrow
\Nv\ emission; such emission is seen in \civ.
Narrow \lya, \Nv, \SIii, and \cii\ absorption are seen at $z$=2.540
and at $z$=2.610 (+4530~\kms).

{\bf SDSS J132333.01$+$004633.8} (J1323; Fig. \ref{f_cands}) 
is at $z$=2.455$\pm$0.038 from our inspection of \Siiii+\ciii;
no DR9Q \ciii\ PCA redshift is available, and the DR9Q full PCA redshift of
$z$=2.376 is spuriously low.
It has troughs in \pv, \Nv, \SIiv, \civ\ and possibly \cii\ that span both
blueshifted and redshifted velocities.
J1323 has blueshifted \civ\ emission but no sign of blueshifted
\SIiv\ emission.
J1323 has two epochs of BOSS spectroscopy, but the second-epoch spectrum
has a considerably lower signal-to-noise ratio than the first, and so no 
notable trough variability is distinguishable between the two epochs.

{\bf SDSS J143945.28$+$044409.2} %        pmf 4780-55682-0992
(J1439) has a DR9Q \ciii\ PCA redshift of $z$=2.492$\pm$0.001, 
which matches the redshift of the peak of its narrow \lya\ emission.
Any \mgii\ emission present is swamped by noise around 9800\,\AA\ (observed);
there is also poor night sky line subtraction near 5892\,\AA\ (observed).
The spectrum of J1439 resembles that of J2157.
J1439 appears to have redshifted absorption in \Cthree, \Nv, \SIiv, and 
\civ\ beginning at its systemic redshift and extending to 
$z=2.605$ ($+9400$~\kms)
in all three lines, and possibly to $z=2.625$ (+11200~\kms) in \civ.
There are also relatively narrow 
absorption systems at $z$=2.465 in \lya, \civ\ and possibly \SIiv,
and at $z$=2.4873 in \lya, \civ, \SIii, \cii, and possibly \Nv;
the latter redshift may mark the short-wavelength end 
of the broad absorption troughs.

{\bf SDSS J144055.59$+$315051.7} (J1440; Fig. \ref{f_cands}),
at a DR9Q \ciii\ PCA redshift of $z$=2.954$\pm$0.015,
has blueshifted and redshifted absorption in 
\ovi, \Nv, \SIii, \cii, \SIiv, and \civ.
The troughs are split by weak emission just longward
of the systemic redshift in each transition.

{\bf SDSS J162805.80$+$474415.6} 
(J1628; Fig. \ref{f_cands2}) at $z=1.5949\pm 0.0019$ \nocite{hw10}({Hewett} \& {Wild} 2010)
has broad \civ\ absorption blueshifted by up to 9500~\kms\ 
and a redshifted \civ\ trough extending to 12400~\kms\ 
in its SDSS spectrum and to 11100~\kms\ in its BOSS spectrum.
We previously suggested the latter trough could instead be either a very high
velocity ($v=46000$~\kms) \aliii\ trough without accompanying
\mgii, or \HeIIsf\ absorption associated with the 
$v=-8000$~\kms\ \civ\ outflow \nocite{unconventional}({Hall} {et~al.} 2004b).
These hypotheses are ruled out by the detection of redshifted 
\SIiv\ absorption, at velocities matching those of the redshifted 
\civ\ absorption, in the BOSS spectrum of this object.
J1628 also exhibits narrow absorption at $z$=1.4967 ($-$11600~\kms)
in C\,{\sc iv}, Al\,{\sc ii}, Fe\,{\sc ii}, and Mg\,{\sc ii}
and narrow Fe\,{\sc ii}+Mg\,{\sc ii} absorption at $z=0.9402$.
No \oii\ emission is detected in the BOSS spectrum of this object.

{\bf SDSS J170953.28$+$270516.6} (J1709; Fig. \ref{f_cands}) % pmf 5014-55717-0763
has strong, narrow \lya\ at $z=3.123\pm 0.003$, consistent with the
DR9Q \ciii\ PCA redshift of $z=3.126$, which we adopt as systemic.
A redshifted trough is seen in \ovi, \Cthree, \Nv, \SIiv\ and \civ,
extending to +4430\,\kms\ (slightly narrower in \SIiv).
Weak, redshifted absorption is seen in \AlIII.

{\bf SDSS J172404.44$+$313539.6} (J1724; Fig. \ref{f_cands}) % pmf 5002-55710-0600
has narrow \lya\ emission at $z$=2.524$\pm$0.001,
but we adopt the DR9Q \ciii\ PCA redshift of $z$=2.516$\pm$0.001 as systemic.
There is absorption which is mostly redshifted in
\Nv, \cii, \SIiv, \civ, \aliii, and possibly \mgii.
The \SIiv\ and \civ\ absorption extends from $z$=2.51 to $z$=2.62
($\Delta v$=9350~\kms), and the \Nv\ absorption is at least that wide.
The \cii\ and \aliii\ troughs are narrower (4200~\kms).
The \mgii\ region is at the far red end of the spectrum,
but there is a suggestion of a trough in the same velocity range
as \cii\ and \aliii\ and a depth of 50\% of the continuum.
Below \lya, there is strong, broad \CTHREE\ and
\pv/\feiii\ $\lambda$1116 emission in this object,
along with redshifted absorption from \ovi, \SIV, \PV\ and \Cthree.
There is a narrow \lya\ absorber at $z$=2.53. 
The apparent narrow \lya\ absorber longward
of that is \cii\ from a narrow intervening system with
\civ, \SIiv\ and possibly \mgii\ at $z$=2.224.
There is also an intervening system with \SIiv, \civ, \aliii, and \mgii\ at 1.9066.

{\bf SDSS J215704.26$-$002217.7} (J2157; Fig. \ref{f_cands})
has a DR9Q \ciii\ PCA redshift of $z=2.240\pm 0.002$,
which we adopt as systemic.
(We do not adopt the DR9Q \mgii\ PCA redshift of $z=2.247\pm 0.002$
because \mgii\ is weak and affected by night sky emission line residuals.)
It has absorption troughs in \Nv, \SIiv\ and \civ\ composed of
absorption beginning at a small blueshift ($v=-1930$~\kms\ in \civ)
and extending smoothly to a large redshift ($v$=9050~\kms\ in \civ)
at generally decreasing depth.
The region of the blueshifted trough at $z=2.229$ either has a different
range of ionization stages present or constitutes a separate, narrow absorption 
system; absorption is seen there in \lya, \SIii\ and \cii\ as well as in
\Nv, \SIiv, and \civ.
There also appears to be redshifted \pv\ and \aliii\ absorption.
The absorption in \aliii\ is narrower than in the other ions,
which is typical for standard BAL quasars which exhibit \aliii\ troughs.
The spectrum exhibits strong, narrow \lya\ emission 
accompanied by narrow \OI+\SIii$\lambda$1304, \HeIIsf, \mgii\ and possibly \civ\ emission.

\section{Objects with candidate redshifted high-ionization absorption} \label{maybe}

\begin{figure*} %%\vspace*{174pt} \makebox[\textwidth]{
\includegraphics[angle=0, width=0.990\textwidth]{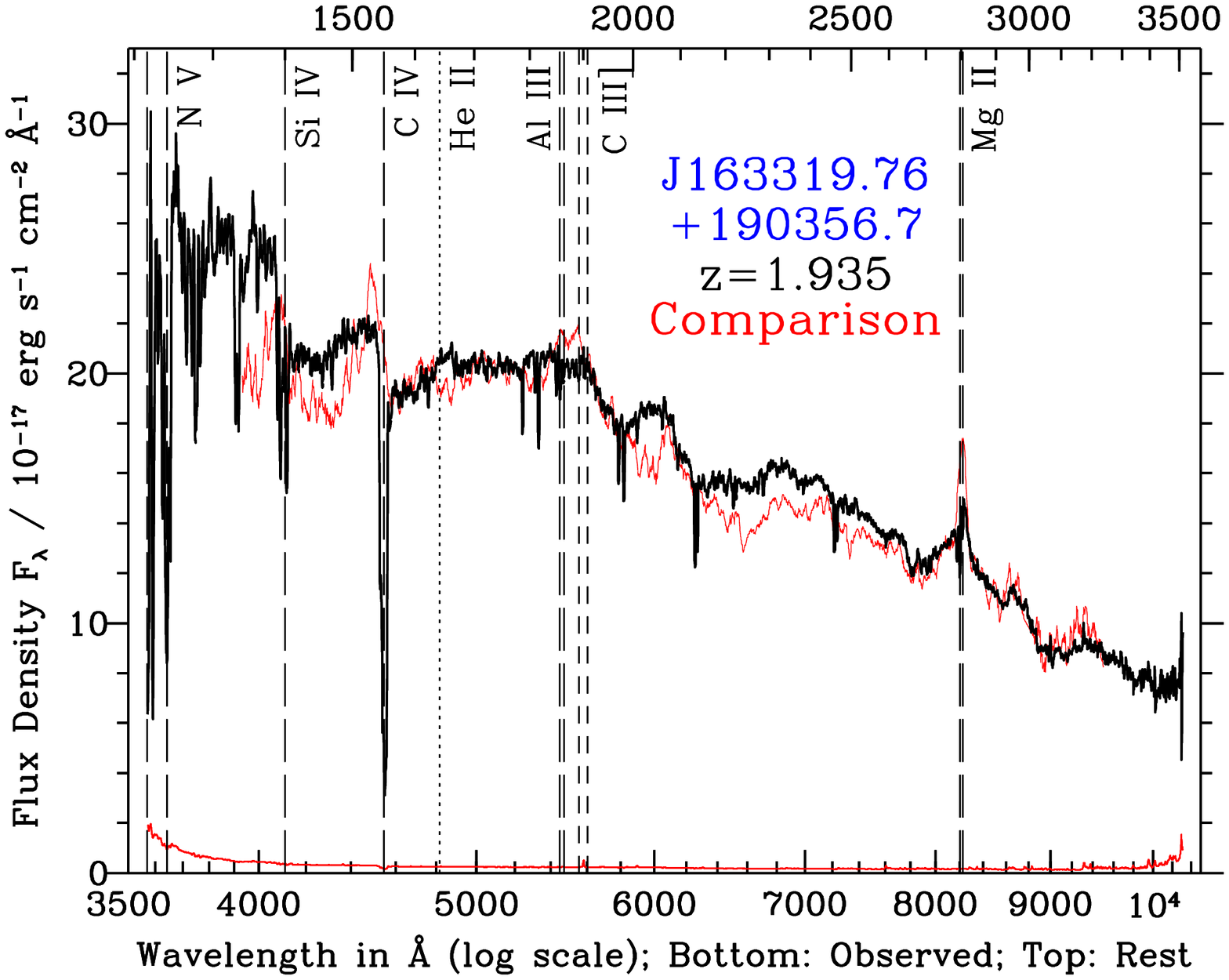} %}
\caption{
SDSS J163319.76+190256.7 (J1633), a quasar with candidate 
redshifted high-ionization absorption.
Vertical lines are the same as in Figure \ref{f_cands}.
Its spectrum (black) is overplotted on the scaled spectrum of the unusual
non-BAL quasar SDSS J113000.64+583248.3 (red); see text. 
}
\label{f_1633} \end{figure*}

\begin{figure*} %%\vspace*{174pt} \makebox[\textwidth]{
\includegraphics[angle=0, width=0.497\textwidth]{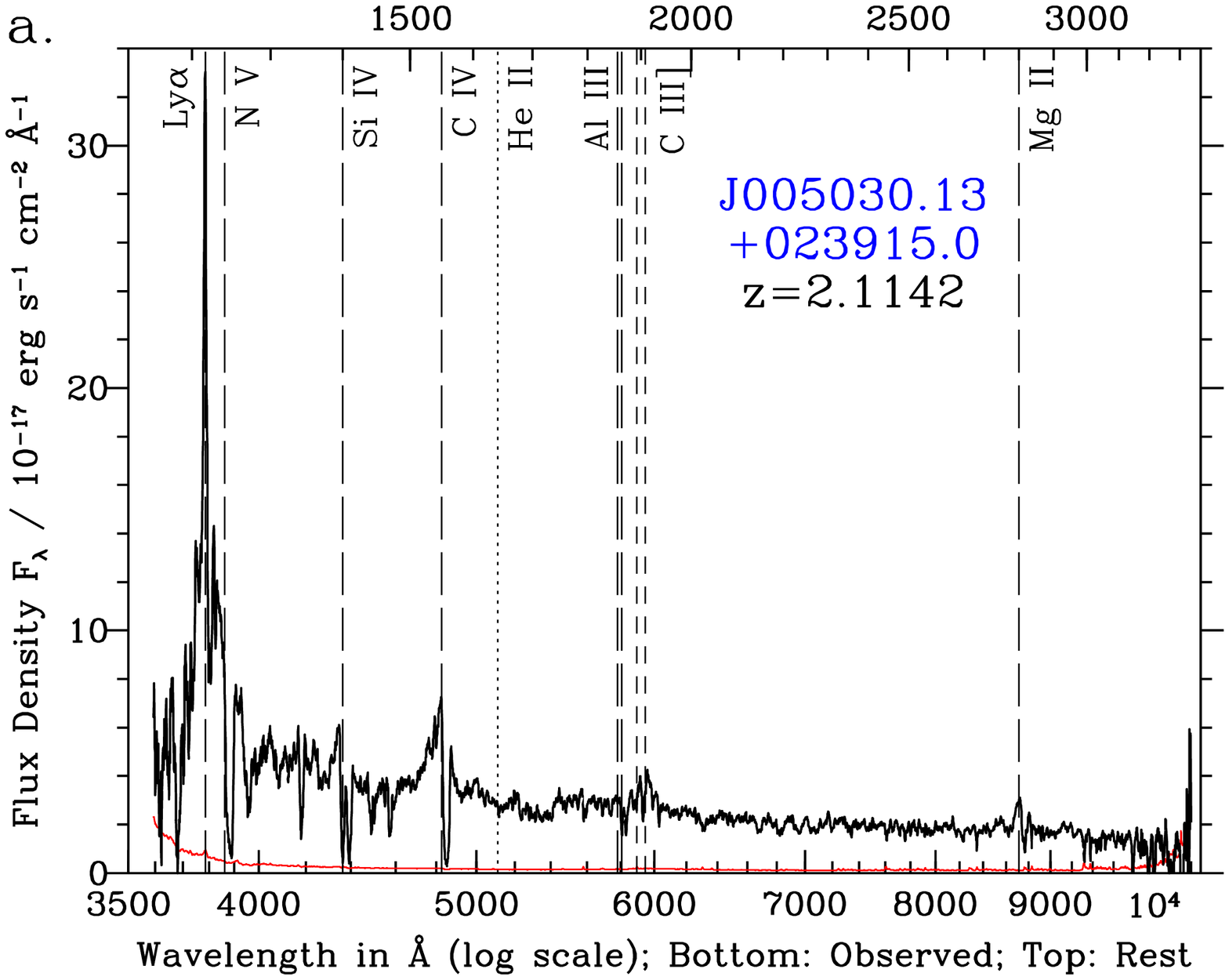} %35--105
\includegraphics[angle=0, width=0.497\textwidth]{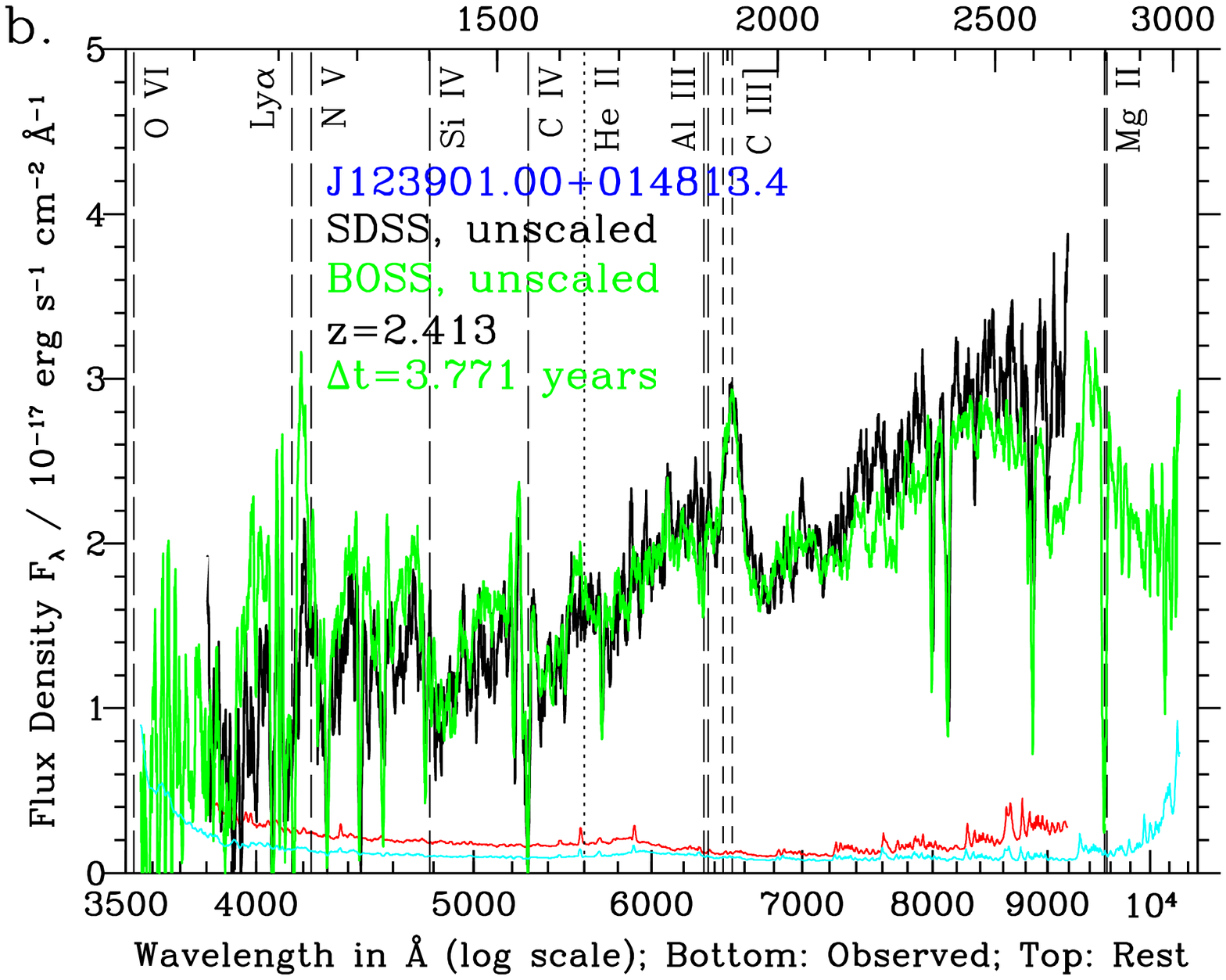} %35--105
\includegraphics[angle=0, width=0.497\textwidth]{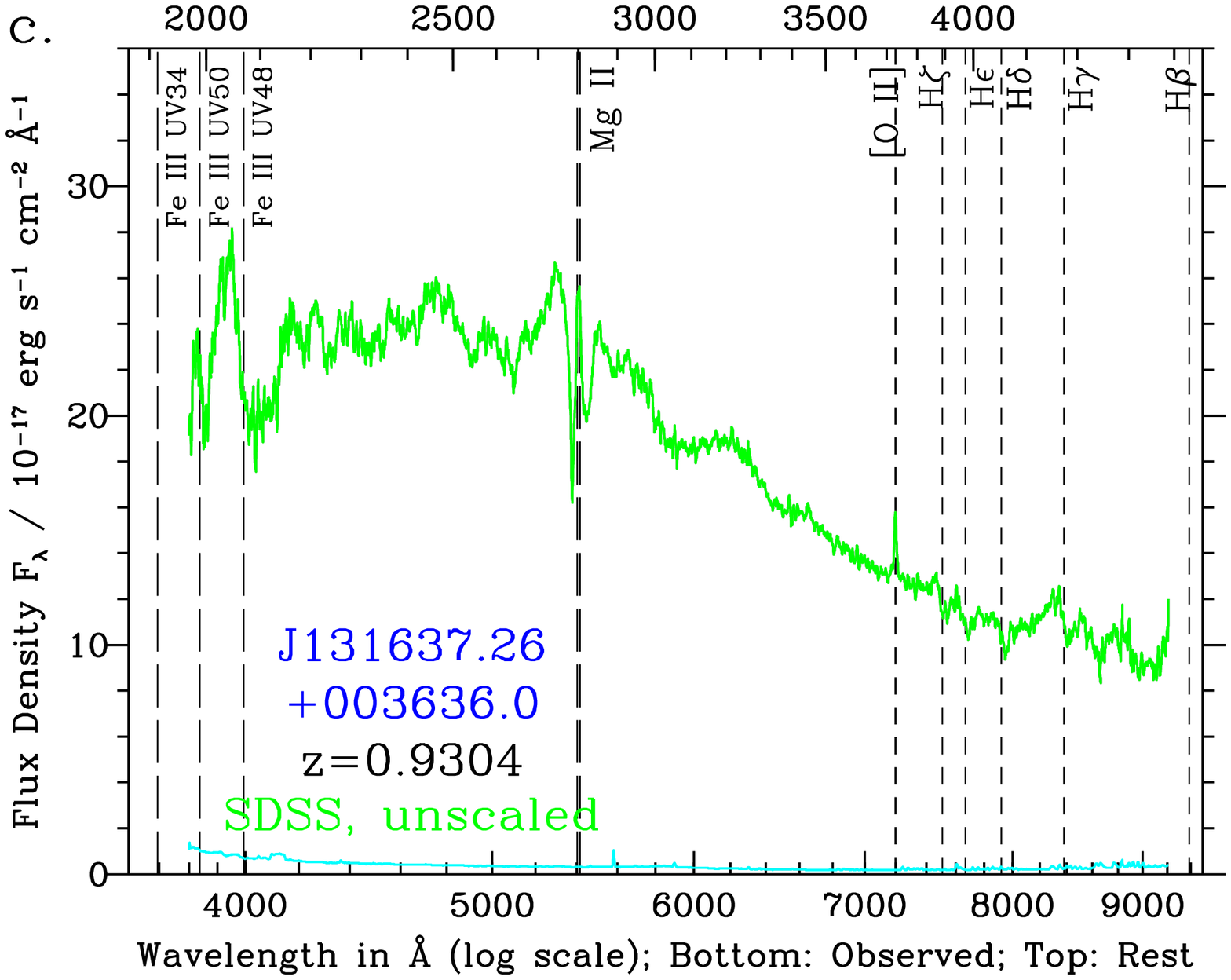} %35--105
\includegraphics[angle=0, width=0.497\textwidth]{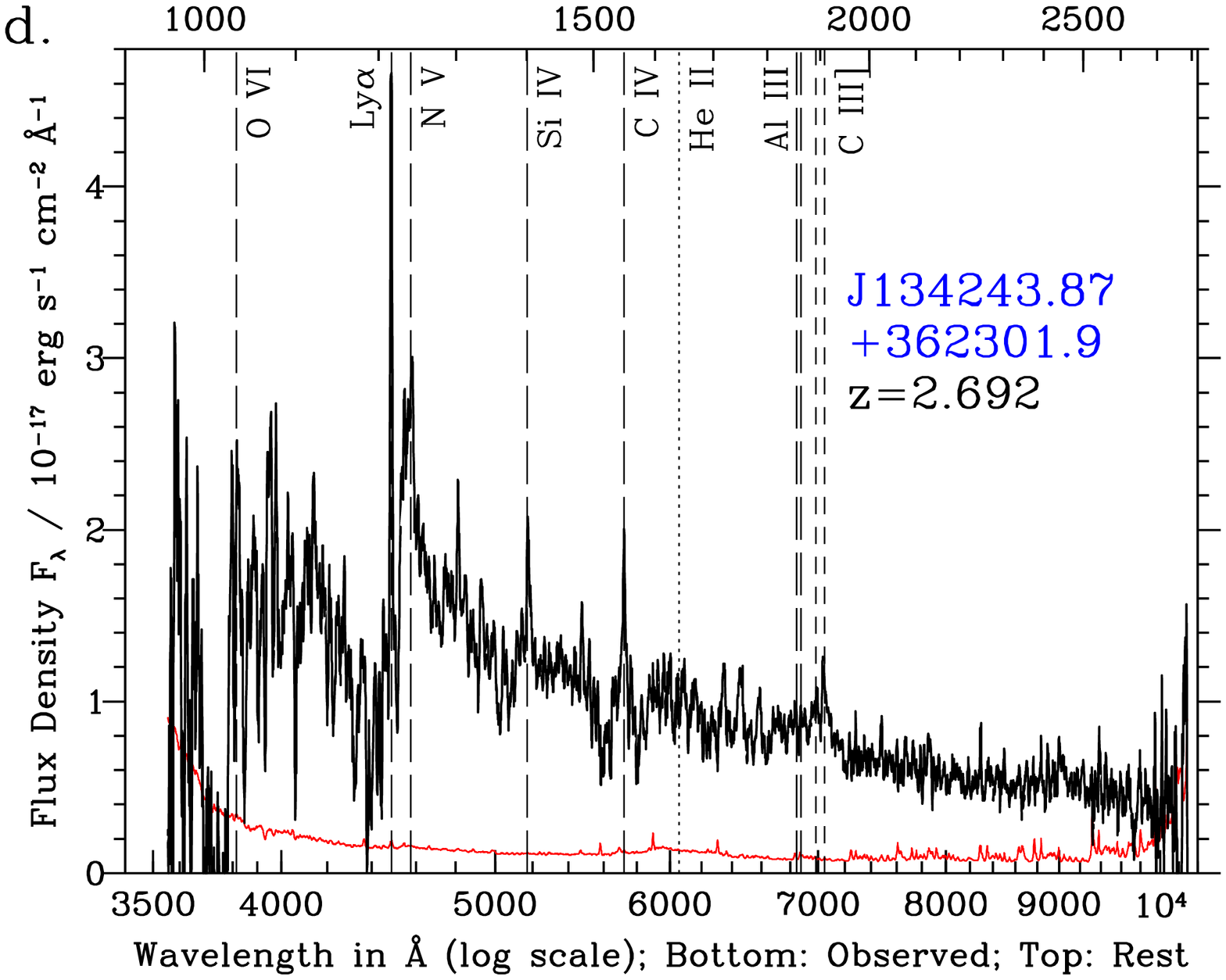} %35--105
\includegraphics[angle=0, width=0.497\textwidth]{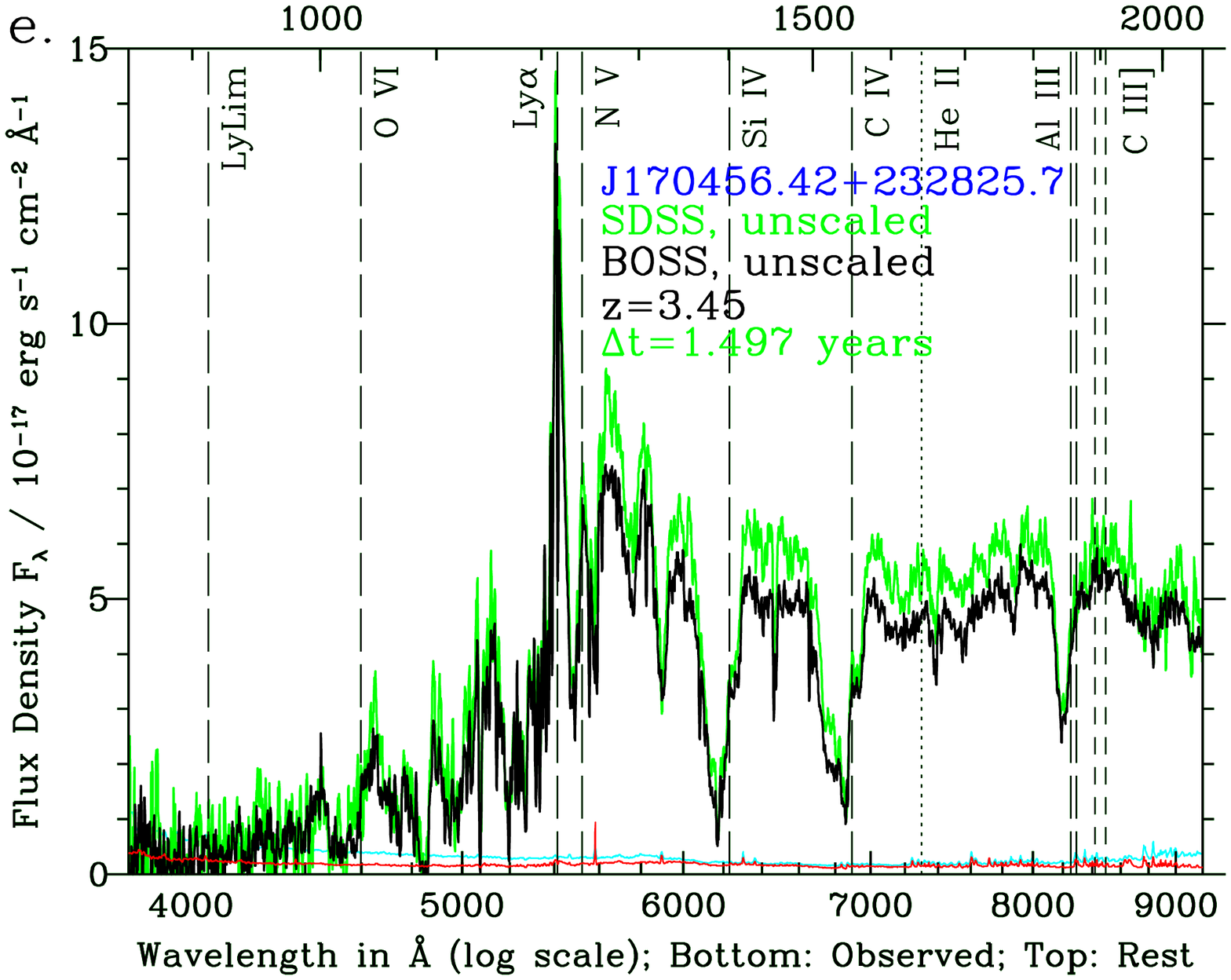}  %38--92
\includegraphics[angle=0, width=0.497\textwidth]{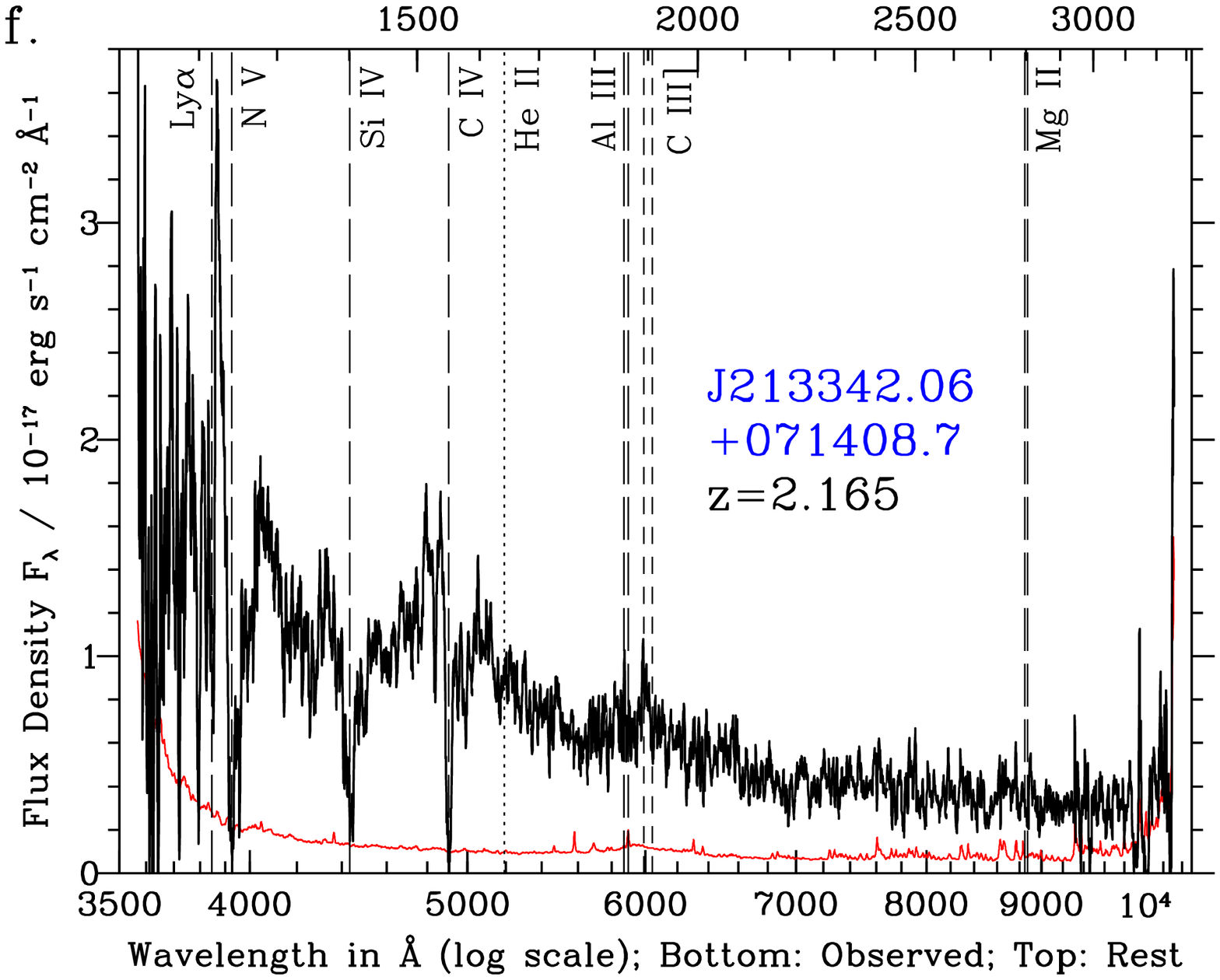} %}
\caption{
Quasars with candidate redshifted high-ionization absorption.
Vertical lines are the same as in Figure \ref{f_cands}.
{\bf (A) Top left:} J0050.
{\bf (B) Top right:} J1239 SDSS spectrum (black) and BOSS spectrum (green).
{\bf (C) Middle left:} J1316 (SDSS spectrum only).
{\bf (D) Middle right:} J1342.
{\bf (E) Bottom left:} J1704.
{\bf (F) Bottom right:} J2133.
}
\label{f_cands3} \end{figure*}

In this section we discuss a number of quasars which have candidate
redshifted high-ionization absorption features in their spectra.
Their spectra are shown in Figure \ref{f_1633} and Figure \ref{f_cands3}.
We discuss the candidate shown in Figure \ref{f_1633} at some length
before discussing the others.

{\bf SDSS J163319.76$+$190856.7} (J1633; Fig. \ref{f_1633}) %(4061-55362-256)
has a DR9Q \mgii\ PCA redshift of $z=1.935\pm 0.001$, 
which we adopt as systemic (no \ciii\ PCA redshift is available).
It has a complex of three narrow absorption systems 
at blueshifts $<$2000~\kms\ in \civ, \Nv, \CII\ and \SIiv.
It also has what appear to be broad troughs extending longward of the rest
wavelengths of \SIiv\ (halfway to the position of \civ) 
and \civ\ (to the position of \heii).
These troughs extend to at least $v$=7300~\kms\ and 
possibly to $v$=14100~\kms\ in \civ.
There may be a redshifted component of \CII\ absorption, 
but there is no clear sign of redshifted absorption in \lya\ or \Nv.
J1633 also exhibits an intervening 
\mgii+\feii\ system at $z=1.2361$ and an intervening \civ\ system at $z=1.4211$.
J1633 was detected in the radio by the FIRST survey \nocite{bwh95}({Becker} {et~al.} 1995)
and has a deconvolved size of 
2.67\arcsec$\times$0.76\arcsec\ with major axis oriented 86.1$^\circ$ E of N.

The reality of the putative redshifted \SIiv\ and \civ\ absorption 
in J1633 is unclear, as follows.
The ultraviolet spectrum of a quasar can be thought of as a roughly power-law
continuum plus strong, broad emission lines and a pseudocontinuum 
of weaker and/or blended broad emission lines, Balmer continuum emission,
\hi\ two-photon emission \nocite{proto}({Hall} {et~al.} 2004a), etc.
The combination of unusual emission-line profiles and strong emission in the
pseudocontinuum can cause gaps between emission complexes to mimic absorption
features.

J1633 is a potential example of this phenomenon.  Its spectrum appears 
explainable by strong \SIiv\ and \civ\ emission which is almost entirely
blueshifted from the systemic wavelength, plus a complex of narrow
\Nv, \SIiv\ and \civ\ absorption systems near the systemic redshift,
plus an unusual spectral shape between \civ\ and \CIII.

Cases of entirely blueshifted \civ\ emission are rare but not unknown
(e.g., Fig. 3 of \nocite{lm04}{Leighly} \& {Moore} 2004).  In particular, the spectrum of J1633
strongly resembles the composite spectrum of `PHL 1811 analog' quasars
presented in Fig. 8 of \nocite{2011ApJ...736...28W}{Wu} {et~al.} (2011).
The UV spectra of PHL 1811 analogs exhibit weak high-ionization emission lines
but strong \feiii\ 2080\,\AA\ and UV \feii\ emission like that seen in J1633.
As an example, the candidate PHL 1811 analog SDSS J113000.64$+$583248.3
shows a strong emission complex between \civ\ and \ciii.

We show in Fig. \ref{f_1633}
the spectrum of J1633 as a thick black line, with a scaled and heavily smoothed
spectrum of the unusual quasar
SDSS J113000.64$+$583248.3 overplotted as a thin red line.
We used a scaling $\propto \lambda^{0.6}$ to roughly account for the effects of
different power-law continuum slopes and reddening in the two objects.
The scaling demonstrates that a quasar without any obvious absorption can
match the continuum shape of J1633 at wavelengths between \civ\ and \mgii.
(This figure also demonstrates that if J1633 lacked narrow absorption troughs,
it would likely qualify as a PHL 1811 
analog according to the criteria of \nocite{2011ApJ...736...28W}{Wu} {et~al.} (2011).)
A slightly shallower scaling would better match the continuum near \SIiv,
indicating that at wavelengths just longward of \civ\ J1633 has weaker emission 
than SDSS J113000.64$+$583248.3 does.
Possible contributors for the strong emission between \civ\ and \ciii\ include
\feii, \heii, \alii, \Niii\ and \aliii.
However, the exact origin is immaterial for our purposes:
we have demonstrated that J1633 can plausibly be explained without invoking
redshifted absorption troughs, and so we exclude it from further discussion
here.  Nonetheless, it would be worth obtaining further spectroscopy
of this object, to shorter rest-frame wavelengths if possible, to 
characterize its spectral shape more accurately
and to search for absorption variability.

Turning to Figure \ref{f_cands3}, we have the candidate
{\bf SDSS J005030.13$+$023915.0} (J0050; Fig. \ref{f_cands3}). %pmf 4306-55584-0846
It has strong, narrow \lya\ emission at $z$=2.1142$\pm$0.0010, 
consistent with the DR9Q \mgii\ PCA redshift of $z$=2.112
and the DR9Q full PCA redshift of $z$=2.118; we adopt the latter as systemic.
(The DR9Q \ciii\ PCA redshift of $z$=2.129 is spuriously large.)
a redshift consistent with the redshift of the peaks of \civ\ and \mgii.
It has a redshifted trough seen in \lya, \Nv, \cii, \SIiv\ and \civ\ from
$v$=0 to +1780 \kms).
The relatively narrow velocity range means that it is either a mini-BAL quasar
or has a complex of unresolved narrow absorption systems along our line of
sight which appear redshifted due to their peculiar velocities
\nocite{2008MNRAS.388..227W}({Wild} {et~al.} 2008).
In fact, within the broad redshifted trough, there is narrow 
\aliii, \feiii, and \mgii\ absorption at $z$=2.129 (+1060 \kms).
The broad absorption does not reach zero depth in \lya, but nearly does in \Nv.
The absorber might be too highly ionized to show strong \lya, or might
not fully cover the \lya\ emission region at the velocities in question.

The line widths and strengths of the absorption features in J0050 
are probably too strong for the absorber to be an intervening system 
with a redshifted peculiar velocity of 1060~\kms\ relative to the quasar.
However, the redshifted velocity is one of the smallest in our sample,
and the \lya, \ciii\ and \mgii\ lines are all affected by absorption,
increasing the uncertainty of this object's systemic redshift so that
a redshift high enough to eliminate the redshifted absorption cannot
be excluded.  Until a more robust systemic redshift is determined
(e.g., through near-infrared spectroscopy of rest-frame optical
narrow emission lines), we consider this quasar merely a candidate
for having redshifted absorption.

{\bf SDSS J123901.00$+$014813.4} %pmf 4754-55649-0294 and pmf 521-52326-0221 (nmf)
(J1239; Fig. \ref{f_cands3}) 
may be a case of a quasar with an unusual continuum shape and
relatively narrow absorption at the systemic redshift.
It has an SDSS spectrum which differs from the BOSS spectrum in having
a redder continuum shape and a weaker \lya+\Nv\ emission line.
J1239 has strong narrow high- and low-ionization absorption 
from a potential damped \lya\ absorber at $z$=2.413$\pm$0.001.
It has a strong 1900\,\AA\ emission complex at a redshift consistent 
with the absorption redshift.
\lya, \Nv, \civ, \heii\ and probably \mgii\ emission are blueshifted
by 3000~\kms\ to $z$=2.379$\pm$0.001.
The spectrum also exhibits a strong UV \feii\ pseudocontinuum.
There appears to be broad redshifted absorption in \civ, \SIiv\ and possibly
\Nv\ extending from at least $v=-2340$\,\kms\ to $v=+1770$\,\kms\ ($\Delta v=4110$~\kms),
which would still extend to redshifted velocities even if we adopted
the \nocite{hw10}{Hewett} \& {Wild} (2010) redshift of 2.4222$\pm$0.0055 for this object.
Our lack of certainty in the reality and extent of these putative redshifted
troughs may be due to the low signal-to-noise ratio of the spectra.  
Nonetheless, as with J1633, 
a combination of blueshifted \SIiv\ and \civ\ emission and strong
\feii\ pseudocontinuum could explain the spectrum of this object without
redshifted absorption troughs, so for now we consider it merely a candidate.

{\bf SDSS J131637.26$-$003636.0} (J1316; Fig. \ref{f_cands3}) at $z=0.9304$
from \oii\ emission \nocite{hw10}({Hewett} \& {Wild} 2010) is our only candidate with \mgii\ absorption.
This object is not a BOSS ancillary target,
but has three SDSS spectra (one of which is very noisy).
There is some flux variability between the two best SDSS spectra, but no 
statistically significant trough variability between the normalized spectra.
The spectrum of this object presented in Figure 3 of \nocite{2001AJ....122..518H}{Hewett}, {Foltz} \&  {Chaffee} (2001)
was taken in between the last two SDSS spectral epochs and confirms the 
absorption features bracketing the systemic redshift of \mgii,  
extending from $-3300$\,\kms\ to $+3800$\,\kms\ in a $\mathsf W$ shape.
The SDSS spectra show that these are not NAL systems; 
both ``troughs'' are smooth over their $\sim$2000\,\kms\ widths.  
The blueshifted trough has been 
included in the BAL quasar samples of \nocite{trump06}{Trump} {et~al.} (2006) and \nocite{gibsonbal}{Gibson} {et~al.} (2009).
We classify this object as having candidate redshifted absorption because we do
not have confirmation of the redshifted trough in other transitions.
In fact, this object was mentioned in \nocite{sdss123}{Hall} {et~al.} (2002), where we suggested that
both apparent \mgii\ troughs could be gaps between \feii\ multiplets seen in
emission, as could the apparent troughs near 4000\,\AA\ observed.  
However, those latter features could be redshifted \feiii\ absorption,
and there may be redshifted Balmer-line absorption as well.  
The high densities required for such absorption \nocite{balmerbal}({Hall} 2007) 
could yield a very useful distance constraint for the absorber in 
this object (see \S\,\ref{dist}), if the absorption is confirmed.

{\bf SDSS J134243.87$+$362301.9} % pmf 3987-55590-0124 
(J1342; Fig. \ref{f_cands3})
has blueshifted troughs in \civ, \SIiv, \lya+\Nv, and \ovi\ (the latter being 
particularly strong relative to \civ).
Our inspection redshift of $z=2.6917\pm 0.0004$ agrees with the DR9 pipeline 
redshift and is equal within the errors to the PCA redshift of $z=2.6915$.
It also appears to have a weak, redshifted \civ\ trough.
However, the significance of that putative trough is low due to the noise
in the spectrum, and there is no redshifted trough apparent in any other ion.
Therefore we consider J1342 as only a candidate for redshifted absorption.

{\bf SDSS J170456.42$+$232825.7} (J1704; Fig. \ref{f_cands3}) 
has \lya\ emission which peaks at $z$=3.4410$\pm$0.0005.
\nocite{hw10}{Hewett} \& {Wild} (2010) give $z$=3.410$\pm$0.009, which is clearly too low a redshift.
There is narrow \lya\ absorption at $z$=3.4500$\pm$0.0005,
consistent with the redshift of the broad \ciii\ line,
and so we adopt that as the systemic redshift.
However, the DR9Q \ciii\ redshift is $z$=3.491,
in which case its absorption troughs would be purely blueshifted.
Examination of the \ciii\ emission region suggests that 
the true systemic redshift lies somewhere between these two values.
Therefore, we consider this object only a candidate for redshifted absorption
until a more robust systemic redshift is available.

Both SDSS and BOSS spectra are available for J1704, which is slightly
fainter in the BOSS epoch.  
There is no significant trough variability between the two epochs.
J1704 has troughs extending across our adopted systemic redshift in \SIiv, 
\civ\ and \aliii\ from at least $v=-9180$\,\kms\ to $v=+2410$\,\kms,
and blueshifted troughs in \aliii, \alii, \cii, \oi/\SIii, \SIii,
\Nv, \Cthree\ and \ovi, with spans of up to $\Delta v=4100\pm 700$~\kms.
Because the redshifted portions of the \civ, \SIiv\ and \aliii\ troughs in this
object extend only to $\sim$+2500~\kms, it is possible that they are blends of 
narrow associated absorbers which appear redshifted due to their peculiar 
velocities \nocite{2008MNRAS.388..227W}(e.g., {Wild} {et~al.} 2008).  That possibility can be 
tested with high resolution spectroscopy.
It is also possible that there is some \feii* absorption in this object 
between \civ\ and \ciii.  However, the wavelengths and strengths of those
\feii* transitions are not a good match to the apparent redshifted absorption
in J1704, which does match \SIiv, \civ\ and \aliii\ in wavelength.

Finally, for {\bf SDSS J213342.06$+$071408.77} 
(J2133; Fig. \ref{f_cands3})
we conservatively adopt a systemic redshift of $z$=2.165
from our inspection of the spectrum 
(the DR9Q PCA and \mpca\ redshifts are $z$$\simeq$2.1).
J2133 has troughs at the adopted systemic redshift,
plus candidate redshifted troughs, in \civ, \SIiv, and \Nv.
It is listed as a candidate primarily because the low signal-to-noise 
ratio of its spectrum makes both its systemic redshift and the reality 
of the putative redshifted troughs uncertain.
For example, while no DR9Q \ciii\ PCA redshift is available 
for this object, its DR9Q full PCA redshift is $z$=2.096,
which if correct would make its troughs even more redshifted.

\section{Spurious candidates} \label{noway}

\begin{figure*} %%\vspace*{174pt} \makebox[\textwidth]{
\includegraphics[angle=0, width=0.81\textwidth]{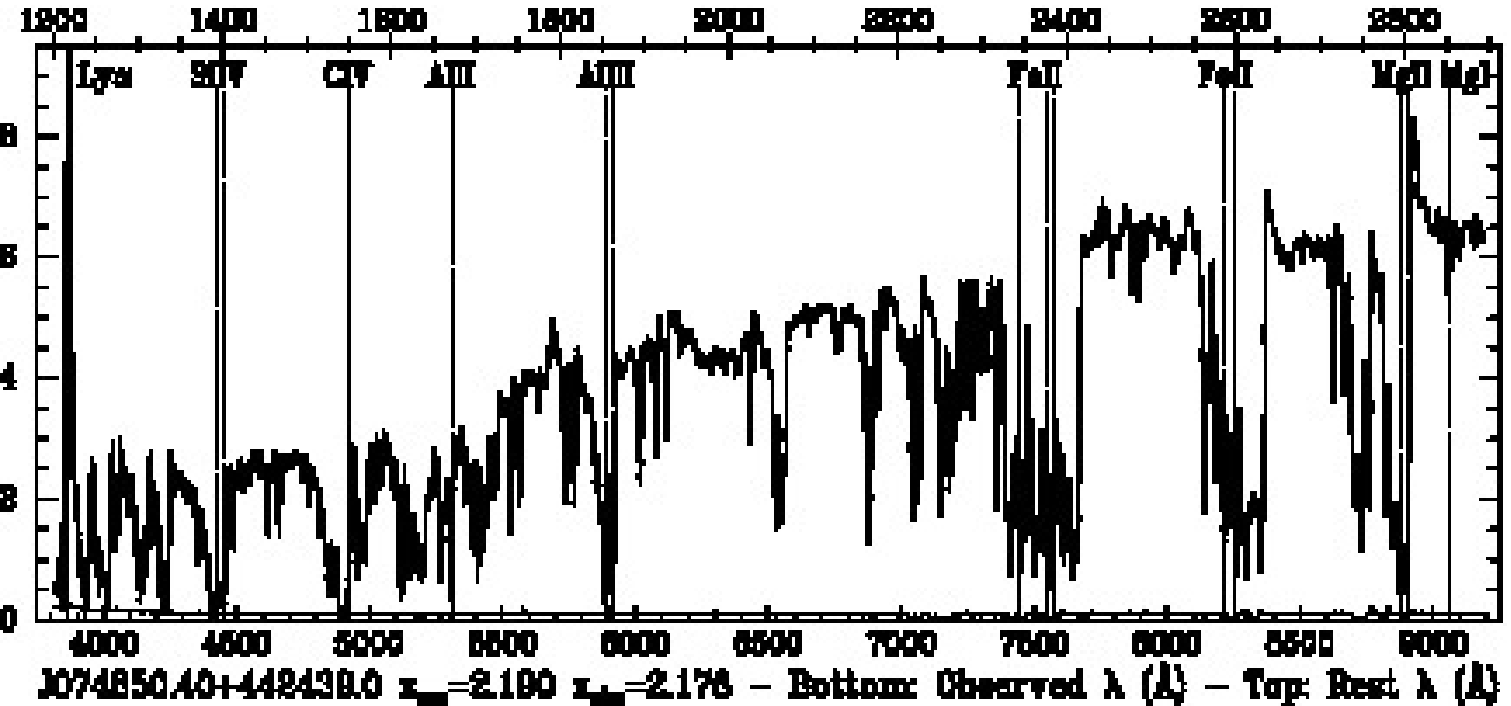} %}
\includegraphics[angle=0, width=0.84\textwidth]{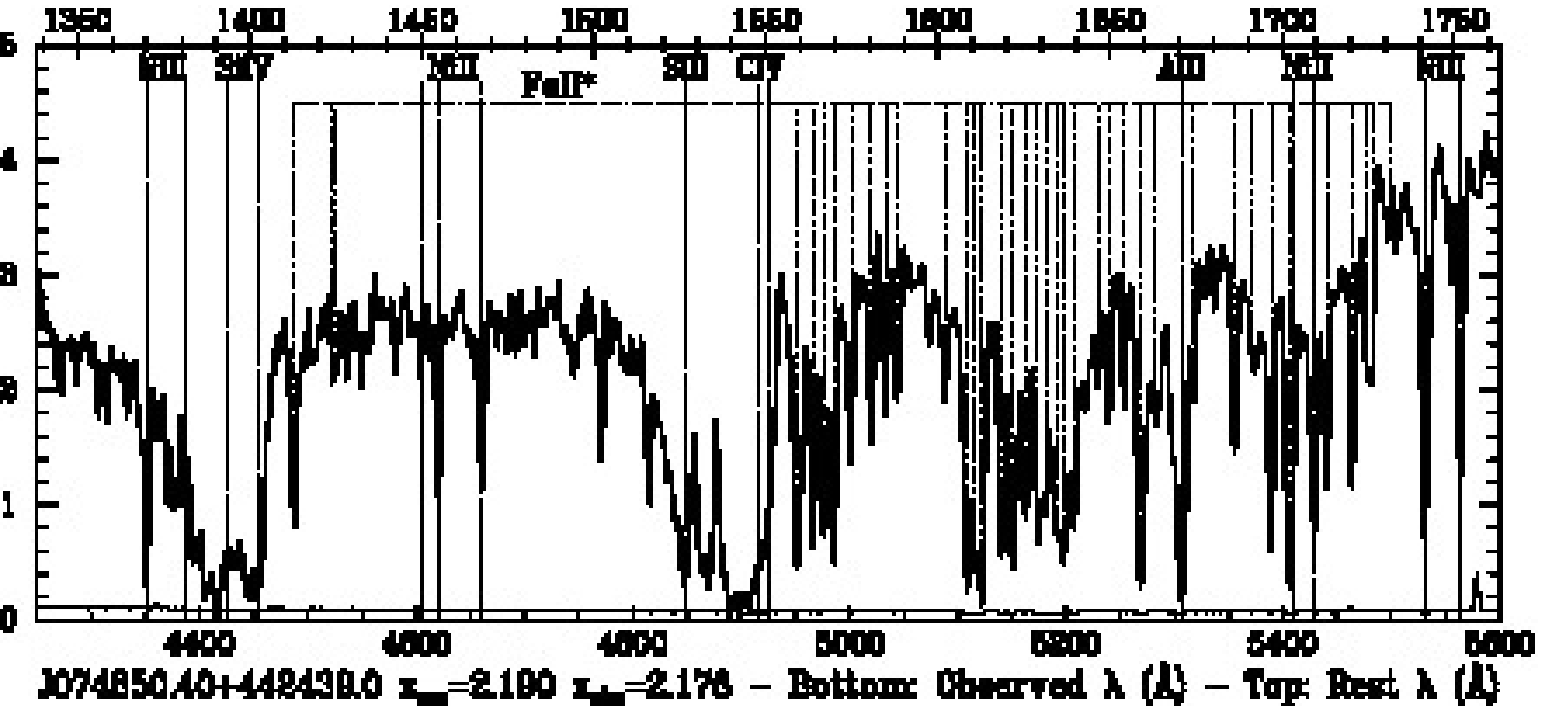} %}
\includegraphics[angle=0, width=0.84\textwidth]{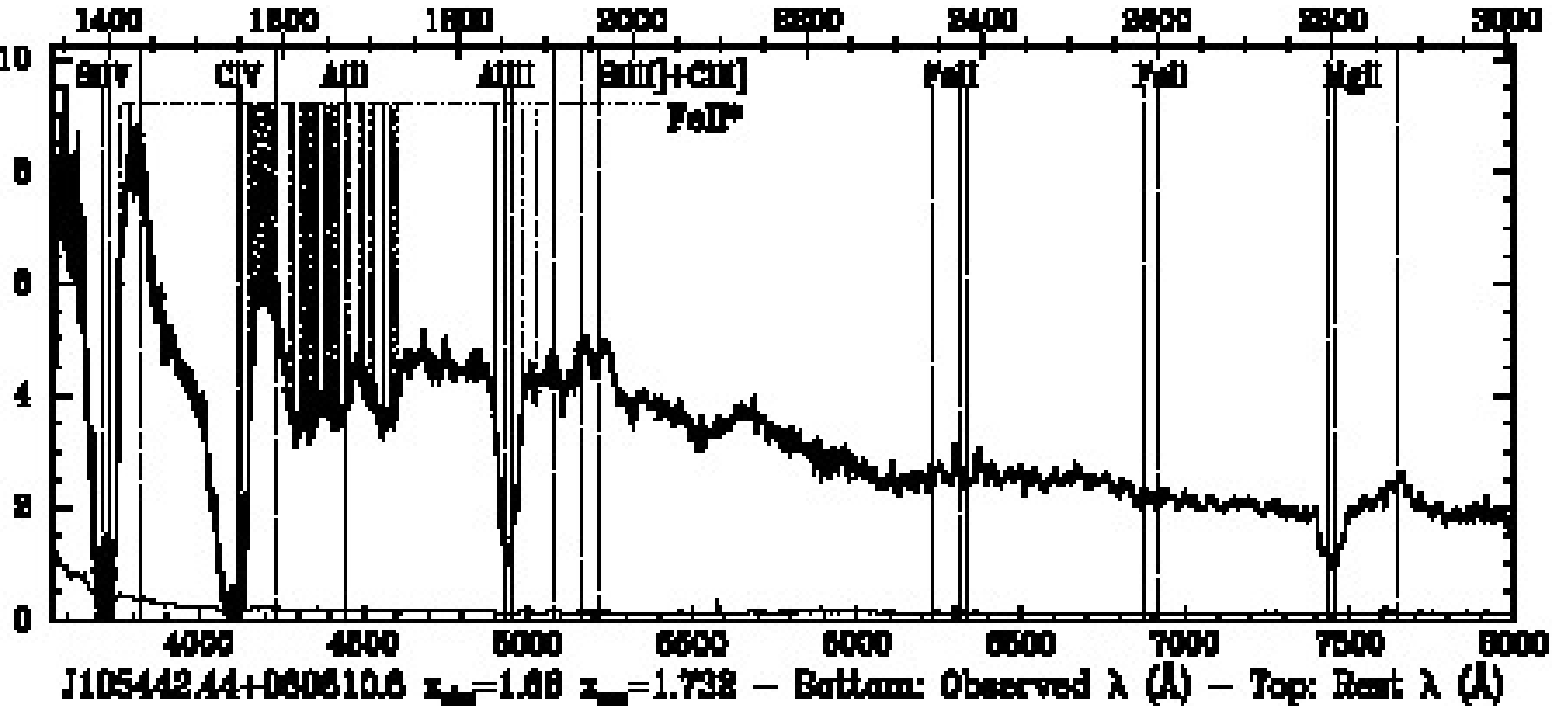} %}
\caption{Quasars with excited \feii\ absorption masquerading as
redshifted high-ionization absorption.  
Each spectrum has been smoothed using a
weighted-average boxcar function and is plotted as 
$F_\lambda$ (in units of 10$^{-17}$ erg s$^{-1}$ cm$^{-2}$ \AA$^{-1}$) vs. $\lambda$.
Top: Coadded spectrum of J0748 from \lya\ to \mgii, with a selection of strong
absorption transitions marked (dashed lines).  \civ\ appears to include 
redshifted absorption but \SIiv\ does not.
Middle: J0748 spectrum between 1350\,\AA\ and 1750\,\AA.  The apparent
redshifted \civ\ absorption is resolved into numerous narrow lines of excited
\feii\ (dotted lines).  There are numerous strong transitions of excited
\feii\ at wavelengths longward of \civ, but few between \SIiv\ and \civ.
Bottom: The spectrum of J1054, with dot-dashed lines showing the emission 
redshift and dashed lines showing the redshift of strongest absorption.
The candidate redshifted \civ\ absorption is very well matched in wavelength
by the same \feii* absorption seen in J0748 (dotted lines).  Absorption atop
the \civ\ emission line can explain its weakness relative to \SIiv.  
}
\label{f_candfe2} \end{figure*}

\begin{figure*} %%\vspace*{174pt} \makebox[\textwidth]{
\includegraphics[angle=0, width=0.85\textwidth]{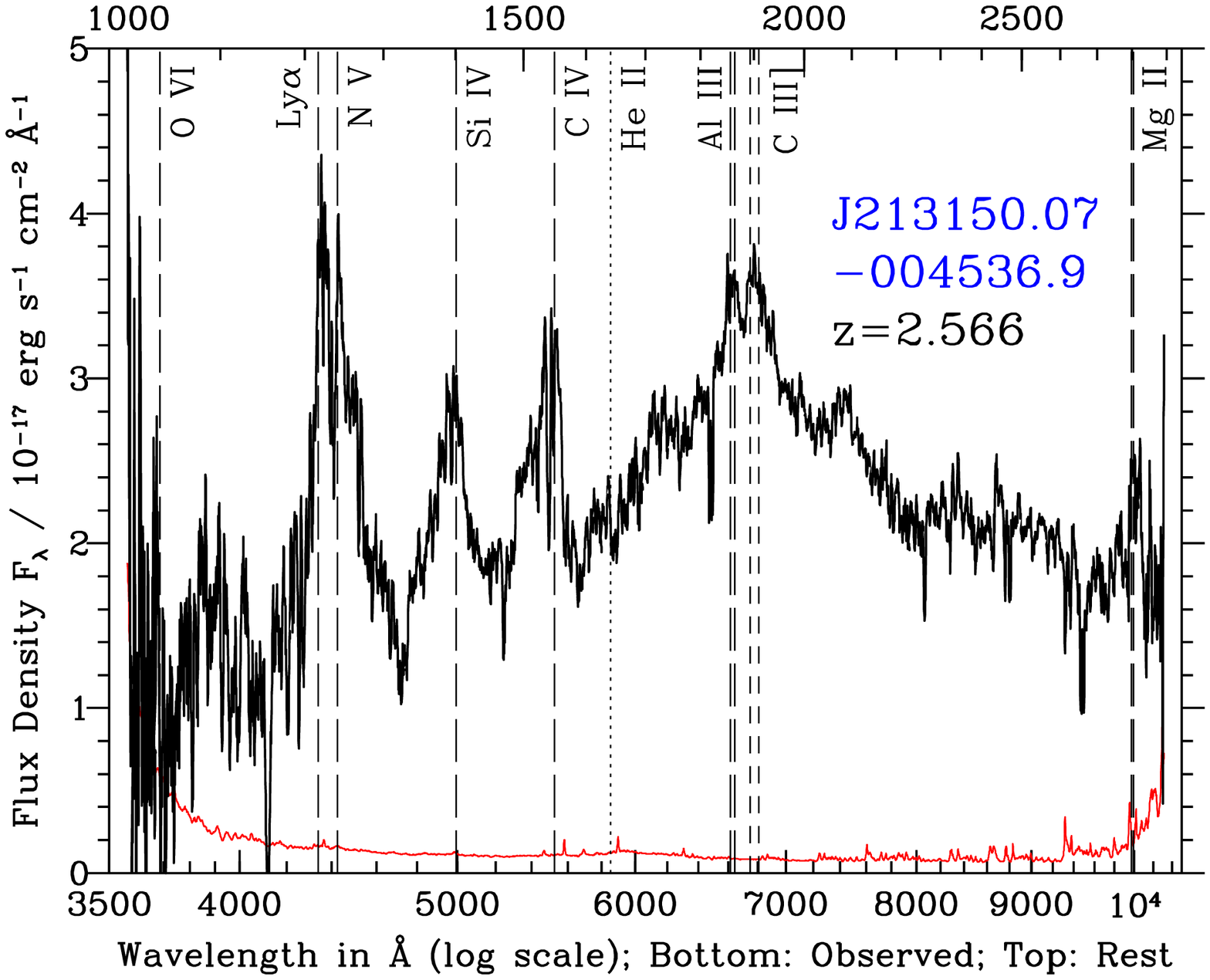} %}
\caption{The BOSS spectrum of SDSS J213150.07$-$004546.9 (J2131).
Vertical lines are the same as in Figure \ref{f_cands}.
J2131 appears to have redshifted absorption in \Nv, \SIiv, and \civ.
However, its spectrum is more likely a combination of strong
\feii\ and \feiii\ emission and a steep reddening curve.}
\label{fj2131rej} \end{figure*}

In this section we discuss a number of quasars which have apparent
redshifted absorption features in their spectra which are not real.

In Figure \ref{f_candfe2} we show spectra of quasars wherein absorption from
excited-state \feii\ (\feii*) resembles redshifted high-ionization absorption.
The quasar SDSS J074850.40+442439.0 (hereafter J0748) is shown in the top
panel.  It is an FeLoBAL with apparent redshifted absorption in \civ\ but
not in \SIiv.  However, closer inspection of its spectrum from 1350\,\AA\ to
1750\,\AA\ (middle panel) reveals that the apparent redshifted \civ\ absorption
is part of a complex of \feii* absorption with many strong transitions between
\civ\ and \ciii\ but only one or two strong transitions between \SIiv\ and
\civ\ (see \S 5.1.1 of \nocite{sdss123}{Hall} {et~al.} 2002).  Such \feii* absorption can blend
together to masquerade as a redshifted \civ\ trough in other quasars such as
SDSS J094633.97$+$365516.8 and SDSS 144424.54+013457.0 (not shown) and
SDSS J105442.44+060610.8 (bottom panel).  In that panel the dotted lines
show \feii* absorption at the redshift of strongest absorption in other
transitions (dashed lines).  Some of the \feii* absorbs the \civ\ emission 
line, weakening it relative to the \SIiv\ emission line, and some absorbs
part of the continuum between rest-frame 1600\,\AA\ and 1750\,\AA.
In this object there is little evidence for additional \feii* absorption
between 2330\,\AA\ and 2630\,\AA\ which should be present, as it arises from
the same lower levels as the observed absorption.  Most likely, the absorption
is present but is blended with \feii\ emission so as to mimic a relatively
smooth continuum.

\feii\ and \feiii\ emission may be at work in SDSS J213150.07$-$004546.9 
(J2131; Figure \ref{fj2131rej}), which we considered, but ultimately rejected,
as a redshifted-absorption candidate.  
This object has a secure redshift from \aliii+\Siiii+\ciii\ emission.
An extrapolation of the continuum
longward of \ciii\ to shorter wavelengths generates the appearance of strong,
redshifted troughs of \civ, \SIiv, and \Nv.  However, the detailed shape and
position of the putative troughs do not agree with that identification.
(E.g., what looks like a moderately broad trough at 4700\,\AA\ in the smoothed
spectrum is actually the blended \SIiv\ doublet of an intervening metal-line 
absorber at $z=2.383$.)
Therefore, the unusual feature in this spectrum is not redshifted absorption
but the huge apparent emission feature from 1700--2300\,\AA.
This feature is somewhat reminiscent of that in SDSS J033810.85+005617.6
(\S 5.3.5 of \nocite{sdss123}{Hall} {et~al.} 2002), but is likely due to some combination of
blended \feii\ and \feiii\ emission and an extinction curve steeper than 
that of the SMC (\S 5.3 of \nocite{sdss123}{Hall} {et~al.} 2002).

Another source of apparent redshifted absorption features is unrecognized
flaws in the data.  Such flaws are a possibility in any pipeline, and spurious
broad features have previously been seen in some SDSS spectra (see, e.g.,
\nocite{rwd}{Hall} {et~al.} 2008).  

Experience has shown that spectra from certain SDSS fibers, including \#374,
are more prone to spurious features.
For example, SDSS J124224.78$+$364537.1
(SDSS plate-MJD-fiber 2022-53827-374 and BOSS 3973-55323-0536)
had an apparent redshifted \civ\ trough in SDSS which was not confirmed by BOSS
spectroscopy.  Inspection of the individual spectroscopic exposures in the SDSS
revealed that the trough was indeed spurious.
The apparent trough is located in the wavelength 
range where the coverage of the red and blue arms of the SDSS spectrographs
overlap.  A decrease in flux at the trough wavelengths was present in all of
the spectra from the blue arm but none of the spectra from the red arm.
The pixels from the blue arm were not rejected, resulting in a spurious trough.
Other examples of spurious troughs involving SDSS fiber \#374 include
SDSS J004418.96$-$090009.4 and SDSS J171216.27+660212.0.

As another example of spurious
features tied to certain fibers, the SDSS spectrum of the BL Lac
object OM~280 (SDSS J115019.21+241753.8, plate-MJD-fiber 2510-53877-086) appears
to show a trough at $\sim$4150\,\AA.  This trough is spurious, likely arising
from a region of problematic flatfielding (e.g., due to a localized variable 
feature in the flatfield) in the CCD in the blue arm of SDSS spectrograph \#1.
This region is visible as a circular depression on the reduced 2-D spectrograms
of that CCD in that wavelength range, centered near fiber 81.  The affected
wavelength region is interpolated over for fibers 78--84, but its effects can
sometimes be seen in other nearby fibers such as \#86 (used for the spectrum of
OM 280), \#76 (producing a similar spurious trough in SDSS J115013.88+234602.1),
and possibly \#73, in which case it could explain the apparent trough in
the SDSS spectrum of the BL Lac object PKS\,0138$-$097 
\nocite{2011arXiv1106.1587S}({Shaohua} {et~al.} 2011).

Lastly, inspection of both spectroscopy and imaging can also identify cases of
apparent redshifted absorption caused by superpositions.  For example, 
the star-quasar blend SDSS J120749.19+082406.7 has an apparent redshifted 
\aliii\ trough from the stellar Mg~b feature.

\section{Absorber motion and the relativistic Doppler shift} \label{rds}

\begin{figure*}
\includegraphics[angle=0, width=0.659\textwidth]{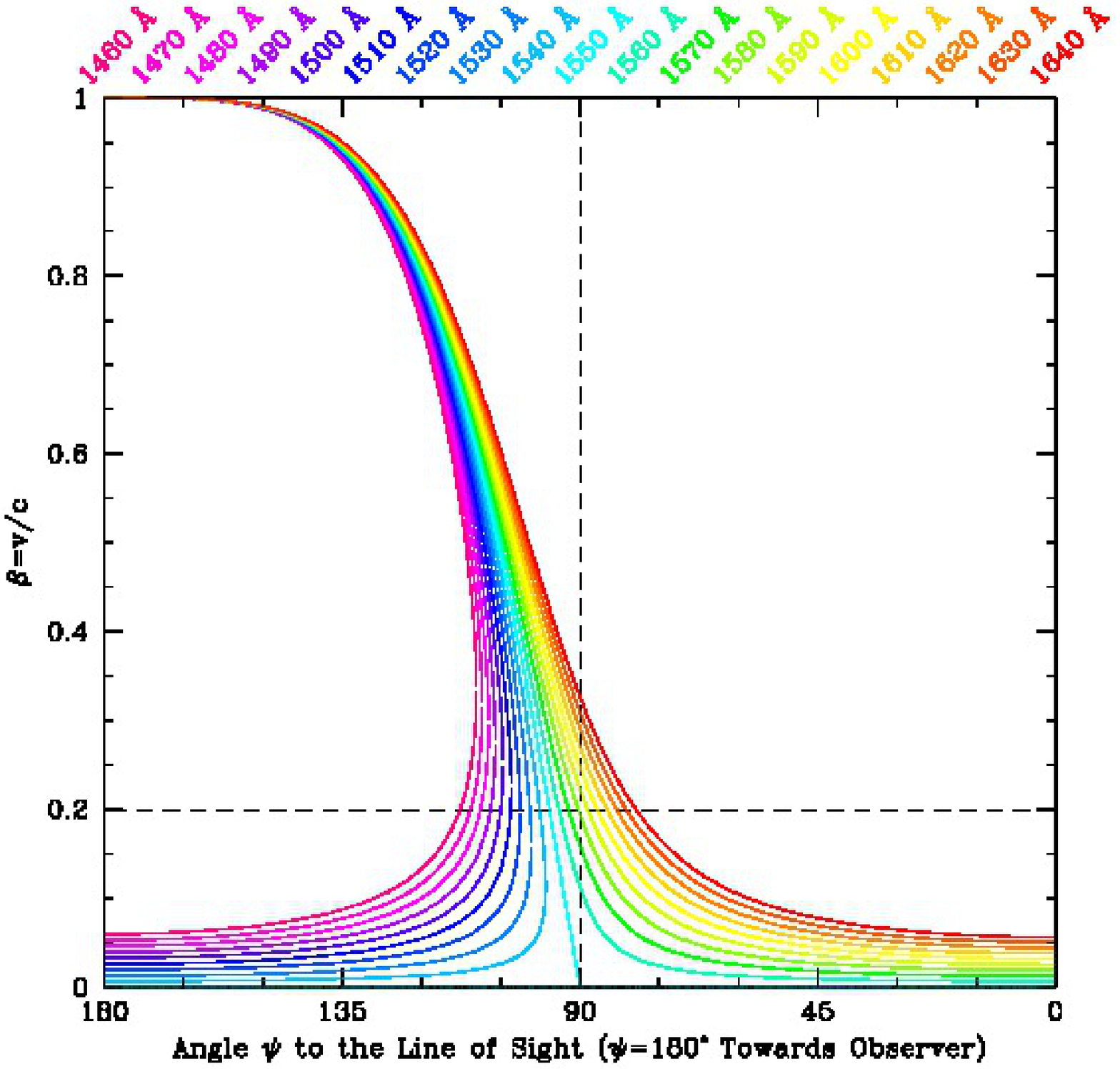}
\caption{As a function of the angle $\psi$ between the
velocity vector of the absorbing gas and the line of sight to the observer,
this plot shows the value(s) of $\beta$ required to absorb at a given 
wavelength near the \CIV\ transition.
Curves are plotted for wavelengths from 1460\,\AA\ (violet)
through 1550\,\AA\ (cyan) to 1640\,\AA\ (red).
Gas with a component of motion away from the observer can only produce
redshifted absorption.
The vertical dashed line shows $\psi=90^\circ$, the direction of motion
exactly transverse to the observer.  The relativistic Doppler effect can be seen
as the increase with $\beta$ of the redshift of photons absorbed by gas moving
at $\psi=90^\circ$.  
Blueshifted absorption can only be produced by gas with a component of motion
toward the observer, but at each angle $\psi$ there are two values of $\beta$
that will absorb at a given blueshifted wavelength.
The horizontal dashed line shows the largest $\beta$ observed
in UV outflows in quasars.}
  \label{figrel}
\end{figure*}

\nocite{1968JRASC..62..105S}{Sher} (1968) discuss the implications of the relativistic
Doppler effect in terms of the radial and total velocities of the source.
We present a similar discussion in terms of the total velocity of the gas
and the angle of its motion relative to our line of sight.
For simplicity we consider an observer at the systemic redshift $z_{sys}$ of
the quasar (but physically distant from it) with the same line of sight 
to the quasar as us.
The only difference between such an observer and us is that the wavelengths
seen on Earth are all a factor of $1+z_{sys}$ longer.

Consider gas moving with speed $v=\beta c$ at an angle $\psi_{obs}$ to the
observer's line of sight in the observer's frame, where $0\leq\beta\leq 1$
and $\psi_{obs}=180^\circ$ points directly toward the observer (in other words,
the observer looks in the direction $\psi_{obs}=0^\circ$ to see the gas).
In its own frame, such gas will absorb at the rest wavelength
of a given transition, $\lambda_{rest}$.
In the observer's frame, those absorbed photons have wavelength
\begin{equation}
\lambda_{obs}=\lambda_{rest}{1+\beta\cos\psi_{obs}\over\sqrt{1-\beta^2}}.
\end{equation}
Following \nocite{fol86}{Foltz} {et~al.} (1986), we rewrite this by defining
\begin{equation}
 R \equiv { \lambda_{rest} \over \lambda_{obs} } =
{ \sqrt{1-\beta^2} \over 1+\beta\cos\psi_{obs} } =
{ \gamma^{-1} \over 1+\beta\cos\psi_{obs} } 
\end{equation}
where $\gamma=1/\sqrt{1-\beta^2}$.
Because $\psi_{obs}>90^\circ$ refers to motion toward the observer, a negative
line-of-sight velocity $\beta\cos\psi_{obs}<0$
indicates motion toward the observer.

Redshifted absorption ($R<1$) 
can be produced by gas moving at relatively low velocity
more or less directly away from the observer ($\psi_{obs}\simeq 0^\circ$).
It can also be produced by gas moving purely in the plane of the sky
($\psi_{obs}=90^\circ$) with $\gamma=R^{-1}$.
That is the relativistic Doppler effect: due to time dilation,
gas moving transversely in the observer's frame
absorbs at a lower frequency (longer wavelength) than stationary gas
(Figure \ref{figrel}).

As there is nothing singular about $\psi_{obs}=90^\circ$,
redshifted absorption can also be produced by gas moving
in directions $\psi_{obs}>90^\circ$ (toward the observer) 
as long as $\gamma$ is large enough that in the observer's frame
the blueshifted velocity of the absorbing gas 
is counteracted by its increased time dilation.

Absorption at the rest wavelength of a transition can be produced by gas
at rest.
It cannot be produced by gas moving away from the observer.
However, absorption at the rest wavelength
can be produced by gas moving toward the observer at any $\beta$,
provided that the gas moves at an angle
\begin{equation}
\psi_{obs}^{rest}=\arccos\left( \sqrt{\beta^{-2}-1} -\beta^{-1} \right).
\end{equation}

Blueshifted absorption can only be produced by gas moving toward the observer
($\psi_{obs} > 90^\circ$).
For any given $R\geq 1$, there is a minimum $\psi_{obs}$ at which gas can be
moving and still produce absorption at $\lambda_{obs}=\lambda_{rest}/R$:
\begin{equation}
\psi_{obs}^{min}=\arccos\left( R^{-1}\sqrt{\beta^{-2}-1} -\beta^{-1} \right).
\end{equation}

However, for any $\psi_{obs}>\psi_{obs}^{min}$, there are
{\sl two} values of $\beta$ which yield absorption at the same $\lambda_{obs}$.
Consider gas moving at a given $\psi_{obs}>\psi_{obs}^{min}$
and photons with $\lambda_{obs}<\lambda_{rest}$.
For $\lambda_{obs}/\lambda_{rest}$ just below unity,
relatively low-velocity gas (for which time dilation is negligible)
can absorb those photons when its motion redshifts them
to the rest wavelength of the transition in the absorbing gas frame.
Extremely high-velocity gas can also absorb those photons because
the highly blueshifted velocity of the absorbing gas in the observer's frame
is counteracted by its large time dilation.
For smaller $\lambda_{obs}/\lambda_{rest}$, there continue to be two velocities
of $\beta$ at which absorption will occur, but time dilation is not negligible
for either of them.
Only for $\psi_{obs}=180^\circ$ does this effect vanish.

As a final aside, we note that detections of relativistic quasar outflows via
highly ionized X-ray absorption troughs, both broad and narrow, have been
reported over the past decade 
\nocite{2009ApJ...706..644C,kb10,2011ApJ...742...44T}(e.g., {Chartas} {et~al.} 2009; {Kaspi} \& {Behar} 2010; {Tombesi} {et~al.} 2011).
While the significance of narrow X-ray absorption line detections has been
questioned, in part due to their variability between different observations
\nocite{2008MNRAS.390..421V}({Vaughan} \& {Uttley} 2008), strong variability would be natural for such
high-velocity outflows and has been observed in Mrk~766 
\nocite{2011MNRAS.410.1027R}({Risaliti} {et~al.} 2011).  Furthermore, an additional 
potential source of variability is that if the outflow varies in $\psi$
even at fixed $\beta$, $\lambda_{obs}$ will vary.  Lastly, a given ratio
of observed energy to rest energy ($E/E_{rest}$)
for such an outflow yields a maximum angle it can have to the line
of sight, which may be a useful constraint for models of such outflows.

\label{lastpage}
\end{document}